\documentclass[11pt,a4paper]{article}
\pdfoutput=1
\usepackage{jheppub}

\usepackage{amsmath,epsfig}
\usepackage{mathrsfs}
\usepackage{amssymb}
\usepackage{mathtools}

\usepackage{graphics}
\usepackage{cancel}
\usepackage{slashed}
\usepackage{url}
\usepackage{hyperref}
\hypersetup{colorlinks,bookmarksopen,bookmarksnumbered,citecolor=blue,linkcolor=blue,pdfstartview=FitH,urlcolor=blue}
\usepackage[normalem]{ulem}
\usepackage{xspace}
\usepackage{dsfont}

\usepackage{multirow}
\usepackage{color}
\usepackage{lscape}
\usepackage{rotating}
\usepackage{float}

\renewcommand{\arraystretch}{1.05}

\usepackage[toc,page]{appendix}

\usepackage{etoolbox}
\appto\appendix{\addtocontents{toc}{\protect\setcounter{tocdepth}{1}}}

\appto\listoffigures{\addtocontents{lof}{\protect\setcounter{tocdepth}{1}}}
\appto\listoftables{\addtocontents{lot}{\protect\setcounter{tocdepth}{1}}}

\definecolor{OliveGreen}{rgb}{0,0.6,0}
\definecolor{morado}{rgb}{0.4706,0.2510,0.5882}

\preprint{
\vspace{-8mm}
\begin{flushright}
IFIC/20-35\\ IFT-UAM/CSIC-20-98\\ FTUAM-20-11\\ FERMILAB-PUB-20-269-ND
\end{flushright}} 

\title{GeV-scale neutrinos: interactions with mesons and DUNE sensitivity}

\vspace{1.0cm}

\author[a,b,1]{Pilar Coloma,}
\note{pilar.coloma@ift.csic.es}
\author[b,c,2]{Enrique Fern\'andez-Mart\'inez,}
\note{enrique.fernandez-martinez@uam.es}
\author[b,c,3]{Manuel Gonz\'alez-L\'opez,}
\note{manuel.gonzalezl@uam.es}
\author[d,4]{\\ Josu Hern\'andez-Garc\'ia,}
\note{garcia.josu.hernandez@ttk.elte.hu}
\author[e,5]{Zarko Pavlovic}
\note{zarko@fnal.gov}

\affiliation[a]{Instituto de F\'{\i}sica Corpuscular, Universidad de Valencia \& CSIC, 
 Edificio Institutos Investigaci\'on, Catedr\'atico Jos\'e Beltr\'an 2, 46980 Spain}
 \affiliation[b]{Instituto de F\'{\i}sica Te\'orica, Universidad Aut\'onoma de Madrid \& CSIC, 
Campus de Cantoblanco, 28049 Madrid, Spain}
\affiliation[c]{Departamento de F\'isica Te\'orica, Universidad Aut\'onoma de Madrid, 
Campus de Cantoblanco, 28049 Madrid, Spain}
\affiliation[d]{Institute for Theoretical Physics, ELTE E\"otv\"os Lor\'and University, P\'azm\'any P\'eter s\'et\'any 1/A, H-1117 Budapest, Hungary}
\affiliation[e]{Fermi National Accelerator Laboratory, Batavia, IL 60510}

\abstract{
The simplest extension of the SM to account for the observed neutrino masses and mixings is the addition of at least two singlet fermions (or right-handed neutrinos). If their masses lie at or below the GeV scale, such new fermions would be produced in meson decays. Similarly, provided they are sufficiently heavy, their decay channels may involve mesons in the final state. 
Although the couplings between mesons and heavy neutrinos have been computed previously, significant discrepancies can be found in the literature. The aim of this paper is to clarify such discrepancies and provide consistent expressions for all relevant effective operators involving mesons with masses up to 2 GeV. Moreover, the effective Lagrangians obtained for both the Dirac and Majorana scenarios are made publicly available as FeynRules models so that fully differential event distributions can be easily simulated. As an application of our setup, we numerically compute the expected sensitivity of the DUNE near detector to these heavy neutral leptons.}

\begin{document}

\maketitle

\section{Introduction}
\label{sec:intro}

The evidence for neutrino masses and mixings from the neutrino oscillation phenomenon demands an extension of the Standard Model (SM) of particle physics so as to accommodate the experimental results. Arguably, the simplest of such extensions is to add fermion singlets to the SM particle content. Indeed, the inclusion of these right-handed neutrinos would make the neutrino sector equivalent to its charged-lepton counterpart and allow for neutrino Yukawa couplings in complete analogy to the other fermions of the SM. However, being complete singlets of the SM gauge group, the novel and distinct option of a Majorana mass term is also open for them. 
This Majorana mass term would not only include a new source of particle number violation, possibly related to the origin of the observed baryon asymmetry of the Universe (BAU), but also include a new energy scale in the Lagrangian, not connected with electroweak symmetry breaking. As such, there is no solid theoretical guideline for the value of this new physics scale. 

An attractive possibility is that the Majorana mass scale is much larger than the electroweak scale, possibly close to the Grand Unification scale, leading to the celebrated type-I Seesaw mechanism~\cite{Minkowski:1977sc,Mohapatra:1979ia,Yanagida:1979as,GellMann:1980vs}. Its most appealing feature is that the smallness of neutrino masses is very naturally explained even with order one Yukawa couplings, since it is inversely proportional to the large Majorana mass of the right-handed neutrinos. Furthermore, the Seesaw mechanism, unlike the SM, is also able to account for the observed BAU via leptogenesis~\cite{Fukugita:1986hr}. Nevertheless, while a very high Majorana mass scale can very naturally accommodate the extreme lightness of neutrino masses, its presence would significantly destabilize the Higgs mass, worsening the Higgs hierarchy problem~\cite{Vissani:1997ys,Casas:2004gh}. Thus, while naturalness arguments favour large Majorana masses to explain the light neutrino masses, lighter scales are instead preferred to accommodate the observed Higgs mass. 

Other variants of the original Seesaw mechanism, such as the inverse~\cite{Mohapatra:1986aw,Mohapatra:1986bd,Bernabeu:1987gr} or linear~\cite{Malinsky:2005bi} Seesaws, naturally explain the lightness of neutrino masses through an approximate lepton number symmetry~\cite{Branco:1988ex,Kersten:2007vk, Abada:2007ux} instead. Thus, they can be realized at lower energy scales without introducing a Higgs hierarchy problem. Low-scale versions of the leptogenesis mechanism are also found to successfully account for the observed BAU~\cite{Akhmedov:1998qx,Asaka:2005pn,Shaposhnikov:2008pf}. Thus, the phenomenology associated to all possible options for the Majorana mass scale should be investigated and compared with experimental observations so as to probe the new physics underlying the observation of neutrino masses and mixings.

The main consequence of lowering the Majorana mass scale in a Seesaw mechanism is that mainly-sterile neutrinos or ``heavy neutral leptons'' (HNLs) appear in the particle spectrum. If sufficiently light, these HNLs will be kinematically accessible to experiments and can thus be produced and searched for. Given the singlet nature of the right-handed neutrinos, their only interactions  are the weak ones, inherited from their left-handed neutrino counterparts via mixing. Thus, the mixing of the HNLs with the electron, muon and tau neutrinos can be probed and constrained as a function of the HNL mass by searching for their production and decay in association with the corresponding charged leptons. These searches range from studying their impact in neutrino oscillations~\cite{FernandezMartinez:2007ms,Antusch:2009pm,Parke:2015goa,Miranda:2016wdr,Ge:2016xya,Blennow:2016jkn,Escrihuela:2016ube,Kosmas:2017zbh,Miranda:2020syh} when they are too light to decay visibly, to collider signals~\cite{delAguila:2008cj,Atre:2009rg,Antusch:2015mia,Deppisch:2015qwa,Antusch:2016ejd,Das:2017nvm,Das:2017zjc,Cai:2017mow,Das:2018hph,Dev:2018kpa,Pascoli:2018heg,Liu:2019qfa} for the highest accessible HNL masses. For even higher masses, their mixing can still be constrained through deviations of unitarity of the PMNS matrix in flavor and electroweak precision observables~\cite{Shrock:1980ct,Shrock:1981wq,Langacker:1988ur,Tommasini:1995ii,Antusch:2006vwa,Antusch:2008tz,Forero:2011pc,Antusch:2014woa,Fernandez-Martinez:2016lgt,Coutinho:2019aiy}. 

For intermediate HNL masses $M$, between the MeV and GeV scales, searches at beam dump experiments or near detectors of neutrino oscillation facilities, where they can be produced via meson decays and detected through their visible decays, can set very stringent constraints \cite{Atre:2009rg,Abada:2007ux,Gorbunov:2007ak,Abada:2016plb,Abada:2018nio,Abada:2018sfh,Ballett:2019bgd,Berryman:2019dme,Abada:2019bac,Krasnov:2019kdc,Bryman:2019ssi,Bryman:2019bjg,Bondarenko:2019yob, Drewes:2015iva, Chrzaszcz:2019inj,Gorbunov:2020rjx}. Indeed, current bounds are even getting near the expectation for the ``vanilla'' type-I Seesaw without a lepton number symmetry protection of the light neutrino masses $m_\nu$, where the mixing scales as $\theta^2 \sim m_\nu/M$. Furthermore, the masses and mixings leading to successful generation of the BAU via low-scale leptogenesis are also accessible through these searches~\cite{Hernandez:2016kel,Abada:2018oly,Ghiglieri:2019kbw}. In this regime, both the production and decay of the HNL depend crucially on its interactions with mesons. While these have been studied previously, significant discrepancies can be found in the literature~\cite{Gorbunov:2007ak,Atre:2009rg,Bondarenko:2018ptm,Ballett:2019bgd} in the branching ratios of the relevant channels. The aim of this work is to clarify such discrepancies and provide a tool for these important searches. With that goal in mind, we derive the effective theory description of the HNL interactions, with particular emphasis on the effective operators involving mesons, which control HNL production and decay via leptonic and semileptonic processes. We do this both for a Majorana HNL as well as for the Dirac scenario, motivated by the inverse and linear Seesaw variants. Furthermore, our results have been collected in two FeynRules~\cite{Alloul:2013bka} models that have been made publicly available (see ancillary files) so that not only the total branching ratios can be computed, but also differential event distributions can be easily simulated by interfacing the output of FeynRules with event generators such as MadGraph5~\cite{Alwall:2014hca}. Finally, while the present work focuses on the low-energy theory, our FeynRules implementation is more general and includes an option to replace all mesons with quarks, so they may also be used to study HNL phenomenology in collider searches at higher energies. 

As an application of our framework, we compute the expected flux of HNLs at the proposed DUNE~\cite{Ballett:2019bgd,Berryman:2019dme} near detector, and compare our full numerical simulation with the approximation of rescaling the massless neutrino fluxes. A significant enhancement due to the larger boost in the beam direction for the HNLs is found. We also compute the expected number of decays inside the DUNE near detector into several decay channels, and use that to estimate the sensitivity of DUNE to the HNL mixing with the charged leptons.

This paper is organized as follows. In Section \ref{sec:full_L} we introduce the Seesaw Lagrangians, both in the Majorana and Dirac cases, and review the weak interactions that the HNL will inherit from the left-handed neutrinos via mixing. In Section \ref{sec:eft} we concentrate on the meson interactions and derive all the relevant effective operators containing HNLs. In Sections \ref{sec:prod} and \ref{sec:decays} we summarize all the relevant production channels and subsequent decays of the HNLs. In Section \ref{sec:dune} we present our results for the expected HNL fluxes at the DUNE near detector together with an estimate of its sensitivity. Finally, in Section \ref{sec:conclusions} we draw our conclusions and summarize the results. 

\section{The full Lagrangian of the theory at high energies}
\label{sec:full_L}

Once the SM is extended with $n$ extra right-handed neutrinos $N_{R}$, Lorentz and gauge invariance allow the inclusion of Yukawa couplings to the lepton doublets ($Y_\nu$) as well as Majorana masses for the heavy singlets ($M$). In the basis where the Majorana mass terms are diagonal, the corresponding Lagrangian reads:
\begin{equation}
\mathcal{L}_{\nu}^{\textrm{mass}} \supset - \sum_{\alpha=e,\mu,\tau} \sum_{j=1}^n Y_{\nu, \alpha j} \overline{L}_{L,\alpha} \tilde{\phi} N_{R,j} - \frac{1}{2} \sum_{j=1}^n M_{j} \overline{N}_{R,j} N^c_{R,j} + \textrm{h.c.} \, ,
\label{eq:type1-lagrangian}
\end{equation}
where $L_{L,\alpha}$ stands for the SM left-handed lepton doublet of flavor $\alpha$, $\phi$ is the Higgs field, $\tilde{\phi} = i \sigma_2 \phi^*$ and $N^c_{R,j} \equiv C \bar N_{R,j}^t$, with $C = i \gamma_0 \gamma_2$ in the Weyl representation we adopt. Once the Higgs develops its vacuum expectation value $v/\sqrt{2}$ upon electroweak (EW) symmetry breaking, the full neutrino mass matrix in the basis $(\nu_L, N^c_R)$ can be written in blocks as:
\begin{equation}
\mathcal{M} = \left( \begin{array}{cc}
\mathbf{0}_{3\times 3} & Y_\nu v/\sqrt{2} \\
Y_\nu^t v/\sqrt{2} & M
\end{array} \right) \, .
\end{equation}
The full unitary rotation $U$ that diagonalizes the mass matrix will have dimensions $(3+n) \times (3+n)$. Neutrino masses are obtained upon diagonalization, as well as the mixing between the active SM neutrinos and the new heavy states introduced. In particular, the spectrum is composed of 3 light ``SM-like'' neutrino mass eigenstates ($\nu_i$), and $n$ heavier and mostly sterile neutrinos ($N_i$). 

Alternatively, and motivated by low-energy Seesaw realizations such as the inverse~\cite{Mohapatra:1986aw,Mohapatra:1986bd,Bernabeu:1987gr} or linear~\cite{Malinsky:2005bi} versions, we will also consider the case in which the extra sterile neutrinos have Dirac (or pseudo-Dirac) masses. In these scenarios, $2n$ extra singlets are added in Dirac pairs $N_{L,j}$, $N_{R,j}$ ($j=1,\dots n$). Neglecting the small lepton-number violating terms (that would eventually source the light neutrino masses), we are left with the following Lagrangian:
\begin{equation}
\mathcal{L}_{\nu}^{\textrm{mass}} \supset - \sum_{\alpha=e,\mu,\tau} \sum_{j=1}^n Y_{\nu, \alpha j} \overline{L}_{L,\alpha} \tilde{\phi} N_{R,j} - \sum_{j=1}^n M_j \overline{N}_{L,j} N_{R,j} + \textrm{h.c.} \,
\label{eq:inverse-lagrangian}
\end{equation}
In this case, the mass matrix in the basis $(\nu_L, N^c_R, N_L)$ would be given by:
\begin{equation}
\mathcal{M} = \left( \begin{array}{ccc}
\mathbf{0}_{3\times 3} & Y_\nu v/\sqrt{2} & \mathbf{0}_{3\times n} \\
Y_\nu^t v/\sqrt{2} & \mathbf{0}_{n\times n} & M \\
\mathbf{0}_{n\times 3} & M & \mathbf{0}_{n\times n}
\end{array} \right) \, .
\end{equation}

Regardless of the Dirac or Majorana character of the heavy neutrinos, the flavor states will thus correspond to a combination of the light and heavy states:
\begin{equation}
\nu_\alpha = \sum_{i=1}^3 U_{\alpha i} \nu_i + \sum_{i=4}^{3+n} U_{\alpha i} N_i  \equiv \sum_i U_{\alpha i} n_i   \, ,
\end{equation}
where we have introduced the mass eigenbasis $n = (\nu , N)$ with index $i$ that runs over the light and heavy mass eigenstates.
The leptonic part of the electroweak Lagrangian can be written as:
\begin{eqnarray}
\mathcal{L}^{\ell}_{\mathrm{EW}} 
& = & \frac{g}{\sqrt{2}}W^+_\mu \sum_{\alpha} \sum_{i} U_{\alpha i}^* \bar n_{i}\gamma^\mu P_L \ell_\alpha + \nonumber \\ 
& + &\frac{g}{4c_w} Z_\mu  \left\lbrace \sum_{i, j} C_{ij}\bar n_{i}\gamma^\mu P_L n_j + \sum_\alpha \bar{\ell}_\alpha\gamma^\mu \left[2s_w^2P_R-(1-2s_w^2)P_L \right]  \ell_\alpha \right\rbrace + \mathrm{h.c.} \,,\nonumber\\
\label{eq:neutrino-currents} 
\end{eqnarray}
where $P_L$ and $P_R$ are respectively the left and right projectors, $c_w \equiv \cos\theta_w$, $s_w \equiv \sin\theta_w$ ($\theta_w$ being the SM weak mixing angle), and 
\begin{equation}
C_{ij} \equiv \sum_\alpha U^*_{\alpha i} U_{\alpha j} \, .
\end{equation} 

The heavy neutrinos can also interact with the quark sector through the charged and neutral current interactions. Thus, the corresponding weak interactions between quarks are reviewed below for convenience: 
\begin{equation}
{\mathcal{L}}^q_{\mathrm{EW}} = \frac{g}{\sqrt{2}} W^{+ \mu} j_{W, \mu}  + \frac{g}{4 c_w} Z^\mu j_{Z, \mu}  + \mathrm{h.c.} \,,
\label{eq:quark_interactions}
\end{equation}
with
\begin{equation}
j_{Z,\mu}  =  \sum_q \bar q \gamma_\mu (T_3^q - 2 Q^q s_w^2) q + \sum_q \bar q \gamma_\mu \gamma_5 (-T_3^q) q \,,
\label{eq:Z-current}
\end{equation}
and 
\begin{equation}
j_{W,\mu} =   \sum_{q=u,c,t} \sum_{q'=d,s,b} V_{q q' } \bar{q} \gamma_\mu P_L q' \,.
\label{eq:W-current}
\end{equation}
Here, $Q^q$ and $T_3^q$ stand for the electric charge and the isospin of quark $q$ in the interaction vertex (from now on the index $q$ will be dropped for simplicity), and $V_{q q'}$ is the corresponding element of the CKM mixing matrix.

Finally, for the derivation of the effective theory in Sec.~\ref{sec:eft} it will be useful to separate both currents in their vector and axial parts. This way, the $Z$ current can be decomposed as
\begin{equation}
j_{Z,\mu}=j_{Z,\mu}^V+j_{Z,\mu}^A \, ,
\label{eq:Z_VA_current}
\end{equation}
with 
\begin{eqnarray}
j_{Z,\mu}^{V} & = & \sum_q \bar{q}   (T_3 - 2Q s_w^2) \gamma_\mu q  \, ,\label{eq:Z-V} \\
j_{Z,\mu}^{A} & = & - \sum_q \bar q \gamma_\mu \gamma_5 T_3 q \,. \label{eq:Z-A}
\end{eqnarray}
Analogously, the $W$ current may be written as
\begin{equation}
j_{W,\mu}=j_{W,\mu}^{V}+j_{W,\mu}^{A}\,,
\label{eq:W_VA_current}
\end{equation}
where its vector and axial parts are given by 
\begin{eqnarray}
 j_{W,\mu}^{V} & =&  \frac{1}{2} \sum_{q=u,c,t} \sum_{q'=d,s,b} V_{q q'}\bar{q} \gamma_\mu  q'  \, ,  \label{eq:W-V}\\
 j_{W,\mu}^{A} & = & - \frac{1}{2} \sum_{q=u,c,t} \sum_{q'=d,s,b} V_{q q'} \bar{q} \gamma_\mu \gamma_5 q' \,. \label{eq:W-A}
\end{eqnarray}

\section{Effective low-energy Lagrangian including mesons}
\label{sec:eft}

In order to compute the production of the heavy neutrinos through meson decays, as well as neutrino decays to lighter mesons, we need to introduce effective interactions between the neutrino and meson fields. 
In this section we derive such interactions, integrating out the $W$ and $Z$ bosons and introducing the relevant meson decay constants and hadronic matrix elements. We compute the amplitudes for low-energy processes involving these vertices, so as to extract the corresponding effective operators. Moreover, FeynRules~\cite{Alloul:2013bka} models with these effective interactions have been made publicly available (as ancillary files to this work), making possible the generation of fully differential event distributions. Note that, while the formalism used in this section is applicable to any number of extra heavy states, only one heavy neutrino has been included in the FeynRules model files for the sake of simplicity. Although the introduction of just one heavy neutrino cannot explain the measured neutrino masses and mixing parameters, such simplified models are useful to study the phenomenology of HNLs, since it will be dominated by the lightest of the extra states.

As a first step, we review the relevant decay constants and matrix elements, and introduce our notation. Throughout this section, the formalism we use is suitable for mesons with masses up to approximately 1 GeV. However, leptonic and semileptonic decays of heavier charmed mesons can constitute a dominant contribution for heavy neutrino production, depending on its mass. Such processes will also be considered here, for the channels with a significant branching ratio into neutrinos (e.g., $D_s \to N \ell$). For even higher neutrino masses (produced typically at collider experiments), a perturbative description of the neutrino decay into quark-antiquark pairs (with subsequent hadronization) would be more suitable. 
 
We adopt a definition of the meson decay constants such that $f_\pi =130$~MeV, namely:
\begin{eqnarray}
\langle 0 | j_{a, \mu}^{A} | P_b \rangle & = & i \delta_{ab} \frac{f_P}{\sqrt{2}} p_\mu \, , \label{eq:fP}   \\
\langle 0 | j_{a, \mu}^{V} | V_b \rangle & = & \delta_{ab} \frac{f_V}{\sqrt{2}} \epsilon_\mu \, , \label{eq:fV}
\end{eqnarray}
for the pseudoscalar ($P$) and vector ($V$) mesons respectively. Here, $p_\mu$ stands for the momentum of the pseudoscalar meson and $\epsilon_\mu$ for the polarization of the vector meson (note that, with this definition, the decay constants $f_V$ have units of $[E]^2$). The corresponding currents $j_{a,\mu}^{A}, j_{a,\mu}^{V}$ are defined as:
\begin{eqnarray}
j_{a, \mu}^{A} & = &  \bar{q} \lambda_a \gamma_\mu \gamma_5 q , \label{eq:axial-currents}\\
j_{a, \mu}^{V} & = &  \bar{q} \lambda_a \gamma_\mu q , \label{eq:vector-currents}
\end{eqnarray}
where
\begin{equation}
q \equiv \left( \begin{array}{c} u \\ d \\ s \end{array} \right) .
\end{equation}
In this notation, the set $\{\lambda_a \}$ corresponds to linear combinations of the eight Gell-Mann matrices (generators of $SU(3)$) plus the identity, normalized such that
\begin{equation}
\label{eq:traces}
{\rm Tr}\left\lbrace \lambda_a \lambda_b \right\rbrace = \frac{\delta_{ab}}{2} \, .
\end{equation}
For convenience, explicit expressions for the generators are provided in Appendix~\ref{app:generators}, while the decay constants most relevant for the effective couplings considered in this work are summarized in Tab.~\ref{tab:Fpi}. 
\begin{table}[htb!]
\renewcommand{\arraystretch}{1.8}
\begin{tabular}[t]{ | c|c | c|c | }
\hline
\multicolumn{2}{|c |}{\textbf{Pseudoscalars}} & \multicolumn{2}{c| }{\textbf{Vectors}}  \\ \hline \hline
$f_{\pi}$ & 0.130 GeV & $f_{\rho}$ & 0.171 GeV$^2$\\
$f_{K}$ &  0.156 GeV     & $f_{\omega}$ & 0.155 GeV$^2$\\
$f_{D}$ &   0.212 GeV   & $f_{\phi}$ &0.232 GeV$^2$\\
$f_{D_s}$ &  0.249 GeV & $f_{K^*}$ & 0.178 GeV$^2$\\\hline
\end{tabular}
\hspace{0.1\textwidth}
\begin{tabular}[t]{ | c|c | c|c | }
\hline
\multicolumn{2}{|c |}{\textbf{Decay constants}} & \multicolumn{2}{c| }{\textbf{Rotation angles}}  \\ \hline \hline
$f_0$ & 0.148 GeV & $\theta_0$ & -6.9$^\circ$\\
$f_8$ &  0.165 GeV    & $\theta_8$ & -21.2$^\circ$\\
\hline
\end{tabular}
\caption{\label{tab:Fpi} \textit{Left.} Decay constants for pseudoscalar and vector mesons, defined as in Eqs.~\eqref{eq:fP} and~\eqref{eq:fV}. The pseudoscalar decay constants are directly taken from Ref.~\cite{Tanabashi:2018oca}, while those for vector mesons have been computed as described in Appendix~\ref{app:decayconstants}. \textit{Right.} Decay constants for the $\eta_0$ and $\eta_8$, and angles that parametrize the rotation to the physical basis, taken from Ref.~\cite{Escribano:2015yup} (see text for details). Note that in Ref.~\cite{Escribano:2015yup} the authors use a different normalization for the current definitions than the one adopted in this work. However, this does not affect our result since they provide their results in terms of the ratios $f_8 /f_\pi$ and $f_0/f_\pi$, which remain unaffected by an overall normalization factor.}
\end{table}

\subsection{Pseudoscalar mesons}

\subsubsection{Neutral mesons: $\pi^0, \eta, \eta^\prime$}
\label{sec:neutral-pseudo}

The quark content of the neutral pseudoscalar mesons will correspond to linear combinations of the diagonal generators $\lambda_0, \lambda_3$ and $\lambda_8$. Substituting the explicit expressions for the generators into Eq.~\eqref{eq:axial-currents} we obtain:
\begin{eqnarray}
j_{3, \mu}^{A } & = & \frac{1}{2} \left[ \bar u \gamma_\mu \gamma_5 u - \bar d \gamma_\mu \gamma_5 d \right]  , \nonumber \\
j_{8, \mu}^{A } & = & \frac{1}{2\sqrt{3}} \left[ \bar u \gamma_\mu \gamma_5 u + \bar d \gamma_\mu \gamma_5 d - 2 \bar s \gamma_\mu \gamma_5 s \right]  , \label{eq:pseudoscalar-mesons} \\
j_{0, \mu}^{A } & = & \frac{1}{\sqrt{6}} \left[ \bar u \gamma_\mu \gamma_5 u + \bar d \gamma_\mu \gamma_5 d 
+\bar s \gamma_\mu \gamma_5 s \right]  . \nonumber 
\end{eqnarray}
The neutral pion can be directly identified with the current $j_{3,\mu}^{A}$, being the neutral member of the SU(2) triplet of pseudo-Goldstone bosons from the flavor symmetry between up- and down-quarks. Conversely, the $\eta$ and $\eta'$ mainly correspond to the currents $j_{8,\mu}^{A}$ and $j_{0,\mu}^{A}$ respectively, although with significant mixing among them, as discussed in detail below. 

These neutral mesons can be produced or decay through neutral current interactions mediated by the $Z$ boson. Thus, in order to obtain their effective interactions with neutrinos we start from the Fermi theory after integrating out the $Z$, inserting the decay constant of the corresponding meson. The $Z$ axial current in Eq.~\eqref{eq:Z-A} can be expressed as a linear combination of the neutral axial currents as:
\begin{equation}
j_{Z, \mu}^{A} = 
-\frac{1}{2}\left( \bar u \gamma_\mu \gamma_5 u - \bar d \gamma_\mu \gamma_5 d - \bar s \gamma_\mu \gamma_5 s \right) =
-\left(  j_{3, \mu}^A + \frac{1}{\sqrt{3}}j_{8, \mu}^A - \frac{1}{\sqrt{6}}j_{0, \mu}^{A}\right)  \, .
\label{eq:axial-Z}
\end{equation}
At low energies, the amplitude, for example, for $\pi^0 \rightarrow  n_i \bar{n}_j$ would read:
\begin{equation}
i\mathcal{M}_{\pi^0 n_i \bar{n}_j} = \frac{ig^2}{4c_w^2M_Z^2}C_{ij}\bar{u}_i\gamma^\mu P_L v_j \langle 0 | j_{Z, \mu}^A | \pi^0 \rangle \,,
\end{equation}
where  $\bar{u}_i$ and $v_j$ are the corresponding spinors for the neutrino mass eigenstates. Substituting the $Z$ current from Eq.~\eqref{eq:axial-Z} and the corresponding hadronic matrix element from Eq.~\eqref{eq:fP}, and introducing Fermi's constant, 
\begin{equation}
\frac{G_F}{\sqrt{2}} =\frac{ g^2 }{8c_w^2 M_Z^2} \, , 
\end{equation}
the amplitude is given by:
\begin{equation}
i\mathcal{M}_{\pi^0 n_i \bar{n}_j} = G_F C_{ij} f_\pi\bar{u}_i\gamma^\mu P_L v_j p_\mu \,,
\label{eq:amppi}
\end{equation}
where $p_\mu$ is the 4-momentum carried by the pion. Translating the momentum into a derivative, we can write down, in configuration space, the effective operator that leads to the amplitude in Eq.~(\ref{eq:amppi}):
\begin{equation}
\mathcal{O}_{\pi^0 n_i \bar{n}_j} =\frac{1}{2}G_F C_{ij} f_\pi \partial_\mu  (\bar n_{i}\gamma^\mu P_L n_j) \pi^0 + \textrm{h.c.}\,
\end{equation}
Furthermore, if all particles are on-shell, it is possible to apply Dirac's equation to obtain Yukawa couplings proportional to the neutrino masses: 
\begin{eqnarray}
\mathcal{O}_{\pi^0 n_i \bar{n}_j} =  \frac{i}{2} G_F C_{ij} f_\pi \bar n_i (m_i P_L - m_j P_R ) n_j \pi^0 + \textrm{h.c.}\,
\end{eqnarray}
Since the coupling is proportional to the masses of the neutrinos, the coupling to the heavy states will dominate the interaction, in complete analogy to the chiral enhancement of the charged pion decay $ \pi \to \mu  \nu_\mu$ versus $\pi \to e \nu_e$.

Similarly, the operators associated to the other neutral pseudoscalar currents, for on-shell particles, can be obtained as:
\begin{eqnarray}
\mathcal{O}_{\eta_0 n_i \bar{n}_j} = -\frac{i}{2}G_F  C_{ij} \frac{f_0}{\sqrt{6}} \bar n_i (m_i P_L - m_j P_R ) n_j \eta_0 + \textrm{h.c.}\, , \\
\mathcal{O}_{\eta_8 n_i \bar{n}_j} = \frac{i}{2}G_F  C_{ij} \frac{f_8}{\sqrt{3}} \bar n_i (m_i P_L - m_j P_R ) n_j \eta_8 + \textrm{h.c.} 
\end{eqnarray}
However, unlike in the $\pi^0$ case, the $\eta$ and $\eta^\prime$ mesons mix significantly and do not correspond exactly with the quark content of the $\eta_8$ and $\eta_0$ defined through the corresponding currents in Eq.~\eqref{eq:pseudoscalar-mesons}. Thus, a change of basis must be performed in order to obtain the effective vertices for the physical states. We adopt the usual parametrization for this change of basis, with two angles, $\theta_0$ and $\theta_8$ (see e.g. Ref.~\cite{Escribano:2015yup}), and define:
\begin{equation}
\left(
\begin{array}{cc}
f_{\eta,8} & f_{\eta,0} \\
f_{\eta^\prime,8} & f_{\eta^\prime,0} 
\end{array} \right) 
=
\left(
\begin{array}{cc}
f_8 \cos\theta_8 & - f_0 \sin\theta_0 \\
f_8 \sin\theta_8 & f_0 \cos\theta_0 
\end{array} \right) \, .
\end{equation}
The values for $f_0, f_8, \theta_0$ and $\theta_8$ have been taken from Ref.~\cite{Escribano:2015yup} and are summarized in Tab.~\ref{tab:Fpi} for convenience. Through this change of basis, the currents for the $\eta$ and $\eta'$ can be obtained as combinations of the $j_{0,\mu}^A,j_{8,\mu}^A$ currents as
\begin{eqnarray}
j_{\eta, \mu}^A= \cos\theta_8j_{8, \mu}^A -  \sin\theta_0 j_{0, \mu}^A \, , \\
j_{\eta^\prime, \mu}^A= \sin\theta_8 j_{8, \mu}^A  + \cos\theta_0 j_{0, \mu}^A \, .
\end{eqnarray}
Therefore, the relevant operators in the mass basis will read 
\begin{eqnarray}
\mathcal{O}_{\eta n_i \bar{n}_j} & = &
 \frac{i}{2} G_F C_{ij} \left[ \frac{\cos\theta_8 f_8}{\sqrt{3}} + 
\frac{ \sin\theta_0 f_0}{\sqrt{6}} \right] \bar n_i (m_i P_L - m_j P_R ) n_j \eta + \textrm{h.c.} \,  , \label{eq:eta_coupling}\\
\mathcal{O}_{\eta^\prime n_i \bar{n}_j} & = &
 \frac{i}{2} G_F C_{ij} \left[ \frac{\sin\theta_8 f_8}{\sqrt{3}} - 
\frac{ \cos\theta_0 f_0}{\sqrt{6}} \right] \bar n_i (m_i P_L - m_j P_R ) n_j \eta^\prime + \textrm{h.c.} \, \label{eq:etap_coupling} 
\end{eqnarray}

\subsubsection{Charged mesons: $\pi^\pm, K^\pm, D^\pm, D^\pm_s$}

The normalized combinations of generators that reproduce the quark content of the $\pi^\pm$ and $K^\pm$ are:
\begin{eqnarray}
j_{\pi^\pm, \mu}^{A} = \frac{1}{\sqrt{2}} \bar q \gamma_\mu \gamma_5 (\lambda_1 \mp i \lambda_2) q \, , \\
j_{K^\pm, \mu}^{A} = \frac{1}{\sqrt{2}} \bar q \gamma_\mu \gamma_5 (\lambda_4 \mp i \lambda_5) q  \, .
\end{eqnarray}
Thus, from Eq.~\eqref{eq:W-A} we get that 
\begin{equation}
j_{W, \mu}^{A} = -\frac{1}{\sqrt{2}}\left( V_{ud} \, j_{\pi^-, \mu}^{A} + V_{us} \, j_{K^-, \mu}^{A} \right) \, .
\label{eq:Wcurrent}
\end{equation}

The amplitude for $\pi^- \to \ell^- \bar{n}$ is obtained after integrating out the $W$ boson, following the same procedure used to derive the effective vertex for the $\pi^0 \to \bar{n} n$ decay in the previous section:
\begin{equation}
i\mathcal{M}_{\pi \ell_\alpha \bar{n}_i } = \frac{ig^2}{2M_W^2}U_{\alpha i}\bar{u}_\alpha \gamma^\mu P_L v_i \langle 0 | j_{W, \mu}^A | \pi^- \rangle \, .
\end{equation}
After introducing the $W$ current defined in Eq.~\eqref{eq:Wcurrent} and evaluating the hadronic matrix element, the amplitude reads:
\begin{equation}
i\mathcal{M}_{\pi \ell_\alpha \bar{n}_i } = \sqrt{2}G_F  U_{\alpha i}V_{ud} f_\pi  \bar{u}_\alpha \gamma^\mu P_L v_i p_\mu\, .
\end{equation}
In the same fashion as before, we translate this amplitude to an effective operator in configuration space, with the 4-momentum $p_\mu$ as a derivative acting on the leptonic current:
\begin{equation}
\mathcal{O}_{\pi \ell_\alpha \bar{n}_i } = \sqrt{2} G_F U_{\alpha i}  V_{ud}  f_\pi \partial_\mu  (\bar \ell_\alpha  \gamma^\mu P_L n_i ) \pi^- + \textrm{h.c.}
\end{equation}
Once again, if all the particles involved are on-shell, it is possible to obtain Yukawa couplings proportional to the fermion masses via Dirac's equation:
\begin{equation}
\mathcal{O}_{\pi \ell_\alpha \bar{n}_i } =  i \sqrt{2} G_F U_{\alpha i} V_{ud} f_\pi \bar \ell_\alpha  (m_\alpha P_L - m_i P_R) n_i  \pi^- + \textrm{h.c.} \, 
\label{eq:effective-pion}
\end{equation}
This procedure can be repeated for the charged kaons, obtaining the same result once the corresponding decay constant and CKM element are introduced:
\begin{equation}
\mathcal{O}_{K \ell_\alpha \bar{n}_i} =  i \sqrt{2} G_F U_{\alpha i} V_{us} f_K \bar \ell_\alpha  (m_\alpha P_L - m_i P_R) n_i  K^- + \textrm{h.c.}\, 
\label{eq:effective-kaon}
\end{equation}
So far we have restricted ourselves to mesons which contain only the three lightest quark flavors. Nevertheless, these results can be generalized to the $D^\pm$ and $D_s^\pm$ mesons. The corresponding effective operators read:
\begin{equation}
\mathcal{O}_{D \ell_\alpha \bar{n}_i} =  i \sqrt{2} G_F U_{\alpha i} V_{cd} f_D \bar \ell_\alpha  (m_\alpha P_L - m_i P_R) n_i  D^- + \textrm{h.c.}\, ,
\end{equation}
\begin{equation}
\mathcal{O}_{D_s \ell_\alpha \bar{n}_i} =  i \sqrt{2} G_F U_{\alpha i} V_{cs} f_{D_s} \bar \ell_\alpha  (m_\alpha P_L - m_i P_R) n_i  D_s^- + \textrm{h.c.}\, 
\end{equation}
\subsection{Vector mesons}

\subsubsection{Neutral mesons: $\rho, \omega, \phi $}
As for the pseudoscalar case, the vector currents associated to the generators can be expressed in terms of the $u$, $d$ and $s$ quarks as
\begin{eqnarray}
j_{3, \mu}^{V} & = & \frac{1}{2} \left[ \bar u \gamma_\mu u - \bar d \gamma_\mu d \right]  , \nonumber \\
j_{8, \mu}^{V} & = & \frac{1}{2\sqrt{3}} \left[ \bar u \gamma_\mu u + \bar d \gamma_\mu d - 2 \bar s \gamma_\mu s \right]  ,
\label{eq:vector-mesons-currents} \\
j_{0, \mu}^{V} & = & \frac{1}{\sqrt{6}} \left[ \bar u \gamma_\mu u + \bar d \gamma_\mu d 
+\bar s \gamma_\mu s \right]  . \nonumber 
\end{eqnarray}
Considering their respective quark contents, the corresponding normalized currents for the $\rho^0$, $\omega$ and $\phi$ mesons are given by: 
\begin{eqnarray}
j_{\rho^0, \mu}^{V} & = & j_{3, \mu}^{V} \, , \nonumber \\
j_{\omega, \mu}^{V} & = & \sqrt{\frac{1}{3}}j_{8, \mu}^{V} + \sqrt{\frac{2}{3}}j_{0, \mu}^{V} \, , \\
j_{\phi, \mu}^{V} & = & - \sqrt{\frac{2}{3}}j_{8, \mu}^{V} + \sqrt{\frac{1}{3}}j_{0, \mu}^{V}  \, . \nonumber 
\end{eqnarray}
The production and decay of the vector mesons take place via the vector component of the $Z$ current, Eq.~\eqref{eq:Z-V}, which can be written as the following linear combination of the vector meson currents:
\begin{equation}
\label{eq:vector-Z}
j_{Z, \mu}^{V} =
\left( 1 - 2 s_w^2\right) j_{\rho^0, \mu}^{V} 
-\frac{2}{3} s_w^2 j_{\omega, \mu}^{V}  
- \sqrt{2}\left( \frac{1}{2} - \frac{2}{3}s_w^2   \right) j_{\phi, \mu}^{V}\, .
\end{equation}
After integrating out the $Z$ boson, the amplitude for the $\rho^0\rightarrow\bar{n}n$ process reads:
\begin{equation}
i\mathcal{M}_{\rho^0 n_i \bar{n}_j}=\frac{ig^2}{4c_w^2M_Z^2}C_{ij}\bar{u}_i\gamma^\mu P_L v_j \left\langle 0\vert j_{Z,\mu}^V\vert\rho^0\right\rangle \, .
\end{equation}
Introducing the vector $Z$ current defined in Eq. \eqref{eq:vector-Z} and evaluating the matrix element according to Eq.~\eqref{eq:fV}, we get:
\begin{equation}
i\mathcal{M}_{\rho^0 n_i \bar{n}_j}=iG_FC_{ij}f_\rho\left( 1 - 2 s_w^2\right) \bar{u}_i\gamma^\mu P_L v_j \epsilon_\mu\, ,
\end{equation}
where $\epsilon_\mu$ is the polarization vector of the $\rho^0$ meson. It is then immediate to extract the effective operator in configuration space:
\begin{equation}
\mathcal{O}_{\rho^0 n_i \bar{n}_j} =  -\frac{1}{2} G_F C_{ij} (1 - 2s_w^2) f_\rho \rho^0_\mu  (\bar n_i \gamma^\mu P_L n_j) + \textrm{h.c.} \, 
\end{equation}
Analogously, for the other two neutral vector mesons we obtain:
\begin{eqnarray}
\mathcal{O}_{\omega n_i \bar{n}_j} & = & \frac{1}{2} G_F C_{ij} \frac{2}{3} s_w^2 f_\omega \omega_\mu (\bar n_i \gamma^\mu P_L n_j )  + \textrm{h.c.} \, , \\
\mathcal{O}_{\phi n_i \bar{n}_j} & = &  \frac{1}{2}G_F C_{ij} \sqrt{2}\left(\frac{1}{2} - \frac{2}{3} s_w^2\right) f_\phi \phi_\mu (\bar n_i \gamma^\mu P_L n_j )  + \textrm{h.c.} \, 
\end{eqnarray}

\subsubsection{Charged mesons: $\rho^\pm, K^{*, \pm}$}

In complete analogy to the charged pseudoscalars, the charged vector meson currents are given by:
\begin{eqnarray}
j_{\rho^\pm, \mu}^{V} = \frac{1}{\sqrt{2}} \bar q \gamma_\mu  (\lambda_1 \mp i \lambda_2) q \, , \\
j_{K^{*,\pm}, \mu}^{V}  = \frac{1}{\sqrt{2}} \bar q \gamma_\mu  (\lambda_4 \mp i \lambda_5) q  \, ,
\end{eqnarray}
and the vector component of the $W$ current from Eq.~\eqref{eq:W-V} can be written as:
\begin{equation}
j_{W, \mu}^{V} = \frac{1}{\sqrt{2}}\left( V_{ud} \, j_{\rho^-, \mu}^{V} + V_{us} \, j_{K^{*,-}, \mu}^{V}\right) \, .
\end{equation}
The computation of the effective operators is done exactly in the same way as for the charged pseudoscalar case. The amplitude for the $\rho^-\rightarrow\bar{n}\ell^-$ process reads:
\begin{equation}
i\mathcal{M}_{\rho^- \ell_\alpha \bar{n}_i}=\frac{ig^2}{2M_W^2}U_{\alpha i}\bar{u}_\alpha \gamma^\mu P_L v_i \langle 0 | j_{W, \mu}^V | \rho^- \rangle = i\sqrt{2}G_F U_{\alpha i}V_{ud}f_\rho \epsilon_\mu \bar{u}_\alpha \gamma^\mu P_L v_i \, .
\end{equation}
Thus, we finally obtain
\begin{equation}
\mathcal{O}_{\rho \ell_\alpha \bar{n}_i}  = -\sqrt{2} G_F U_{\alpha i} V_{ud}  f_\rho \rho^-_\mu (\bar \ell_\alpha  \gamma^\mu P_L n_i )  + \textrm{h.c.} \, ,
\label{eq:effective-rho}
\end{equation}
and, equivalently, for the $K^{*,\pm}$ meson we get
\begin{equation}
\mathcal{O}_{K^{*} \ell_\alpha \bar{n}_i}  = - \sqrt{2} G_F U_{\alpha i} V_{us}  f_{K^*} K^{*,-}_\mu (\bar \ell_\alpha  \gamma^\mu P_L n_i ) + \textrm{h.c.} \, 
\end{equation}

\subsection{Semileptonic meson decays}

Some mesons exhibit non-negligible branching ratios for semileptonic decay channels into neutrinos, charged leptons and lighter mesons. These can even dominate over the two-body leptonic decays if the mass of the heavy neutrino is not large enough to sufficiently enhance the latter, and thus must be taken into account. 

After integrating out the $W$ boson, the amplitude for the $P\rightarrow D\bar{n}\ell$ decay (where $P$ and $D$ stand for generic parent and daughter mesons, respectively) reads:
\begin{equation}
i\mathcal{M}_{P D \ell_\alpha \bar{n}_i}=\frac{ig^2}{2M_W^2}U_{\alpha i}\bar{u}_\alpha\gamma^\mu P_L v_i \left\langle D \vert j^V_{W,\mu}\vert P\right\rangle \, ,
\end{equation}
where $j_{W,\mu}^V$ is defined in Eq.~\eqref{eq:W-V}. This hadronic matrix element is usually expressed in terms of two form factors, $f_+$ and $f_-$ \cite{Lubicz:2017syv}:
\begin{equation}
\label{eq:fplus_fminus}
\left\langle D\vert j_{W,\mu}^V \vert P\right\rangle =\frac{1}{2}V_{qq^\prime}\left( p_\mu f_+(q^2) + q_\mu f_-(q^2)\right)  \, ,
\end{equation}
where $V_{qq^\prime}$ is the CKM element corresponding to the quarks which interact with the $W$ in the hadronic transition, while $p_\mu\equiv p_\mu^D+p_\mu^P$ is the sum of the 4-momenta of the parent and daughter mesons and $q_\mu\equiv p_\mu^D-p_\mu^P$ is the 4-momentum transfer between them. Thus, the amplitude can be written as:
\begin{equation}
i\mathcal{M}_{P D \ell_\alpha \bar{n}_i}=i\sqrt{2}G_F V_{qq^\prime}U_{\alpha i}\bar{u}_\alpha\gamma^\mu P_L v_i\left( p_\mu f_+(q^2) + q_\mu f_-(q^2)\right)\, .
\end{equation}
In what follows, it becomes convenient to express this in terms of the 4-momenta of the daughter meson, $p_\mu^D$, and of the leptonic pair, $p_\mu^{n\ell}$:
\begin{equation}
i\mathcal{M}_{P D \ell_\alpha \bar{n}_i}=i\sqrt{2}G_F V_{qq^\prime}U_{\alpha i}\bar{u}_\alpha\gamma^\mu P_L v_i\left[2 p_\mu^Df_+(q^2)+p_\mu^{n\ell}\left(f_+(q^2)-f_-(q^2) \right) \right] \, .
\end{equation}
Note that we have not specified the electric charges of the involved mesons. In fact, this amplitude describes all the processes allowed by charge conservation ($P^-\rightarrow D^0\bar{n}\ell^-$ and $P^0\rightarrow D^+\bar{n}\ell^-$, as well as their CP-conjugates). However,  it should be stressed that, even though electromagnetic contributions to these hadronic form factors are generally small, in some cases the numerical parameters they contain might be slightly different depending on the charge of the mesons, since they come from fits to different datasets.

From this amplitude it is possible to extract the corresponding effective operator in configuration space, writing the 4-momenta as derivatives:
\begin{eqnarray}
\mathcal{O}_{P D \ell_\alpha \bar{n}_i} 
& = & -i\sqrt{2}G_FV_{qq^\prime}U_{\alpha i}\left[2f_+(q^2)\bar{\ell}_\alpha\gamma^\mu P_L n_i\left(\partial_\mu \phi_D \right)\phi^\dagger_P  + \right. \nonumber \\ 
& + & \left. \left(f_+(q^2)-f_-(q^2) \right) \partial_\mu(\bar{\ell}_\alpha\gamma^\mu P_L n_i)\phi_D \phi^\dagger_P \right] + \textrm{h.c.} \, ,
\label{eq:semilep}
\end{eqnarray}
where $\phi_P$ and $\phi_D$ are the parent and daughter meson fields, respectively. Once more, if the involved fields are on-shell, it is possible to apply Dirac's equation and substitute the derivative acting on the leptonic current by terms proportional to their masses. The resulting operator reads:
\begin{eqnarray}
\mathcal{O}_{P D \ell_\alpha \bar{n}_i} = \sqrt{2}G_FV_{qq^\prime}U_{\alpha i}\bar{\ell}_\alpha
 & \left[\left(f_+(q^2)-f_-(q^2) \right)  (m_\alpha P_L-m_i P_R) \phi_D  \right. \nonumber \\
& \left.  - 2if_+ (q^2) (\partial_\mu \phi_D) \gamma^\mu P_L  \right] n_i\phi^\dagger_P + \textrm{h.c.} 
\end{eqnarray}

\subsubsection{Form factors}
\label{subsec:formfactors}
Many parametrizations for the hadronic form factors are available in the literature, most
of which are given in terms of $f_+$ and $f_0$. The former was defined together with $f_-$ in Eq.~\eqref{eq:fplus_fminus}, while the latter can be related to $f_+$ and $f_-$ via
\begin{equation}
f_0(q^2)=f_+(q^2)+\frac{q^2}{M_D^2-M_P^2}f_-(q^2) \, .
\end{equation}

The semileptonic decays we will be mostly interested in are $K\rightarrow \pi n \ell$ and $D\rightarrow K n \ell$. For the former we employ a linear parametrization, as in Ref.~\cite{Bijnens:1994me}, according to which 
\begin{equation}
f_{+,0}^{K\pi}(q^2)=f_+^{K\pi}(0)\left[1+\lambda^{K\pi}_{+,0}\frac{q^2}{M_{\pi^+}^2} \right] \, .
\end{equation}
Conversely, in the case of the $D\rightarrow Kn\ell$ decay we make use of a ``pole'' parametrization~\cite{Lubicz:2017syv}:
\begin{eqnarray}
f_+^{DK}(q^2) & = & \frac{f_+^{DK}(0)+c_+^{DK}(z-z_0)(1+\frac{z+z_0}{2})}{1-\frac{q^2}{M^2_{D_s^*}}} \, , \label{eq:fpDK} \\
f_0^{DK}(q^2) & = & f_+^{DK}(0)+c_0^{DK}(z-z_0)\left(1+\frac{z+z_0}{2} \right) \, ,
\end{eqnarray}
where 
\begin{eqnarray}
z & = & \frac{\sqrt{t_+-q^2}-\sqrt{t_+-t_0}}{\sqrt{t_+-q^2}+\sqrt{t_+-t_0}}\,, \\
z_0 & \equiv & z(q^2=0)\,,
\end{eqnarray}
with
\begin{eqnarray}
t_+ & = & (M_D+M_P)^2, \\
t_0 & = & (M_D+M_P)\left(\sqrt{M_D}-\sqrt{M_P} \right) ^2\,.
\end{eqnarray}
The values used for the form factor parameters are summarized in Tabs.~\ref{tab:paramtableD} and~\ref{tab:paramtableK}.
\begin{table}[ht!]
\begin{center}
\renewcommand{\arraystretch}{1.8}
\begin{tabular}{|c|c|c|}
\hline 
$f_+^{DK}(0)$~\cite{Lubicz:2017syv}& $c_+^{DK}$~\cite{Lubicz:2017syv}&  $c_0^{DK}$~\cite{Lubicz:2017syv}  \\
\hline
\hline
$0.7647$&$-0.066$&$-2.084$ \\
\hline
\end{tabular}
\end{center}
\caption{Parameters entering our form factor definitions for the semileptonic $D\rightarrow K n \ell$ decays.
\label{tab:paramtableD} }
\end{table}
\begin{table}[ht!]
\begin{center}
\renewcommand{\arraystretch}{1.8}
\begin{tabular}{|c|c|c|c|}
\hline
\textbf{PD} &$f^{\mathrm{PD}}_+(0)$~\cite{Aoki:2019cca} &$\lambda^{\mathrm{PD}}_+$~\cite{Tanabashi:2018oca}&$\lambda^{\mathrm{PD}}_0$~\cite{Tanabashi:2018oca} \\
\hline
\hline
 $K^\pm\pi^0$ &\multirow{2}{*}{$0.9749$}&$0.0297$&$0.0195$ \\
\cline{1-1}
\cline{3-4}
 $K^0\pi^\pm$ &&$0.0282$&$0.0138$\\
\hline
\end{tabular}
\end{center}
\caption{Parameters entering our form factor definitions for the semileptonic $K\rightarrow \pi n \ell$ decays. Note that we make use of different parameters for the decays of charged and neutral kaons, following Ref.~\cite{Tanabashi:2018oca}. 
\label{tab:paramtableK} }
\end{table}

We have included these form factors in our FeynRules model, and numerically checked with MadGraph5~\cite{Alwall:2014hca} that our implementation reaches an agreement of at least a 95\% with the measured branching ratios for the SM decay channels $K \to \pi \nu \ell$ and $D \to K \nu \ell$~\cite{Tanabashi:2018oca}. For convenience we provide two separate implementations for such couplings, as explained in detail in Appendix~\ref{app:feynrules}.

\section{Production of Heavy Neutral Leptons from meson decays}
\label{sec:prod}

In this section we provide the expressions for the production of a heavy neutrino $N_4$ of mass $M_4$ via meson decays. We have computed them employing the Feynman rules derived from the effective operators obtained in Sec.~\ref{sec:eft}, and verified their agreement with the simulations generated via MadGraph5 using the implementation of our model in FeynRules. In order to do so, we have diagonalized explicitly the full mass matrix and expressed the ensuing mixing matrix in terms of the original Yukawa couplings. Further details on this diagonalization can be found in Appendix~\ref{app:matrices}. 

\subsection{Two-body leptonic decays}

The generic expression for the leptonic decay of a charged pseudoscalar meson $P$ of mass $m_P$ is given by~\cite{Gorbunov:2007ak,Atre:2009rg,Helo:2010cw,Abada:2013aba,Bondarenko:2018ptm}
\begin{eqnarray}
\Gamma(P^\pm \rightarrow N_4 \ell^\pm_\alpha)&=& \displaystyle \dfrac{G_F^2 m_P^3}{8\pi}f_P^2\vert U_{\alpha 4}\vert^2 \vert V_{q q'}\vert^2 \lambda^{1/2}(1, y_4^2, y_\alpha^2) \left(y_4^2+y_\alpha^2-\left(y_4^2-y_\alpha^2 \right)^2\right) \,,
\label{eq:width_P_N_l}
\end{eqnarray}
where the values of $f_P$ are given in Tab.~\ref{tab:Fpi}, and we have defined $ y_4 \equiv M_4/m_P$, $ y_\alpha \equiv m_{\ell_\alpha}/m_P$, and 
\begin{equation}
\label{eq:lambda}
\lambda(a,b,c) = a^2 + b^2 + c^2 - 2ab -2bc -2ac \, .
\end{equation} 

\subsection{Three-body semileptonic decays}

The decay width for the semileptonic decay of a parent pseudoscalar meson $P$ into a daughter pseudoscalar $D$, a charged lepton $\ell_\alpha$ and a heavy neutrino $N_4$ is given by~\cite{Gorbunov:2007ak,Helo:2010cw,Abada:2013aba,Bondarenko:2018ptm} 
\begin{eqnarray}
\Gamma(P \rightarrow D N_4 \ell^\pm_\alpha)&=& \displaystyle \dfrac{G_F^2 m_P^5}{64\pi^3} C_D^2 \vert U_{\alpha 4}\vert^2 \vert V_{q q'}\vert^2 \left(I_1^{PD} +I_2^{PD} +I_3^{PD} \right)  \,,
\label{eq:width_P_D_N_l}
\end{eqnarray}
where $C_D=1$ in all cases under consideration, except for  $K^\pm \rightarrow \pi^0 N_4 \ell^\pm_\alpha$, for which $C_D = \frac{1}{\sqrt{2}}$. The integrals $I_i^{PD}$ are expressed in terms of the form factors $f^{PD}_+(q^2)$ and $f^{PD}_0(q^2)$ defined in Sec.~\ref{subsec:formfactors}.
\begin{eqnarray}
I_1^{PD}&=& \int_{(y_\alpha+y_4)^2}^{(1-y_D)^2}{\frac{dz}{3z^3}\vert f^{PD}_+\left( z m_P^2 \right)\vert^2 \lambda(1,y^2_D,z)^{3/2}\lambda(z,y^2_4,y_\alpha^2)^{3/2}}  \,, \\
I_2^{PD}&=& \int_{(y_\alpha+y_4)^2}^{(1-y_D)^2}{\frac{dz}{2z^3}\vert f^{PD}_+\left( z m_P^2 \right)\vert^2 \lambda(1,y^2_D,z)^{3/2}\lambda(z,y^2_4,y_\alpha^2)^{1/2}g(z)}  \,, \\
I_3^{PD}&=& \int_{(y_\alpha+y_4)^2}^{(1-y_D)^2}{\frac{dz}{2z^3}\vert f^{PD}_0\left( z m_P^2 \right)\vert^2 \lambda(1,y^2_D,z)^{1/2}\lambda(z,y^2_4,y_\alpha^2)^{1/2}g(z)\left(1-y_D^2 \right)^2}  \,, 
\label{eq:integrals_P_D_N_l}
\end{eqnarray}
where $\lambda(a,b,c)$ is defined in Eq.~\eqref{eq:lambda}, $y_D \equiv m_D/m_P$ and
\begin{equation}
g(z)=z\left(y_4^2+y_\alpha^2 \right)-\left( y_4^2-y_\alpha^2 \right)^2 \, .
\end{equation}
 
\section{Decays of Heavy Neutral Leptons into SM particles}
\label{sec:decays}

\subsection{Two-body decays}

Here we provide general expressions for the decay widths of a heavy neutrino $N_4$ of mass $M_4$ into final states including pseudoscalar and vector mesons separately. We have computed them employing the Feynman rules derived from the effective operators obtained in Sec.~\ref{sec:eft}, and verified their agreement with the simulations generated via MadGraph5 using our model implementation in FeynRules. Throughout this section, we will neglect the masses of the light neutrinos for simplicity. 

\subsubsection{Pseudoscalar mesons}

The generic expression for the heavy neutrino decay width into a neutral pseudoscalar meson $P$ is given by
\begin{eqnarray}
\Gamma(N_4\rightarrow P \nu)&=& \displaystyle\sum_{j} \dfrac{G_F^2 M_4^3}{32\pi}f_P^2\vert C_{4j}\vert^2\left(1-x_P^2\right)^2 \,,
\label{eq:width_N_P_nu}
\end{eqnarray}
where we have defined $x_P \equiv m_P/M_4$, and 
\begin{equation}
f_P = \left\{ \begin{array}{c l }
f_\pi  & \quad \mathrm{for } \; P = \pi^0 \, , \\[3mm]
\dfrac{\cos\theta_8 f_8}{\sqrt{3}}+\dfrac{\sin\theta_0 f_0}{\sqrt{6}}  & \quad \mathrm{for } \; P = \eta \, , \\[3mm]
\dfrac{\sin\theta_8 f_8}{\sqrt{3}}-\dfrac{\cos\theta_0 f_0}{\sqrt{6}} & \quad \mathrm{for } \; P = \eta^\prime \, ,
\end{array} \right. 
\label{eq:feta-fetap} 
\end{equation}
according to the parametrization used to describe the $\eta-\eta^\prime$ mixing in Sec.~\ref{sec:neutral-pseudo}. Using the parameters provided in Tab.~\ref{tab:Fpi}, this leads to the ``effective decay constants'' $f_\eta \simeq 81.6~\mathrm{MeV}$ and $f_{\eta'} \simeq - 94.6~\mathrm{MeV}$.
Finally, note that the sum $\sum_j$ in Eq.~\eqref{eq:width_N_P_nu} runs over the three light neutrino mass eigenstates, since they cannot be individually identified. However, at leading order in $U_{\alpha 4}$, this is equivalent to a sum running over the three active flavors, since 
\begin{equation}
 \sum_j \vert C_{4j}\vert^2 = \sum_{j,\alpha,\beta} U^*_{\alpha 4} U_{\alpha j} U_{\beta 4} U^*_{\beta j} = \sum_{\alpha,\beta} U^*_{\alpha 4} U_{\beta 4} (\delta_{\alpha \beta} - U_{\alpha 4} U^*_{\beta 4}) \simeq \sum_\alpha \vert U_{\alpha 4}\vert^2 \, .
\end{equation}

On the other hand, the decay width into a charged pseudoscalar meson $P^\pm$ is given by
\begin{equation}
\Gamma(N_4\rightarrow P^\pm \ell^\mp_\alpha) = 
\dfrac{G_F^2 M_4^3}{16\pi} f_P^2\vert U_{\alpha 4}\vert^2 \vert V_{q q'}\vert^2
\lambda^{1/2}(1, x_P^2, x_\alpha^2) \left[1-x_P^2 - x_\alpha^2 \left(2+ x_P^2 - x_\alpha^2 \right) \right] \,,
\label{eq:width_N_P_l}
\end{equation}
where $x_\alpha  \equiv m_{\ell_\alpha}/M_4$, and the relevant meson decay constants $f_P$ are provided in Tab.~\ref{tab:Fpi}.

\subsubsection{Vector mesons}

In the case of neutral vector mesons, the decay width reads:
\begin{eqnarray}
\Gamma(N_4\rightarrow V \nu)&=& \displaystyle\sum_{j} \dfrac{G_F^2 M_4^3}{32\pi m_V^2}f_V^2 g_V^2 \vert C_{4j}\vert^2\left(1+ 2 x_V^2\right)\left(1-x_V^2\right)^2 \,,
\label{eq:width_N_V_nu}
\end{eqnarray}
with $x_V \equiv m_V/M_4$, and where we have again summed over all light neutrinos in the final state. The values for the decay constants $f_V$ are given in Tab.~\ref{tab:Fpi}, while expressions for $g_V$ in terms of the weak mixing angle are provided in Tab.~\ref{tab:gV}.
\begin{table}[ht!]
\begin{center}
\renewcommand{\arraystretch}{1.8}
\begin{tabular}{|c|c|c| }
\hline
$N_4\rightarrow \rho^0 \nu$ & $N_4\rightarrow \omega \nu$ &$N_4\rightarrow \phi \nu$\\  \hline \hline 
$1-2 s_w^2$  & $-\dfrac{2 s_w^2}{3}$ & $-\sqrt{2}\left(\dfrac{1}{2}-\dfrac{2 s_w^2}{3}\right)$  \\[2mm]
\hline
\end{tabular}
\end{center}
\caption{Expressions $g_V$ entering the heavy neutrino decay widths into neutral vector mesons, Eq.~(\ref{eq:width_N_V_nu}).}
\label{tab:gV}
\end{table}

On the other hand, for the decays into charged vector mesons we get
\begin{eqnarray}
\Gamma(N_4\rightarrow V^\pm \ell^\mp_\alpha)  =  \dfrac{G_F^2 M_4^3}{16\pi m_{V^\pm}^2}f_V^2& &\vert U_{\alpha 4}\vert^2 \vert V_{q q'}\vert^2  \lambda^{1/2}(1, x_V^2, x_\alpha^2) \times \nonumber \\
& &\left[\left(1-x_V^2\right)\left(1+2x_V^2\right) + x_\alpha^2\left( x_V^2 + x_\alpha^2 - 2 \right)\right] 
\,,
\label{eq:width_N_V_l}
\end{eqnarray}
where the decay constants $f_V$ are again summarized in Tab.~\ref{tab:Fpi}.

\subsection{Three-body decays}

Heavy neutrinos may also decay into three body final states either purely leptonically or semileptonically. The latter include $N_4 \to \pi^+ \pi^0 \ell^- $, $N_4 \to \pi^0 \pi^0 \nu$ and $N_4 \to K^+ \pi^0 \ell^- $. However, their respective contributions are dominated by $N_4 \to \rho^+ \ell^- $, $N_4 \to \rho^0 \nu$ and $N_4 \to K^{*,+}  \ell^- $ respectively, already included in the previous section. This can be seen from the data from $\tau$ decays, since the hadronic matrix elements involved in the semileptonic decays would be the same. Indeed, the branching ratio of $\tau^- \to \nu \pi^- \pi^0$ is $25.49\%$, while the contribution which does not correspond to $\tau^- \to \nu \rho^-$ is negligible: $\left(3.0 \pm 3.2 \right)\cdot 10^{-3}$~\cite{Tanabashi:2018oca}. We will thus review here only the three-body purely leptonic decays $N_4 \to \ell \ell \nu$ and $N_4 \to \nu \nu \nu$, taken from Refs.~\cite{Gorbunov:2007ak,Atre:2009rg,Helo:2010cw,Bondarenko:2018ptm}. 

The invisible decay of the heavy neutrino reads~\cite{Gorbunov:2007ak,Atre:2009rg,Helo:2010cw,Bondarenko:2018ptm}
\begin{eqnarray}
\Gamma(N_4\rightarrow \nu \nu \nu) &= & \displaystyle\sum_{j} \vert C_{4j}\vert^2\dfrac{G_F^2 M_4^5}{192\pi^3 } \, ,
\label{eq:width_N_nu_nu_nu}
\end{eqnarray}
where we have summed over all possible light neutrinos in the final state.

For the three-body decays involving charged leptons in the final state, we will distinguish between two cases. If the heavy neutrino decays into two leptons of the same flavor $\beta$, there are both $W$ and $Z$ mediated diagrams contributing to the amplitude. The total decay width can be expressed as~\cite{Gorbunov:2007ak,Atre:2009rg,Helo:2010cw,Bondarenko:2018ptm}
\begin{equation}
\Gamma(N_4\rightarrow \nu \ell^-_\beta \ell^+_\beta) =  
\sum_{\alpha} \vert U_{\alpha 4} \vert^2 \dfrac{G_F^2 M_4^5}{192\pi^3}
\left[ \left( C_1 + 2 s_w^2 \delta_{\alpha \beta} \right) f_1(x_\beta) + \left( C_2 + s_w^2 \delta_{\alpha \beta} \right)  f_2(x_\beta) \right] \,,
\label{eq:width_N_la_la}
\end{equation}
where 
\begin{equation}
C_1= \frac{1}{4} \left( 1- 4 s_w^2 + 8 s_w^4 \right), \quad C_2 = \frac{1}{2} \left(-s_w^2+2 s_w^4 \right) \, , 
\end{equation}
and we have defined the functions
\begin{eqnarray}
f_1(x) & = & (1 - 14 x^2 - 2 x^4 - 12 x^6) \sqrt{1 - 4 x^2} +  12 x^4 (x^4 - 1) L (x) \, , \\
f_2(x) & = & 4 \left[x^2 (2 + 10 x^2 - 12 x^4) \sqrt{1 - 4 x^2} + 6 x^4 (1 - 2 x^2 + 2 x^4) L(x)\right] \, , 
\end{eqnarray}
with
\begin{equation}
L(x) = \ln{\left(\dfrac{1 - 3 x^2 - (1 - x^2) \sqrt{1 - 4 x^2}}{x^2 (1 + \sqrt{1 - 4 x^2})}\right)}\,.
\label{eq:log_fun}
\end{equation}

On the other hand, the decay of the heavy neutrino into two leptons of different flavor is only mediated by the $W$ interaction. In the limit in which one of the charged lepton masses can be neglected, the corresponding decay width simplifies to~\cite{Gorbunov:2007ak, Bondarenko:2018ptm}
\begin{eqnarray}
\Gamma(N_4\rightarrow \nu \ell^-_\alpha \ell^+_\beta) & \simeq & 
\vert U_{\alpha 4}\vert^2 \dfrac{G_F^2 M_4^5}{192\pi^3}
\left(1 - 8 x_M^2 + 8 x_M^6 - x_M^8 - 12 x_M^4 \ln(x_M^2)\right)\, ,
\label{eq:width_N_la_lb}
\end{eqnarray}
where $x_M=\mathrm{max}\left\lbrace x_\alpha,x_\beta\right\rbrace $. Note that this expression corresponds to a Dirac neutrino decay; for Majorana neutrinos there would be a second contribution proportional to $ |U_{\beta 4}|^2$ since there are two diagrams allowed, each of them proportional to a different mixing matrix element.

\subsection{Decays to 4 or more bodies}
\label{subsec:multimesons}

Finally, for HNL masses above 1~GeV, the appropriate description of the hadronic final states transitions from the effective theory with the different meson resonances included in the previous sections, to quark production in the final state with subsequent hadronization, more suitable for perturbative QCD. For reference, the $\tau^-$, with its $1.78$~GeV mass, is precisely at the transition region. Indeed, it shows a $10.8 \%$ and $25.5 \%$ branching ratio to the $\nu_\tau \pi^-$ and $\nu_\tau \rho^-$ channels respectively. But also a $9.3 \%$ branching ratio to both $\nu_\tau \pi^- 2 \pi^0 $ and to $\nu_\tau 2 \pi^- \pi^+$ and even $4.6 \%$ and $1.0 \%$ branching ratios to $\nu_\tau 2 \pi^- \pi^+ \pi^0$ and $\nu_\tau \pi^- 3\pi^0$ respectively~\cite{Tanabashi:2018oca}. These last decay modes, with three or more mesons in the final state, are more suitably described from the underlying quark interactions with a subsequent correction to account for the hadronization process:
\begin{equation}
1 + \Delta_\mathrm{QCD} \equiv \frac{\Gamma \left( \tau \to \nu_\tau + \mathrm{hadrons} \right)}{\Gamma_{tree} \left( \tau \to \nu_\tau + u + \bar{d} \right)+\Gamma_{tree} \left( \tau \to \nu_\tau + u + \bar{s} \right)}
\end{equation}
with~\cite{Gorishnii:1990vf}
\begin{equation}
\Delta_\mathrm{QCD} = \frac{\alpha_s}{\pi}+5.2\frac{\alpha_s^2}{\pi^2}+26.4\frac{\alpha_s^3}{\pi^3}.
\label{eq:QCDcorrection}
\end{equation}

We adopt the same approach as Ref.~\cite{Bondarenko:2018ptm} (see also~\cite{SHiP:2018xqw,Bondarenko:2019yob}) and use Eq.~(\ref{eq:QCDcorrection}) to account for the hadronization of the HNL decays $N_4 \to \ell_\alpha u \bar{d}$ and $N_4 \to \ell_\alpha u \bar{s}$ for HNL masses above 1~GeV. We also apply the same correction to the neutral current decays $N_4 \to \nu q \bar{q}$ with $q=u,d,s$. However, we add a phase space suppression factor $\sqrt{1- 4 m_K^2/M_4^2}$ for the $N_4\to \nu s \bar{s}$ channel since it would otherwise overestimate its importance for $M_4 =1$~GeV, where the phase space prevents two $K$ in the final state. For the running of $\alpha_s$ we follow the dedicated review in Ref.~\cite{Tanabashi:2018oca}. The difference between these fully inclusive hadronic final states and the HNL decays to specific mesons discussed above will provide an estimate of the HNL decays to 3 or more mesons. We have tested this procedure for the $\tau$ decays and reached good agreement with its tabulated branching ratios.

\begin{figure}[htb!]
\centering
\includegraphics[width=0.48\textwidth]{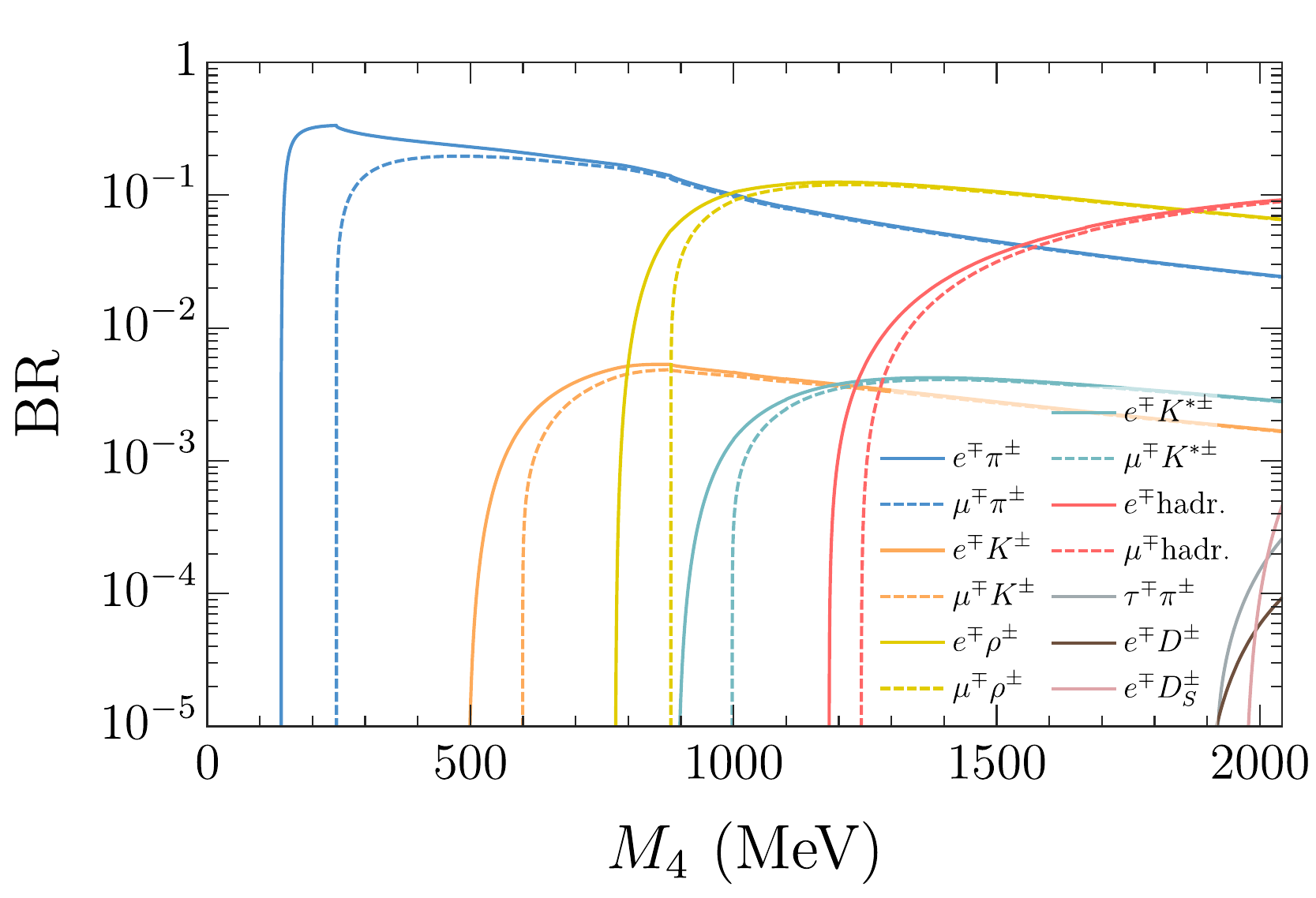}
\includegraphics[width=0.48\textwidth]{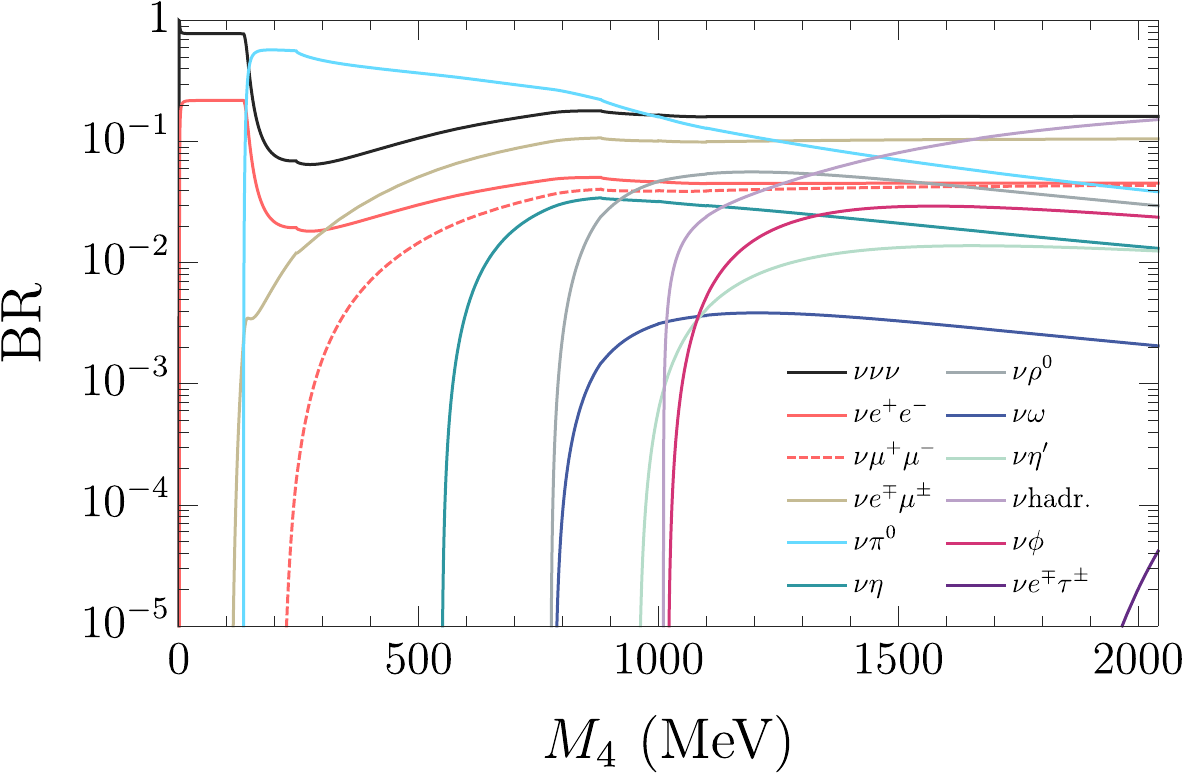}
\caption{Branching ratios for the heavy $N_4$ as a function of its mass, obtained under the assumption of same mixing to all flavors ($\vert U_{e 4}\vert^2 = \vert U_{\mu 4}\vert^2 = \vert U_{\tau 4}\vert^2$). Left (right) panels correspond to decays without (with) light neutrinos in the final state. The decay channels into semileptonic final states are not shown, as their branching ratio is expected to be negligible for the range of masses considered here. }
\label{fig:branching_ratios_deg}
\end{figure}
\begin{figure}[htb!]
\centering
\includegraphics[width=0.48\textwidth]{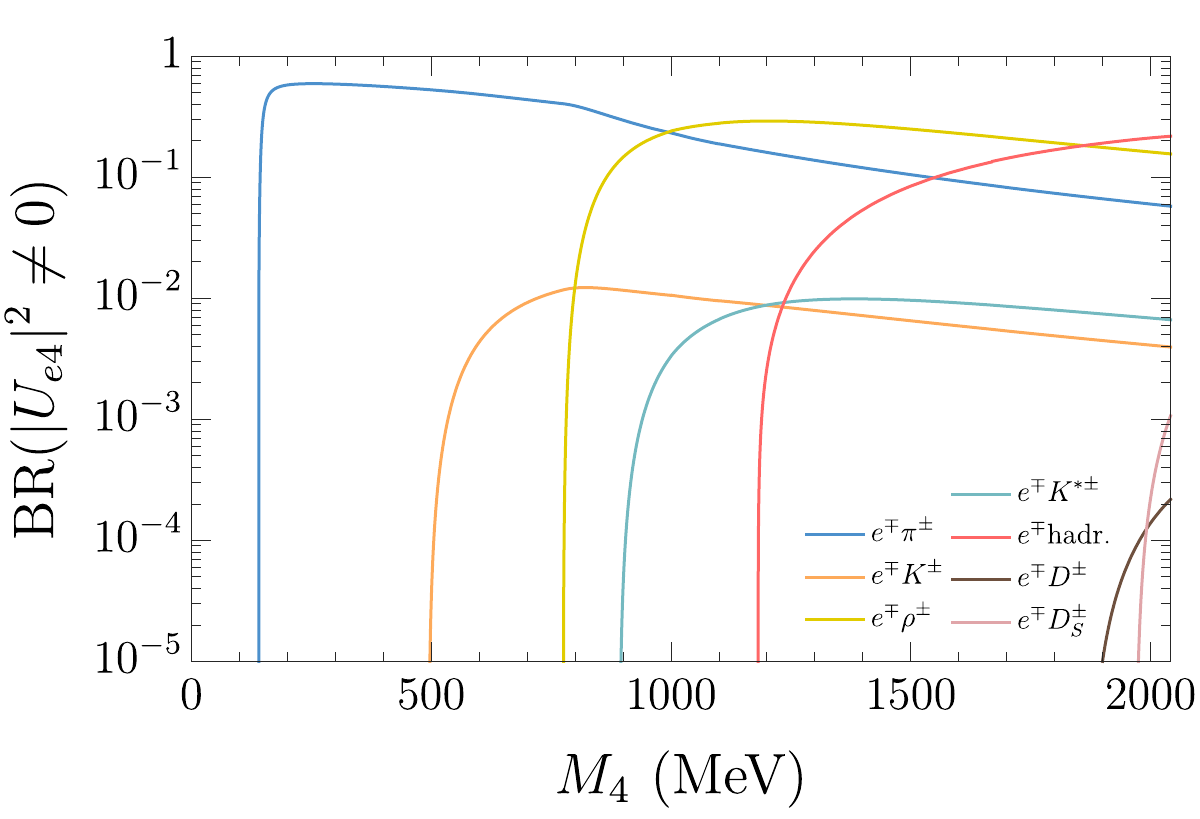}
\includegraphics[width=0.48\textwidth]{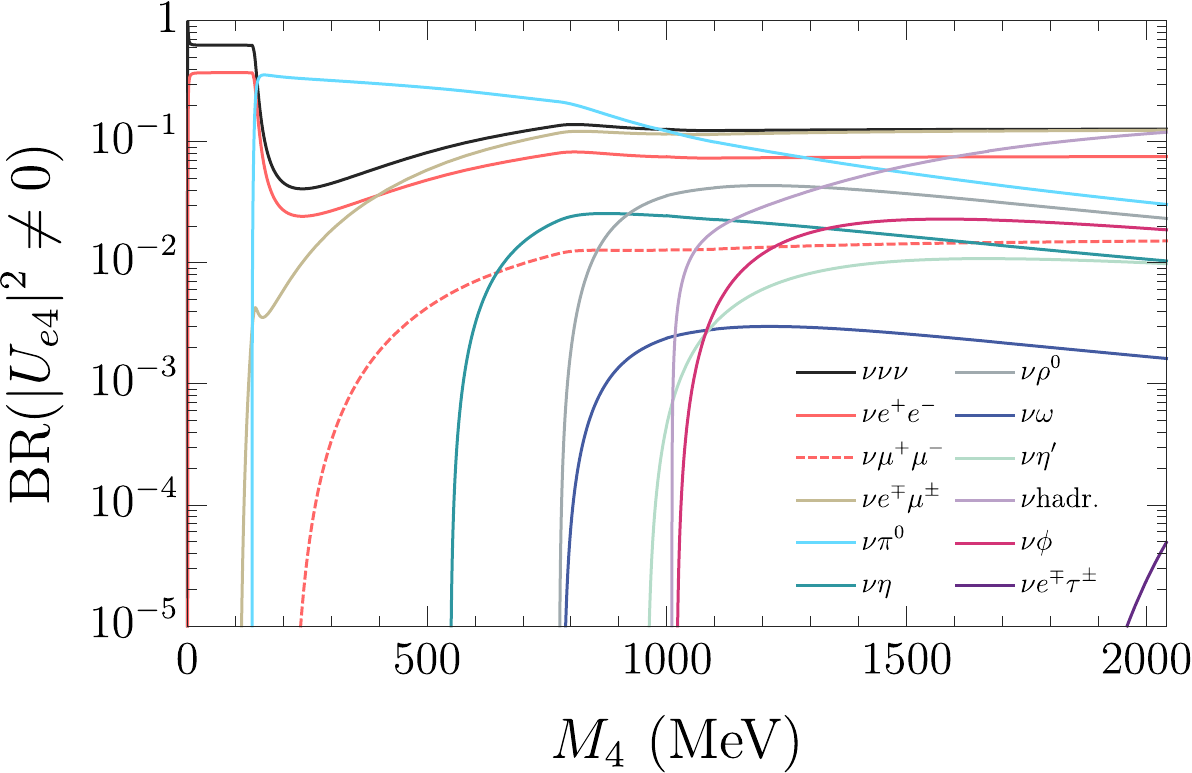}
\includegraphics[width=0.48\textwidth]{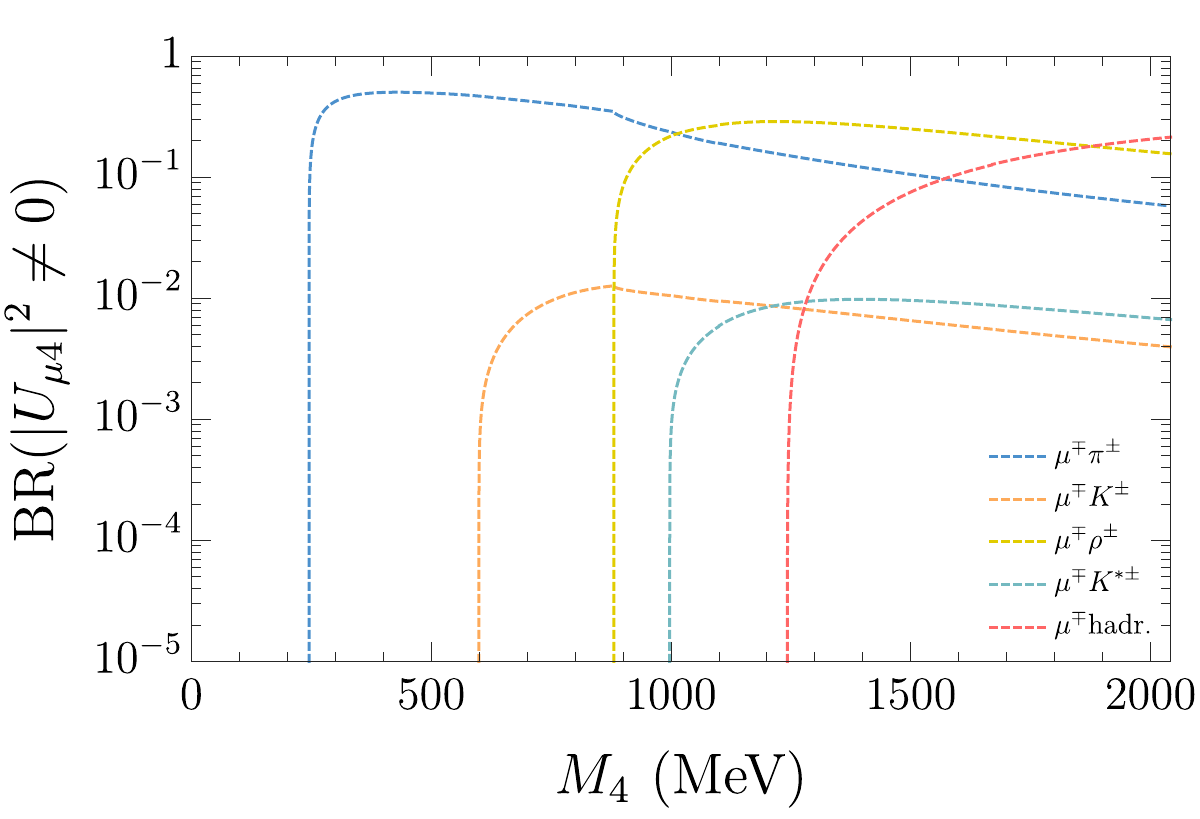}
\includegraphics[width=0.48\textwidth]{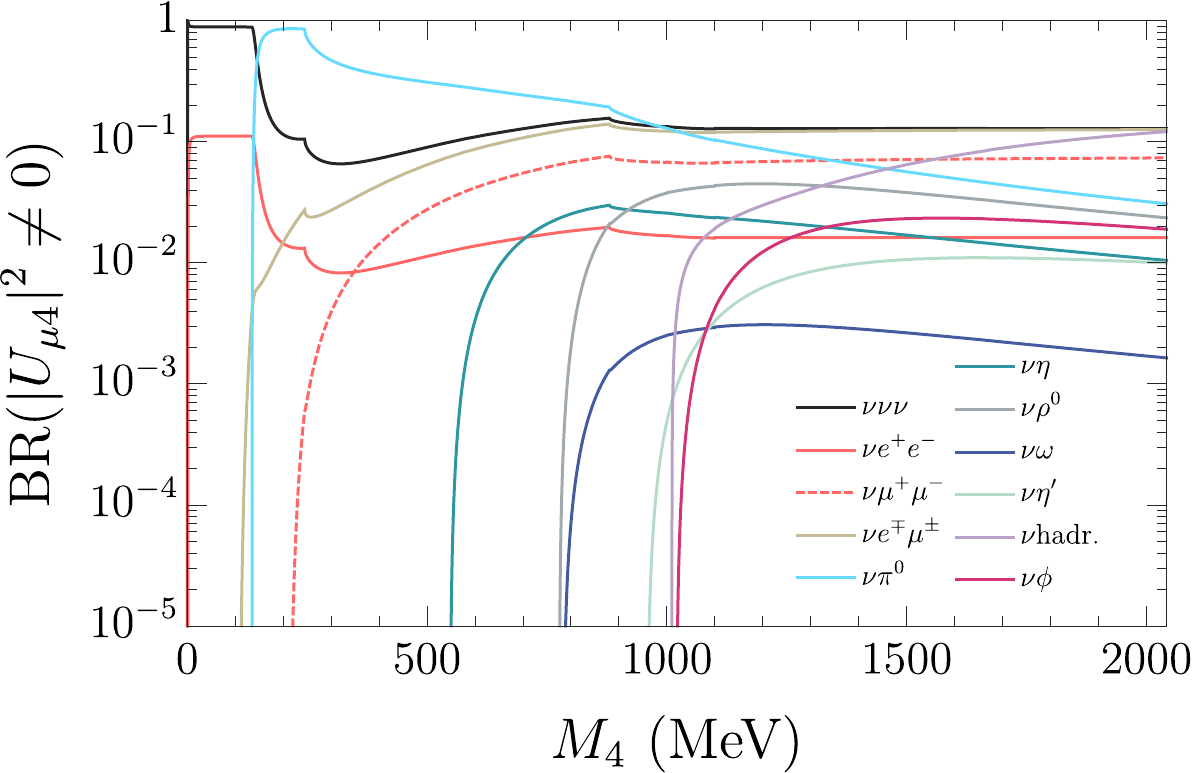}
\includegraphics[width=0.48\textwidth]{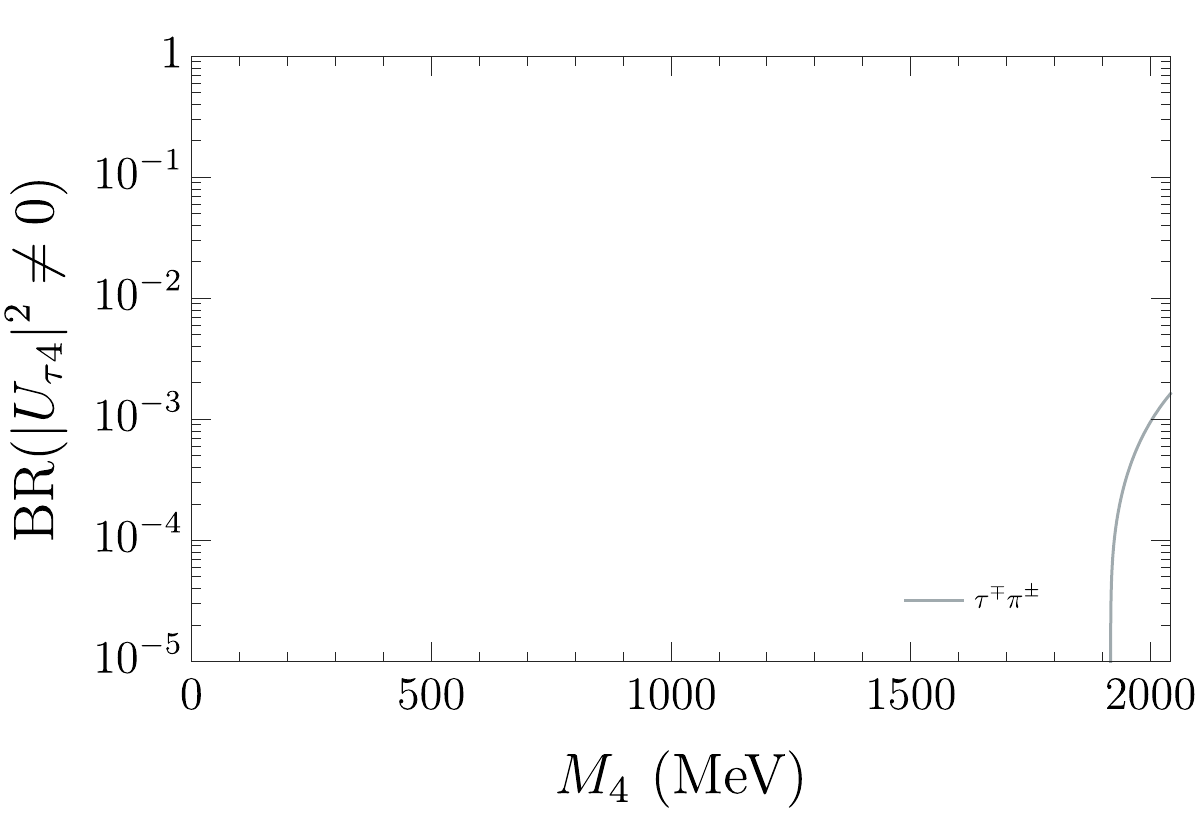}
\includegraphics[width=0.48\textwidth]{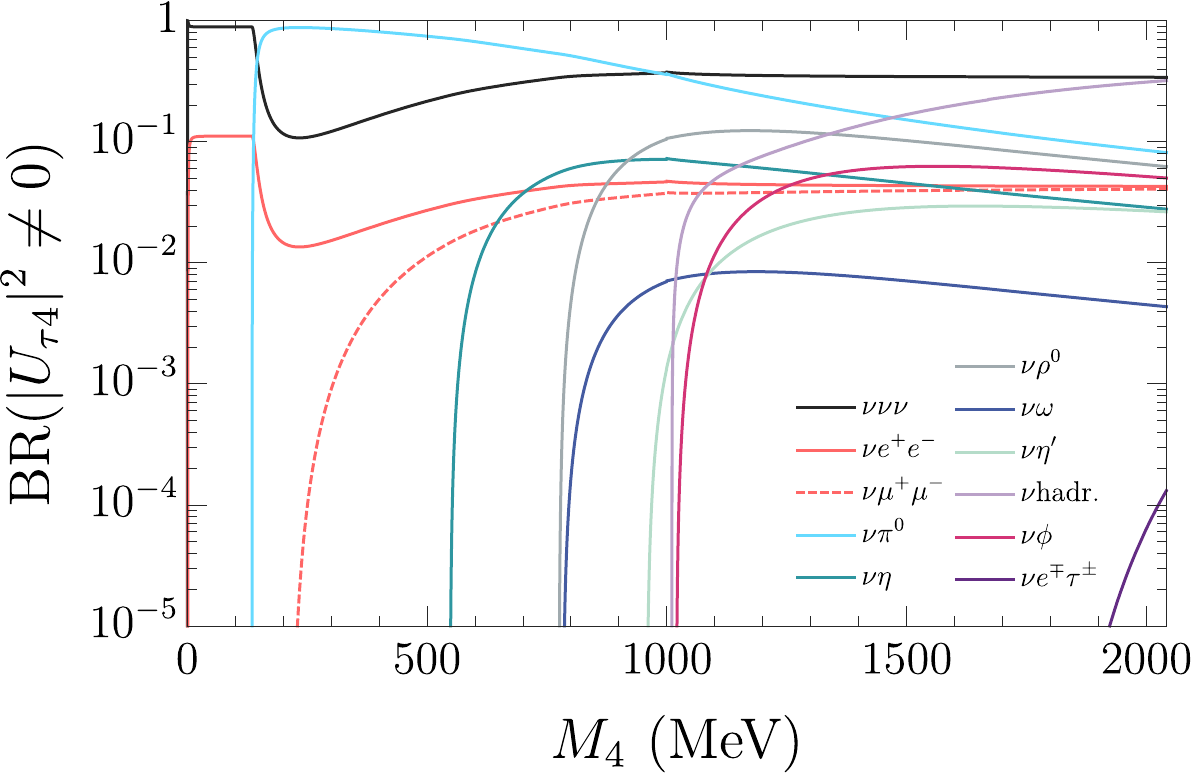}
\caption{Branching ratios for the heavy $N_4$ as a function of its mass, obtained under the assumption that only its mixing to one lepton flavor is non-zero, as indicated by the labels in each row. Left (right) panels correspond to decays without (with) light neutrinos in the final state. The decay channels into semileptonic final states are not shown as their branching ratio is expected to be negligible in the range of masses considered here.}
\label{fig:branching_ratios_flav}
\end{figure}
Figures~\ref{fig:branching_ratios_deg} and~\ref{fig:branching_ratios_flav} show the branching ratios for the different decay channels of the heavy neutrino, as a function of its mass, for two different cases: degenerate mixings to all lepton flavors ($\vert U_{e 4}\vert^2 = \vert U_{\mu 4}\vert^2 = \vert U_{\tau 4}\vert^2$), and in the case when only one of the mixing matrix elements is non-zero. The labels $\ell^\mp$hadr. and $\nu$hadr. stand for $N_4$ decays, mediated by charged and neutral currents respectively, with 3 or more mesons in the final state.

\subsection{Discrepancies with previous literature}
\label{subsec:discrepancies}

The decay widths of a HNL into mesons, neutrinos and leptons have been derived several times in previous literature; for an incomplete list see e.g. Refs.~\cite{Shrock:1980ct, Bondarenko:2018ptm, Ballett:2019bgd, Atre:2009rg, Gorbunov:2007ak, Helo:2010cw}. Here we summarize the main discrepancies and differences found between our results and some of these works:
\begin{enumerate}
\item Overall, we find a relatively good agreement with Ref.~\cite{Bondarenko:2018ptm} for the meson decay constants and for most vertices involving heavy neutrinos, with the exception of the couplings to $\omega$ and $\phi$ mesons, for which we find different expressions in terms of $\sin^2\theta_w$ (see our Tab.~\ref{tab:gV}, in comparison with Tab.~9 in Ref.~\cite{Bondarenko:2018ptm}). 
\item We find that the expressions in Ref.~\cite{Gorbunov:2007ak} for the HNL decay into vector mesons have an extra factor 2 with respect to our results, both for the neutral and the charged channels. Also, their expressions for the decay into neutral vector mesons seem not to include a dependence on $\sin^2\theta_w$ (see our definitions for $g_V$ in Tab.~\ref{tab:gV}). Finally, there are significant differences in the values reported  in Ref.~\cite{Gorbunov:2007ak} for the neutral pseudoscalar meson decay constants $f_\eta$ and $f_{\eta'}$.
\item We find that the expressions for the HNL decay into a light neutrino and a neutral pseudoscalar meson in Refs.~\cite{Atre:2009rg,Helo:2010cw} have an extra factor $2$ in the denominator, as the authors of Ref.~\cite{Bondarenko:2018ptm} pointed out. 
\item We also find significant discrepancies with the HNL decays to neutral vector mesons in Ref.~\cite{Atre:2009rg}, which were also already pointed out in Ref.~\cite{Bondarenko:2018ptm}.
\item Regarding Ref.~\cite{Ballett:2019bgd}, which was published more recently, we again find some discrepancies on the branching ratios for vector mesons: our results show a significantly higher branching ratio for the decay channels $N_4 \to \rho \ell$, and lower branching ratios for the $N_4 \to \omega \nu $ and $N_4 \to \phi \nu$ decays (as can be seen from the comparison between our Fig.~\ref{fig:branching_ratios_deg} and their Fig.~1). These discrepancies can be partially explained by the different couplings we obtain for the neutral vector meson couplings (see our Tab.~\ref{tab:gV}, in comparison with the values given below Eq.~(3.12) in Ref.~\cite{Ballett:2019bgd}) and possibly by the different values used for the corresponding decay constants. Indeed, there are also significant discrepancies in the literature for the choices of the vector meson decay constants. We thus clarify our choice in Appendix~\ref{app:decayconstants}.
\end{enumerate}

\section{Heavy Neutral Leptons at DUNE}
\label{sec:dune}

In the remainder of this work, we use the effective theory derived in the previous sections to compute the expected heavy neutrino flux at the DUNE near detector (ND), as well as the expected number of decays for different channels. From now on we will assume a Dirac HNL; in the Majorana case, the heavy neutrino decay widths, and thus the number of events, would increase in a factor of 2. In all our calculations, we consider a ND geometry as described in the DUNE Technical Design Report (TDR)~\cite{Abi:2020wmh}. The ND complex will be located 574~m downstream from the neutrino beam source, and will include three primary detector components: a liquid Argon Time Projection Chamber (LArTPC) called ArgonCube; a high-pressure gaseous TPC surrounded by an electromagnetic calorimeter (ECAL) in a 0.5~T magnetic field, called the Multi-Purpose Detector (MPD); and an on-axis beam monitor called System for on-Axis Neutrino Detection (SAND). We will consider the detector volume corresponding to the MPD\footnote{To be specific, we consider a cylinder of 5~m diameter and 5~m length, as described in Ref.~\cite{neardet}. We also consider a tilt angle $\alpha = 0.101$ due to the beam inclination with respect to the horizontal~\cite{tilt}. } for which the beam-induced background is smaller given its  lower density. 

All calculations presented in this section have been performed using the nominal beam configuration and luminosity envisioned for DUNE~\cite{Abi:2020wmh}: 120~GeV protons and $1.1\cdot 10^{21}$ protons on target (PoT) per year, divided equally into positive and negative horn focusing modes, which yields a total of $7.7\cdot 10^{21}$ PoT over 7 years of data taking, which is expected to start in 2027. The simulation of the meson production in the target has been done as follows. For pions and kaons, we use the results of the detailed GEANT4~\cite{Agostinelli:2002hh,Allison:2006ve,Allison:2016lfl} based simulation (G4LBNF) of the LBNF beamline developed by the DUNE collaboration~\cite{Abi:2020wmh}. The simulation includes a detailed description of the geometry, including the 1.5~m long target, three focusing horns, decay region, and surrounding shielding. The DUNE collaboration provides both neutrino and antineutrino mode predictions, generated for a 120~GeV primary proton beam. For positive horn focusing mode (PHF) we use the results of the full simulation to calculate the predicted event rate at the DUNE ND, while for negative horn focusing (NHF) mode we scale the event rates from PHF mode based on the flux ratios between $\pi^-/\pi^+$ and $K^-/K^+$ as predicted by G4LBNF.

However, G4LBNF does not include the production of $D$, $D_s$ and $\tau$ leptons. Thus, in this case Pythia (v 8.2.44)~\cite{Sjostrand:2014zea} was used to create a pool of events and predict production rates for proton collisions at various momenta, and a GEANT4-based simulation was subsequently used to predict proton inelastic interactions with 120~GeV primary protons impinging on the target. For each inelastic interaction, we randomly pick a Pythia event from the pool of events generated at the corresponding momentum, with a weight proportional to the rate predicted by Pythia. In doing this, we neglect the effect of the magnetic horns since these heavy particles decay very promptly and, therefore, it is safe to assume that their production will be similar for the PHF and NHF modes. The average number of parent mesons and $\tau$ leptons per PoT produced in the target\footnote{The production rates reported in Tab 3.1 of version 1 of Ref.~\cite{Berryman:2019dme} are a factor 2-3 smaller. The reinteractions that we take into account lead to higher production of low energy pions, which do not have a significant impact in the final sensitivity given their low collimation. For the heavier mesons the discrepancy is due to the use of different Pythia versions. The authors of Pythia looked into the issue and confirmed a bug was introduced in version 8.240 of Pythia that led to the lower rates found in Ref.~\cite{Berryman:2019dme}.} are listed in Tab.~\ref{tab:parent_PoT}.

\begin{table}[htb!]
\begin{center}
\begin{small}
\renewcommand{\arraystretch}{1.8}
\begin{tabular}[t]{ |c| c| c| c| c| c| }
\hline
 & $\pi$ & $K$ & $\tau$ & $D$ & $D_s$  \\
\hline
\hline
$P^+$/PoT & 6.3 & 0.75 & $2.1 \cdot 10^{-7}$ & $1.2\cdot 10^{-5}$ & $3.3 \cdot 10^{-6}$ \\ \hline
$P^-$/PoT & 5.7 & 0.33 & $3.0 \cdot 10^{-7}$ & $1.9\cdot 10^{-5}$ & $4.6 \cdot 10^{-6}$\\
\hline
\end{tabular}
\caption{\label{tab:parent_PoT} Average number of positive and negative parent mesons and $\tau$ leptons $P$ per PoT produced in the target. }
\end{small}
\end{center}
\end{table}

Tab.~\ref{tab:HNL_production} summarizes the different HNL production channels that have been included in our analysis. This set contains the dominant leptonic and semileptonic decays into heavy neutrinos of the parent mesons ($\pi$, $K$, $D$ and $D_s$) produced in the target. Moreover, since the $D$ and $D_s$ decay very promptly and have sizable branching ratios to $\tau$ leptons, a significant $\tau$ production rate is expected. This provides an additional production mechanism for HNL masses below the $\tau$ mass controlled by $\vert U_{\tau 4}\vert^2$, allowing DUNE to significantly improve the sensitivity to this more elusive mixing matrix element. All decay modes of the $\tau$ could allow to produce a HNL in the final state, provided that it is kinematically allowed. Nevertheless, we have opted to conservatively consider only the $\tau$ decay modes $\tau^-\to \rho^- N_4$, $\tau^-\to \pi^- N_4$ and $\tau^-\to \ell_\alpha^- N_4 \bar{\nu}$. Unlike for the production from meson decays, we did not provide their explicit expressions in Sec.~\ref{sec:prod}, since they would be the same as for the corresponding $N_4$ decays in Eqs.~(\ref{eq:width_N_V_l}), (\ref{eq:width_N_P_l}) and (\ref{eq:width_N_la_lb}), respectively.   The decays with 3 or more mesons in the final state have been neglected since the phase space is reduced for the production of a massive particle and the simulation of the HNL kinematics is more challenging for these channels (see Sec.~\ref{subsec:multimesons}).
\begin{table}[htb!]
\begin{small}
\renewcommand{\arraystretch}{1.8}
\begin{tabular}[t]{ |c c c| }
\hline
\textbf{Parent} & \textbf{2-body decay} & \textbf{3-body decay} \\
\hline
\hline
$\pi^+ \to$ & $e^+ N_4$ &  \textemdash\\
        & $\mu^+ N_4$ & \\
\hline
$K^+ \to$& $e^+ N_4$ & $\pi^0 e^+ N_4$ \\
      & $\mu^+ N_4$ & $\pi^0 \mu^+ N_4$ \\
\hline
$\tau^- \to$ & $\pi^- N_4$ & $e^- \overline{\nu} N_4$ \\
         &     $\rho^-  N_4$        & $\mu^- \overline{\nu} N_4$ \\
\hline
\end{tabular}
\hspace{0.05\textwidth}
\begin{tabular}[t]{ |c c c| }
\hline
\textbf{Parent} & \textbf{2-body decay} & \textbf{3-body decay} \\
\hline
\hline
$D^+ \to$ & $e^+ N_4$ & $e^+ \overline{K^0} N_4$ \\
      & $\mu^+ N_4$ & $\mu^+ \overline{K^0} N_4$\\
      & $\tau^+ N_4$ & \\
\hline
$D_s^+ \to$ & $e^+ N_4$ & \textemdash \\
        & $\mu^+ N_4$ & \\
        & $\tau^+ N_4$ & \\
\hline
\end{tabular}
\caption{\label{tab:HNL_production} List of 2-body and 3-body decays into HNLs, for the parent particles considered in this work. The decay channel $\tau^- \to \pi^- \pi^0 N_4$ is simulated via the approximation $\tau^-\to \rho^-  N_4\,, ~\rho^- \to \pi^- \pi^0$ as discussed in the text.}
\end{small}
\end{table}

Once we have obtained an expected flux of HNL entering the detector, this is then matched to the 22 different decay modes into SM particles studied in Sec.~\ref{sec:decays} (and shown in Figs.~\ref{fig:branching_ratios_deg} and~\ref{fig:branching_ratios_flav}) according to their corresponding branching ratios to obtain the expected signal at the detector. 

In the remainder of this section we first illustrate the impact on the detector acceptance due to the boost of the HNL, and then we compute the expected number of heavy neutrino decays inside the DUNE ND to estimate its sensitivity.

\subsection{The effect of the HNL mass on the detector acceptance}
\label{subsec:boost}

\begin{figure}[htb!]
\centering
\includegraphics[width=0.49\textwidth]{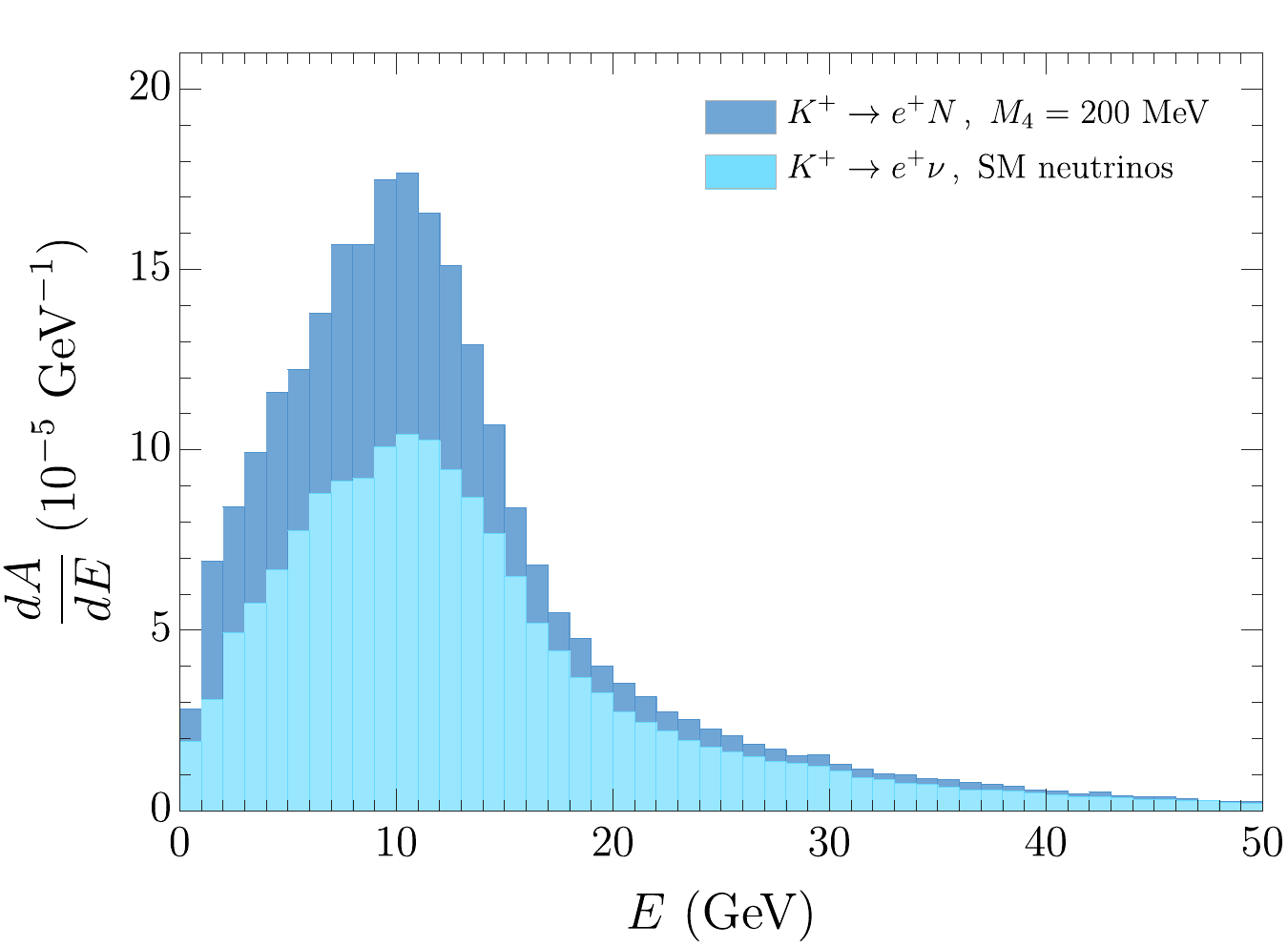}
\includegraphics[width=0.49\textwidth]{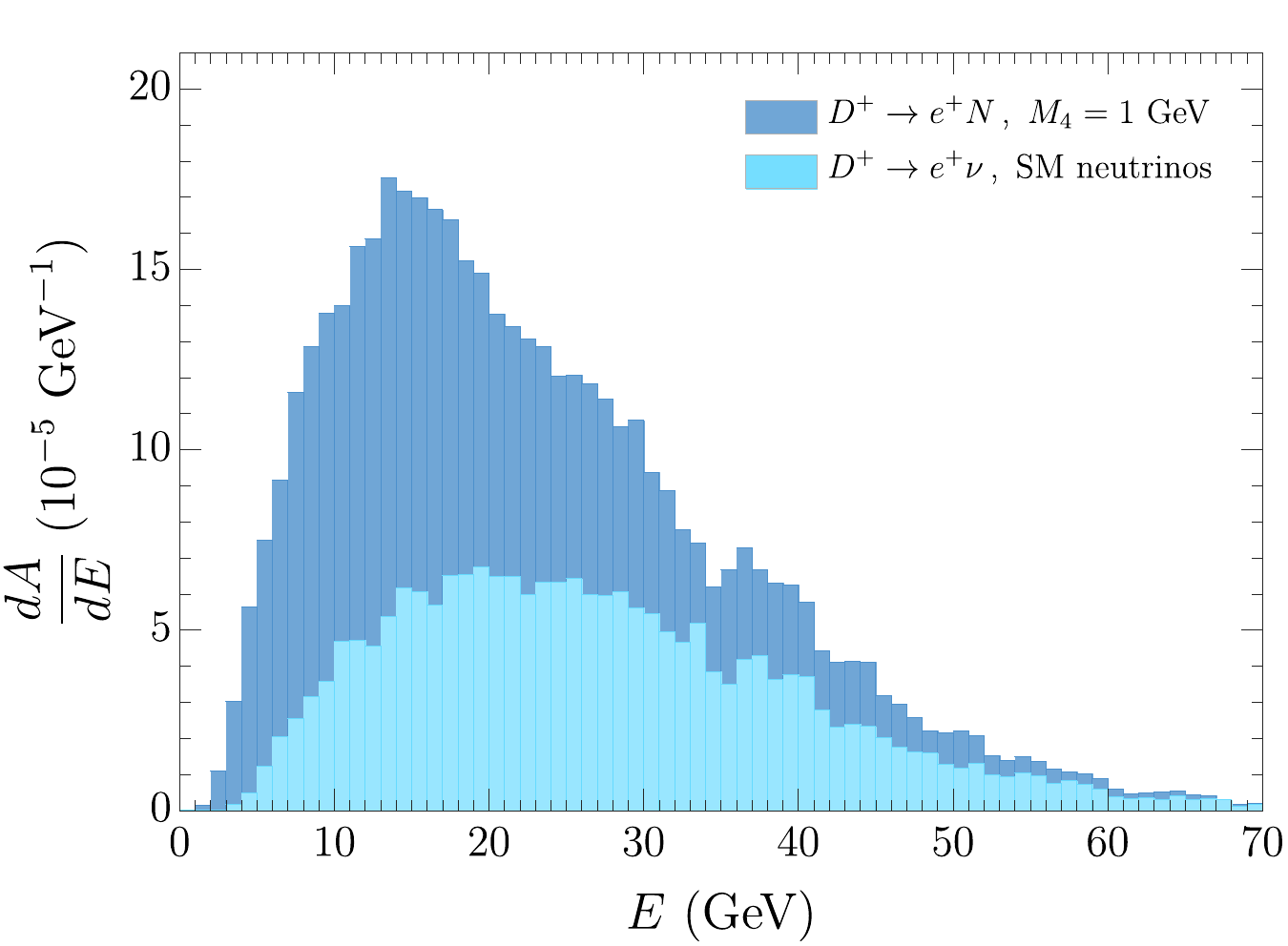}
\caption{Detector acceptance (dark blue in both panels) as a function of the neutrino energy, for neutrinos with a mass of 200~MeV,  produced from $K^+ \to e^+ N_4$ decays (left panel), and neutrinos with a mass of 1~GeV, produced from $D^+ \to e^+ N_4$ decays (right panel). In both panels, the light blue histogram shows the detector acceptance when the neutrino mass is set to zero.}
\label{fig:energy_spectrum}
\end{figure}

The effect of the boost in the beam direction is more efficient for particles with smaller velocities. Therefore, a larger detector acceptance is obtained when the effect of the heavy neutrino mass is properly included in the flux simulations, compared to estimations of the HNL flux based on the massless neutrino distributions. In order to illustrate the effect of the boost on the detector acceptance, we simulate the heavy neutrino flux at the DUNE ND from meson decays, and we compare it to the result obtained for the light neutrino flux. Our results are shown in Fig.~\ref{fig:energy_spectrum}, for neutrinos produced from kaon decays (left panel, where $M_4 = 200~\mathrm{MeV}$) and from $D$ decays (right panel, where $M_4 = 1~\mathrm{GeV}$). 
As can be seen in this figure, the increase in acceptance is considerable: up to a factor of two for 200~MeV neutrinos from kaon decays, and up to a factor of three for 1~GeV neutrinos coming from $D$ decays. The effect of the boost will also lead to a distortion in the expected spectra due to the different dependence of the detector acceptance with the neutrino energy, which can be seen from the comparison of the shape of the light and dark histograms in each panel. The net result is a relative increase in the number of neutrinos at low energies that enter the ND, given their smaller velocities and hence stronger collimation.   

\begin{figure}[htb!]
\centering
\includegraphics[width=0.6\textwidth]{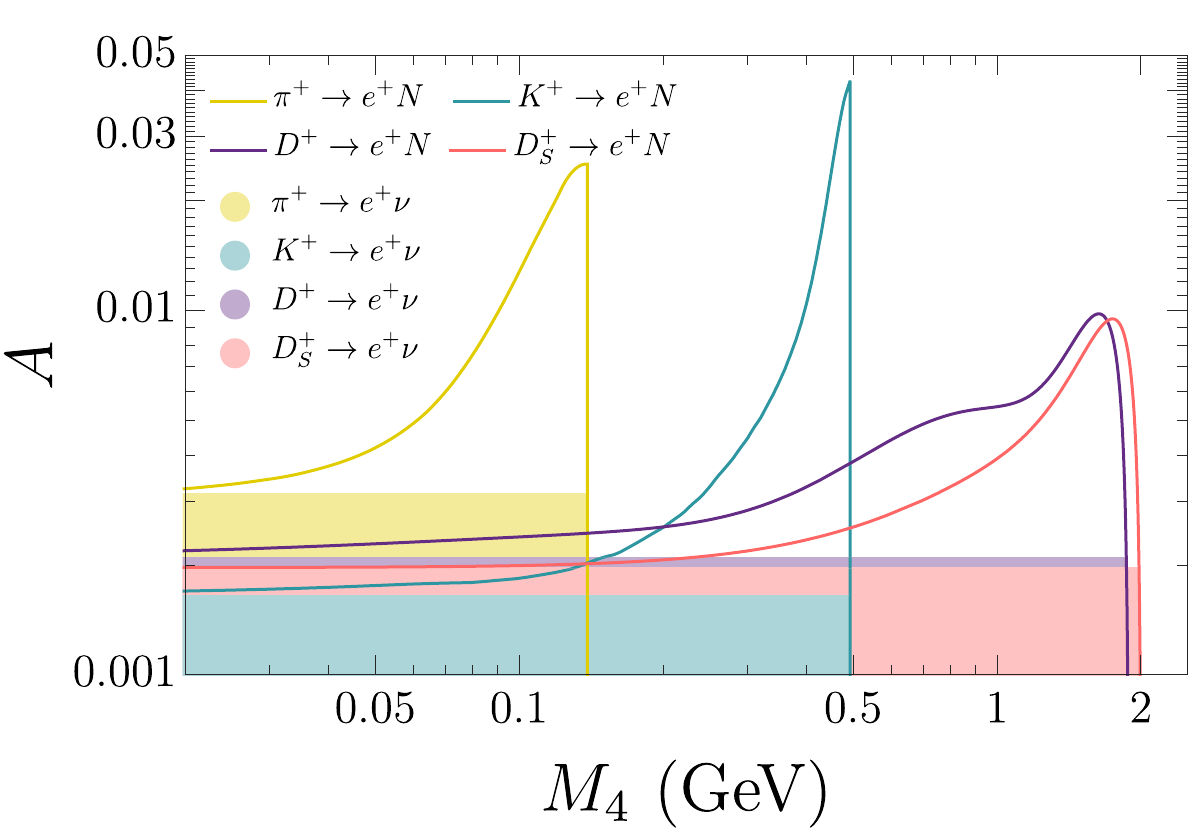}
\caption{Total detector acceptance as a function of the heavy neutrino mass. For reference, the total acceptance of the detector in the light neutrino case is indicated by the shaded regions: $3.2\cdot 10^{-3}$ for neutrinos produced from pion decays, $1.7\cdot 10^{-3}$ for neutrinos from kaon decays, $2.1\cdot 10^{-3}$ for neutrinos from $D$ decays, and $2.0\cdot 10^{-3}$ for neutrinos from $D_s$ decays.}
\label{fig:events}
\end{figure}

Finally, Fig.~\ref{fig:events} shows the total detector acceptance, after integrating over the neutrino energy, as a function of the HNL mass. Note that the acceptance is expected to be different depending on the parent meson that produced the neutrino due to the effect of the horns: while pions are typically very well-focused at a long-baseline experiment, this is not the case for heavier mesons, which are not only harder to focus due to their larger masses but also decay much faster. This effect is most significant for $D$ and $D_s$ mesons, which decay very promptly and therefore are practically unaffected by the horn focusing system. In Fig.~\ref{fig:events} we show different lines for neutrinos obtained from different meson decays, as indicated by the labels. As can be seen, as the heavy neutrino mass approaches the production threshold and its velocity decreases accordingly, the acceptance grows very rapidly given the stronger boost in the beam direction. Notice however that also the phase space is decreasing and hence the number of total HNL events will also be reduced. 

\subsection{Expected sensitivity to HNL decays}

Once the flux of heavy neutrinos $d\phi_N/dE_N$ that reach the ND has been computed numerically as a function of the neutrino energy $E_N$, the total number of expected neutrino decays into a given decay channel $c$ inside the DUNE ND can be expressed as
\begin{equation}
N_c ({\rm{ND}})={\rm{BR}}_c \times \int dE_N P(E_N) \frac{d\phi_N}{dE_N} \, ,
\end{equation}
where BR$_c$ is the branching ratio of the corresponding decay channel and $P(E_N)$ stands for the probability of the heavy neutrino decaying inside the ND (which depends on the boost factor and therefore on the neutrino energy):
\begin{equation}
P(E_N)=e^{-\frac{\Gamma  L}{\gamma \beta} } \left(1- e^{-\frac{\Gamma \Delta \ell_{\rm det}}{\gamma \beta }}\right) \, .
\label{eq:prob}
\end{equation}
Here, $\Gamma$ is the total decay width of the heavy neutrino in its rest frame, while $\gamma = E_N/M_4$, $\beta = |\vec{p_N}| / E_N$ and $\vec{p_N}$ stands for the neutrino momentum. $L$ is the distance between the HNL production and the ND while $\Delta \ell_{\rm det}$ is the length of the HNL trajectory inside the detector. 

From Eq.~\eqref{eq:prob} it is easy to see that the neutrino must be sufficiently long-lived to reach the ND, or otherwise the number of decays will be exponentially suppressed. This will be the case for large enough energies and small enough matrix elements $U_{\alpha 4}$, which correspond to the most interesting region of the parameter space. In this limit, $\Gamma L \ll \gamma \beta$, and the decay probability can be further approximated as 
\begin{equation}
P(E_N) \approx \frac{\Gamma \Delta \ell_{det}}{\gamma\beta} \, .
\label{eq:prob-approx}
\end{equation}
Nevertheless, this approximation does not hold anymore for large masses and mixings, and thus we will employ Eq.~\eqref{eq:prob} to compute the probability of the HNL decaying inside the detector.

Given that the neutrino flux entering the detector will be directly proportional to its aperture, and that the probability for the neutrino to decay inside the ND is proportional to $\Delta \ell_{det}$, it is easy to see that the sensitivity to heavy neutrino decays will scale with the volume of the ND.

\begin{figure}[htb!]
\centering
\includegraphics[width=0.49\textwidth]{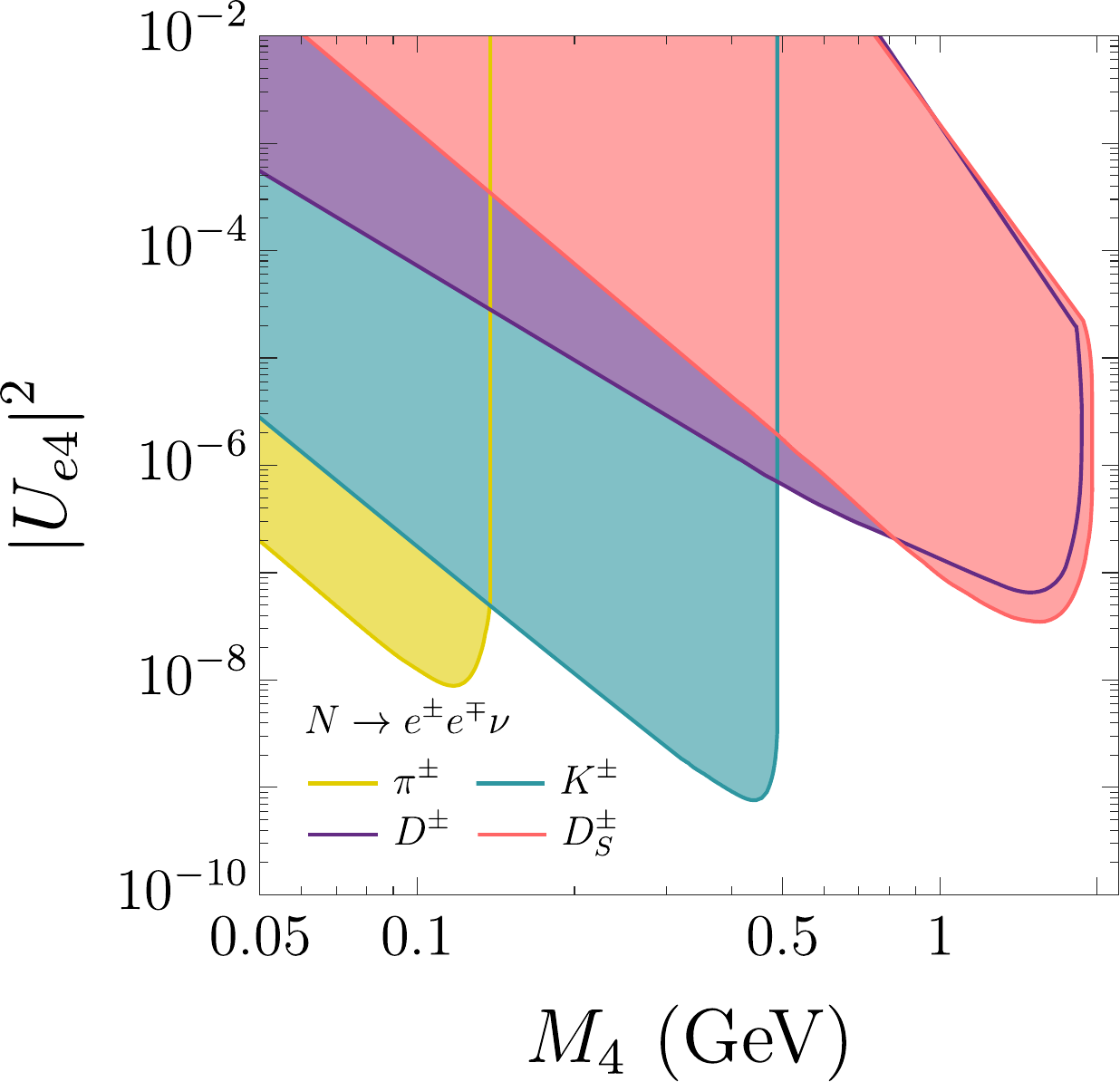}
\caption{90\% CL contour for the expected sensitivity to the mixing $|U_{e4}|^2$ as a function of the heavy neutrino mass for a total amount of $ 7.7\cdot 10^{21}$ PoT. The different regions show the contributions obtained when the heavy neutrinos are produced from the decays of a given parent meson, as indicated by the labels.  \label{fig:result-per-parent} }
\end{figure}

In order to compute the final sensitivity to HNLs, a detector simulation should be performed, including relevant background contributions from SM neutrino interactions in the ND. Such a fully detailed detector simulation is beyond the scope of this work, where we rather show an estimate to the sensitivity of DUNE as an application of the methods derived in the previous sections. The main source of background for this search comes from neutrino interactions in the detector volume, and is very significant. In Ref.~\cite{Ballett:2019bgd} the background rates for Argon were estimated at $\sim 3\cdot 10^5~\rm{events/ton/}10^{20}~\rm{PoT}$. Fortunately, SM neutrino events present a very different topology than that of heavy neutrino decays, and a series of kinematic cuts can heavily reduce the expected background and bring it down to a negligible level. This was the case, for example, for the T2K near detector HNL search performed in Ref.~\cite{Abe:2019kgx} (which also used a gas TPC). Therefore, following Refs.~\cite{Ballett:2019bgd,Abe:2019kgx}, hereafter we will assume that this is achievable and show our expected sensitivity contours to heavy neutrino decays under the assumption of no background. We also assume that the cuts applied to reduce the background will translate into similar signal efficiencies in our case as those obtained in Ref.~\cite{Abe:2019kgx}. Although the efficiency will eventually depend on the mass of the HNL and the considered decay channel (see Fig. 4 in Ref.~\cite{Abe:2019kgx}), here we use 20\% as an educated guess. Finally, for our sensitivity contours we estimate the 90~\% confidence level (CL) sensitivity on the signal following the Feldman and Cousins~\cite{Feldman:1997qc} prescription for a Poisson distribution with no background and under the hypothesis of no events being observed, which corresponds to the expected number of signal events being smaller than 2.44.  

\pagebreak

Before showing our sensitivity contours, we show in Fig.~\ref{fig:result-per-parent} an example to illustrate the relative importance of the different HNL production mechanisms on the results. In this example, we show the contours obtained under the assumption that the heavy neutrino mixes primarily with the $e$ sector. The different regions show the contributions obtained when the heavy neutrinos are produced from the decays of a given parent meson, as indicated by the labels. The signature in this case would be electron-positron pairs, corresponding to the decay $N_4 \to \nu e^+ e^-$. As can be seen, for $M_4 < m_\pi$ the leading production mechanism is $\pi^\pm$ decay. For masses in the region $m_\pi < M_4 < m_K$, $K^\pm$ dominates and, in fact, the sensitivity contour reaches lower values of $U_{e4}$ at the best point (in spite of the smaller number of kaons produced, when compared to the number of pions). The reason for this is that for $M_4 < m_\pi$ the heavy neutrino becomes very long-lived, leading to a reduced number of decays inside the detector and a consequent reduction in sensitivity. On the other hand, in the heavy mass region ($M_4 > m_K$) the heavy neutrino is predominantly produced from either $D$ or $D_s$ meson decays (although there is a subdominant contribution from $\tau$ decays). While the $D_s$ is heavier (and therefore more difficult to produce) than $D$ mesons, its decay to heavy neutrinos is mediated by the CKM element $V_{cs}$ instead of $V_{cd}$. This compensates for the reduced meson production rate and, as a result, the sensitivity in this region is dominated by $D_s$ decays. Finally, the different slope as a function of $M_4$ for the $D$ contribution is simply due to the fact that, unlike for the $\pi$, $K$ and $D_s$ decays, the $D$ meson production of $N_4$ is dominated by three-body decays instead of two-body (see Sec.~\ref{sec:prod}).

\begin{figure}[H]
\centering
\includegraphics[height=5.748cm,keepaspectratio]{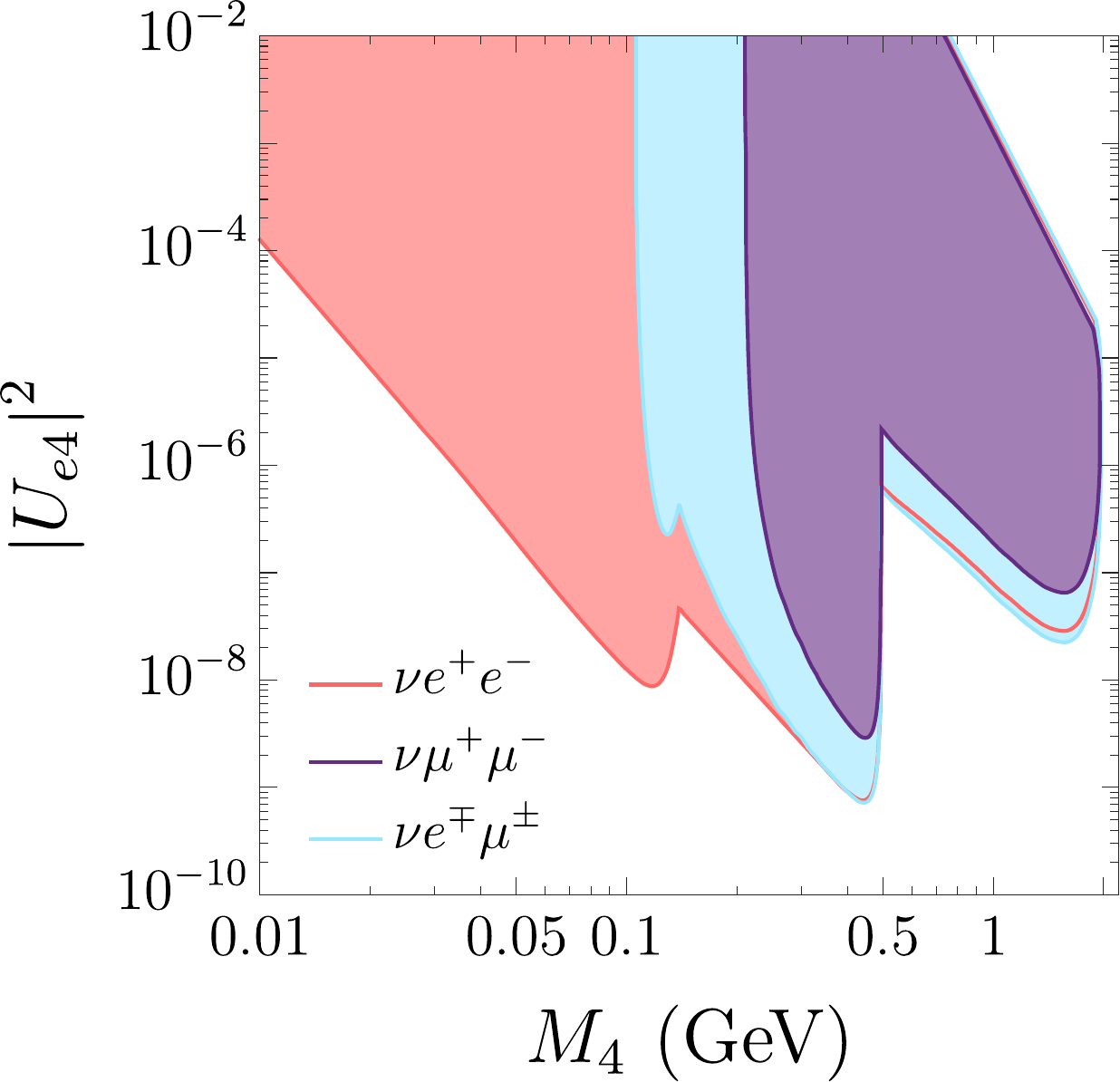} \hspace{-0.52 cm}
\includegraphics[height=5.56cm,keepaspectratio]{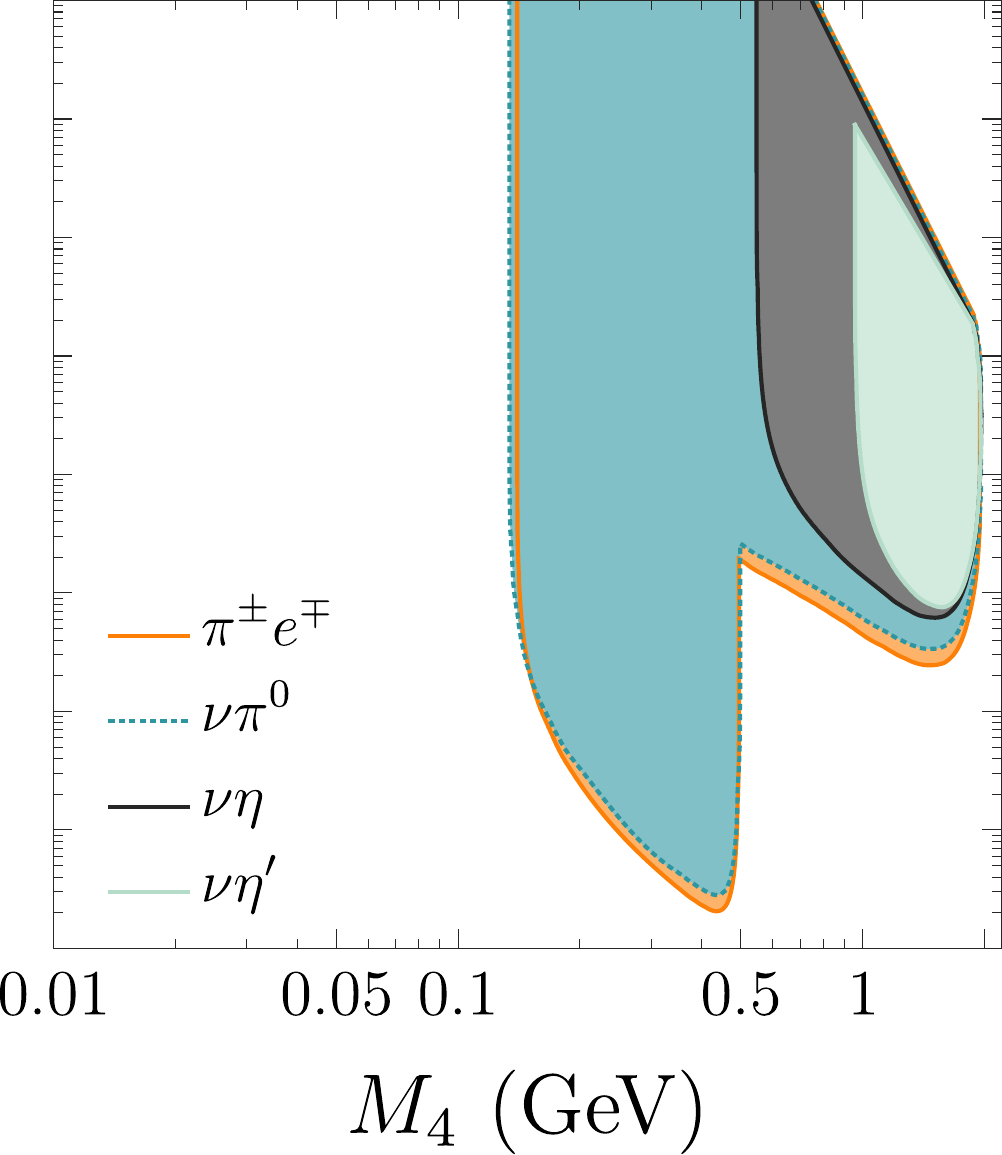} \hspace{-0.52 cm}
\includegraphics[height=5.56cm,keepaspectratio]{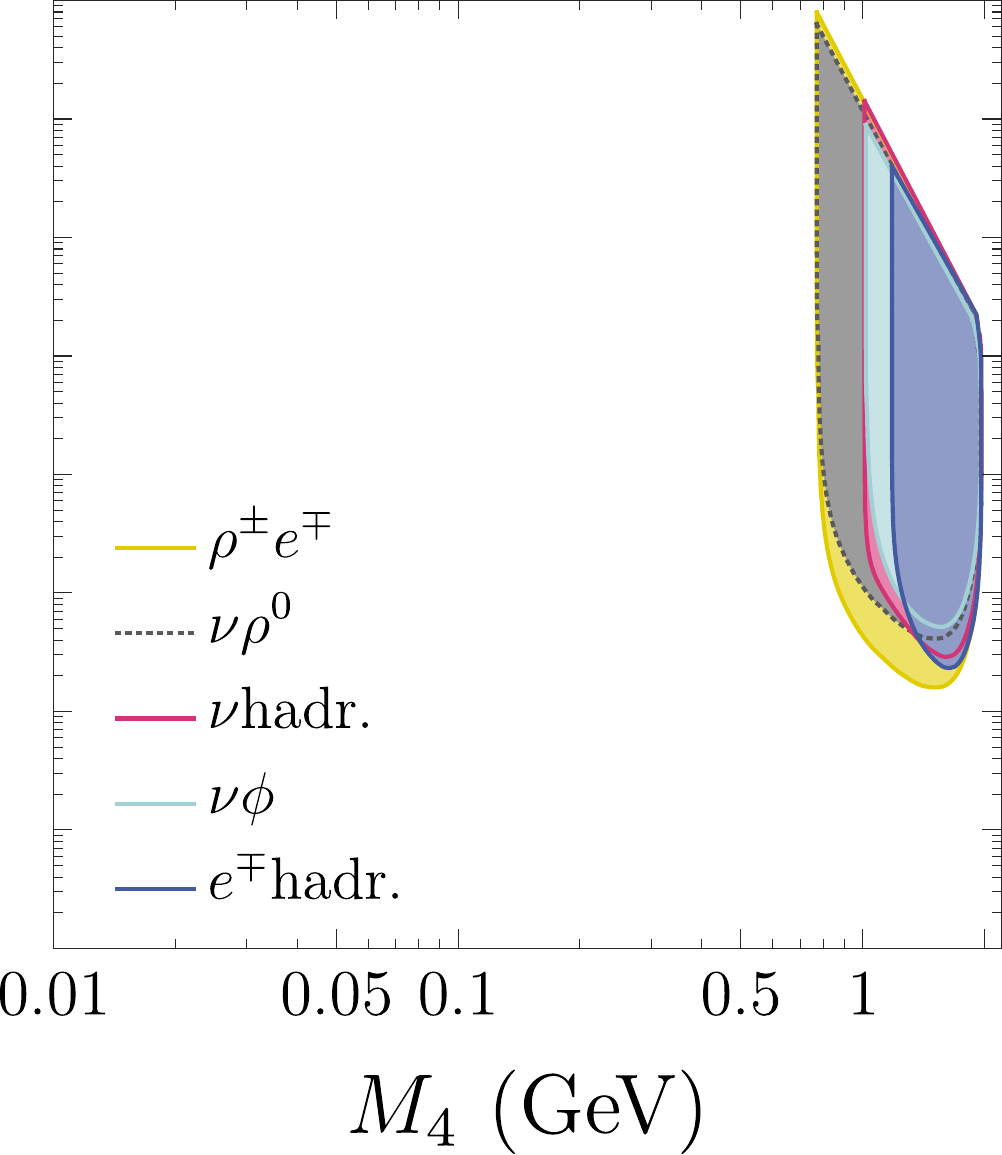}
\includegraphics[height=5.748cm,keepaspectratio]{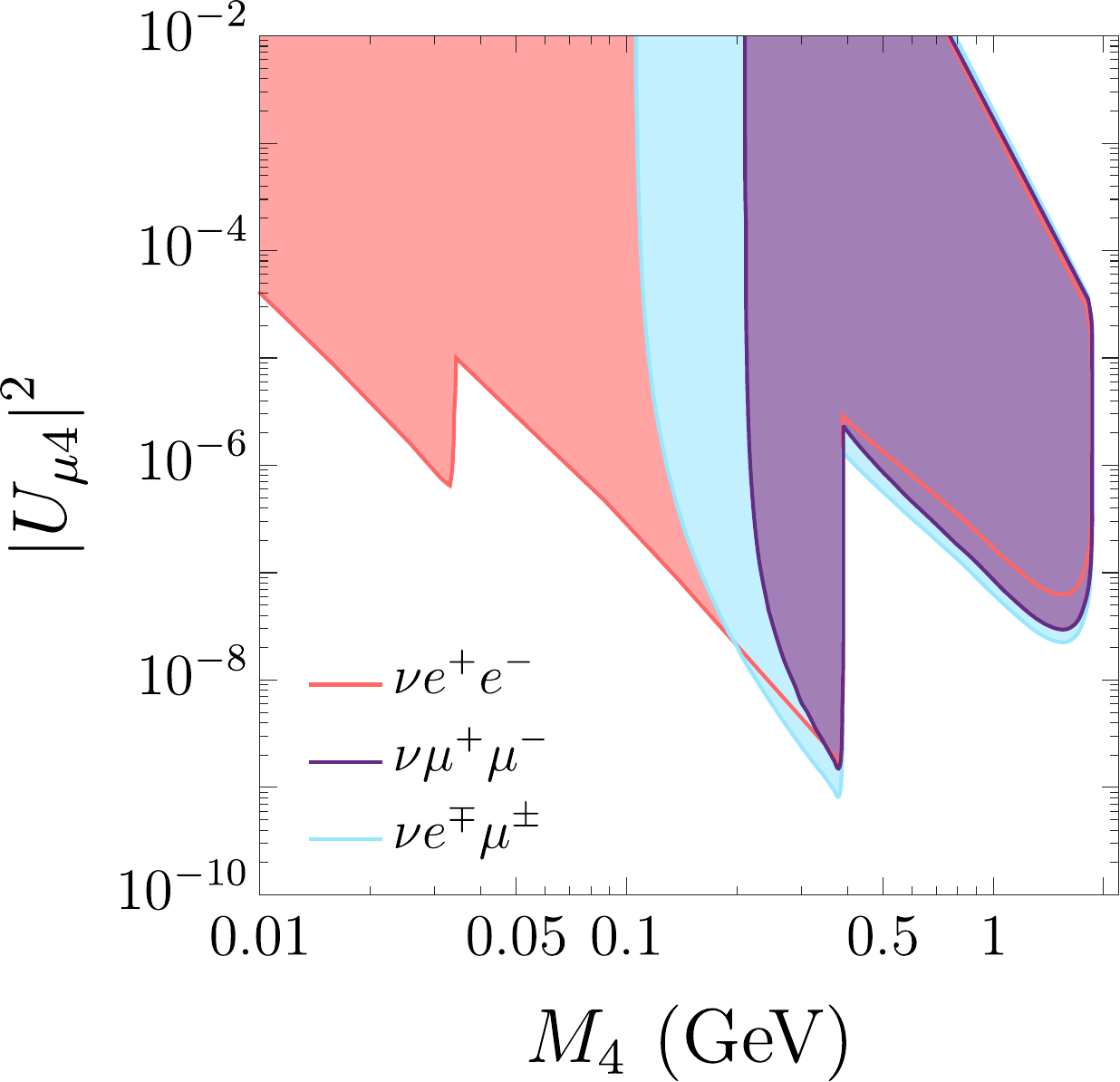} \hspace{-0.52 cm}
\includegraphics[height=5.56cm,keepaspectratio]{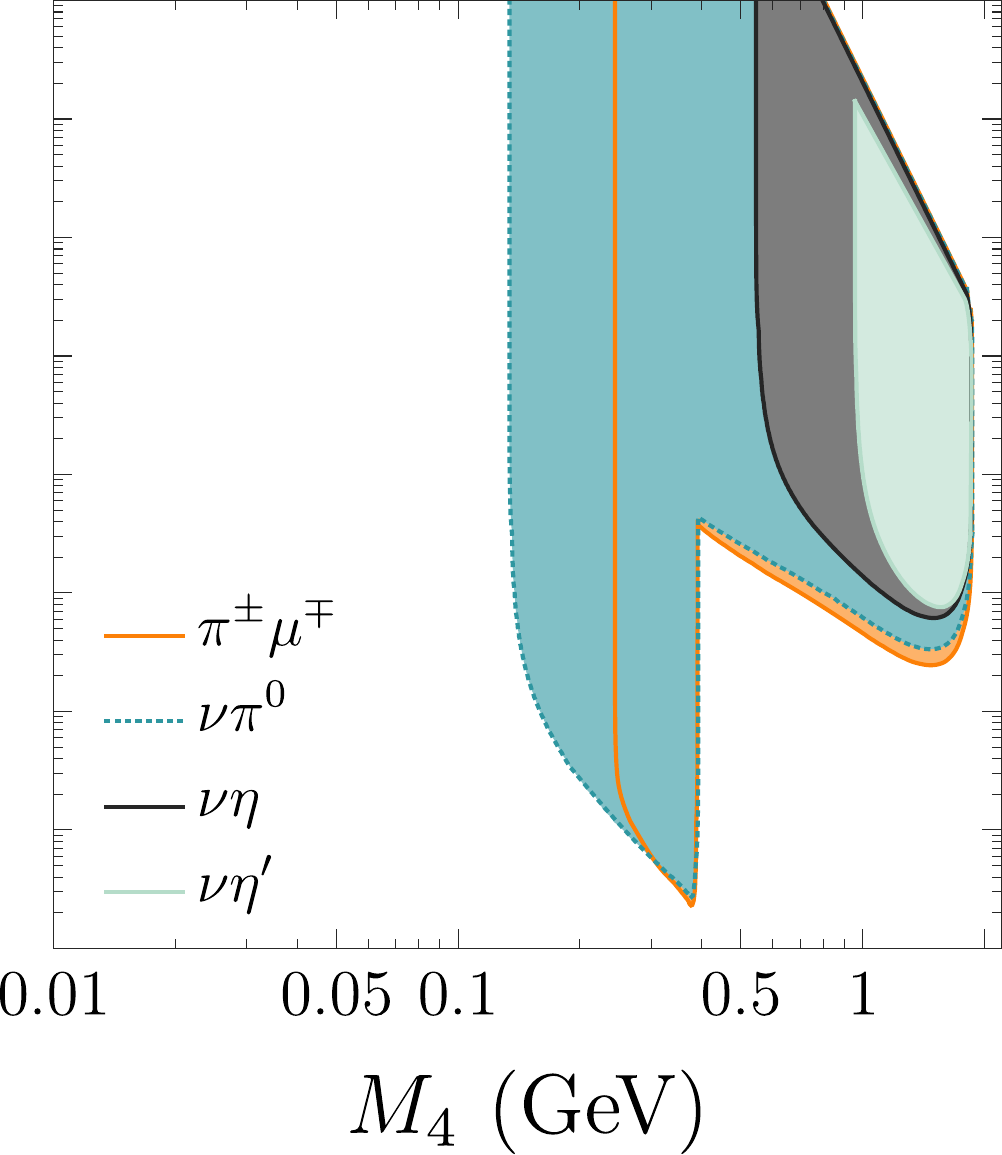} \hspace{-0.52 cm}
\includegraphics[height=5.56cm,keepaspectratio]{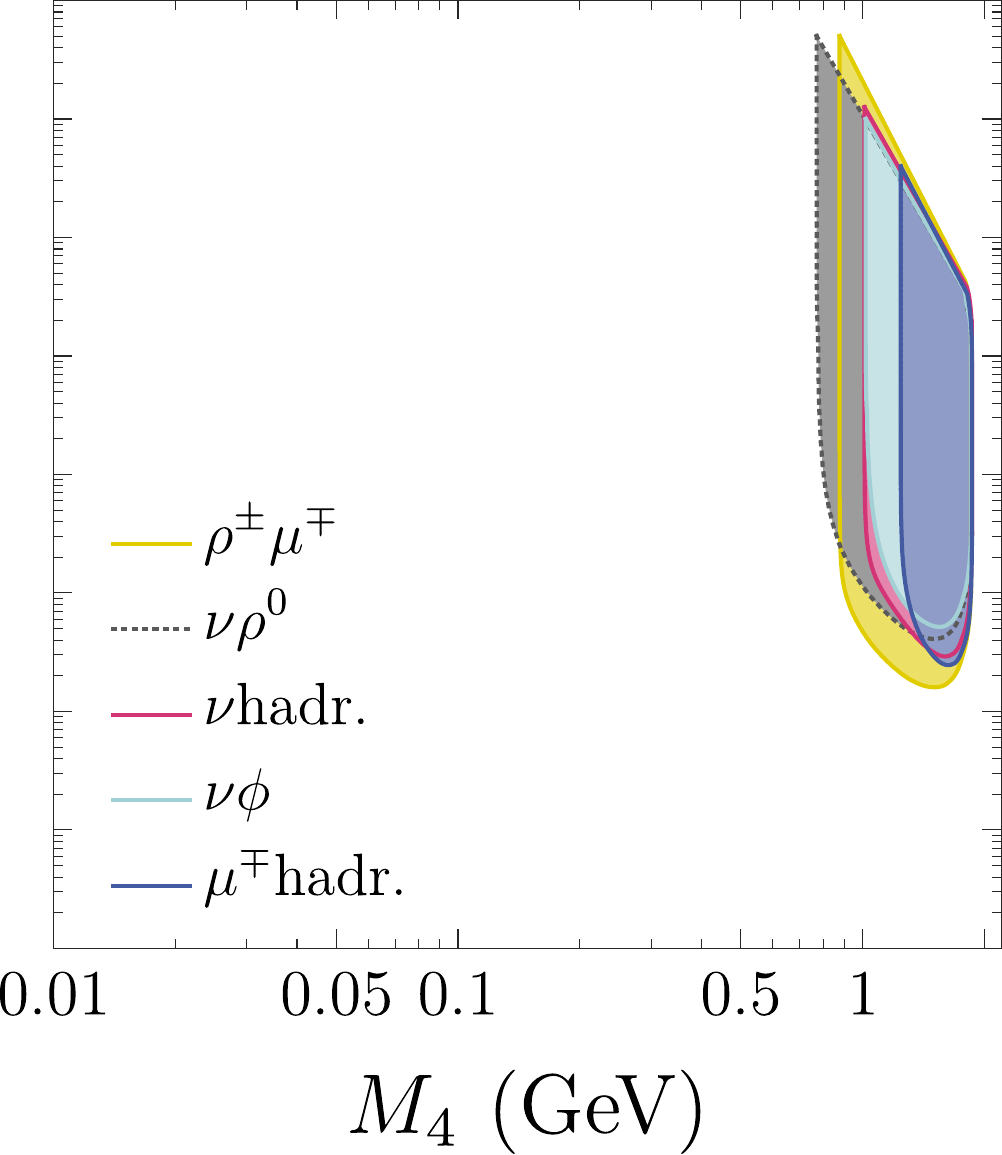}
\includegraphics[height=5.748cm,keepaspectratio]{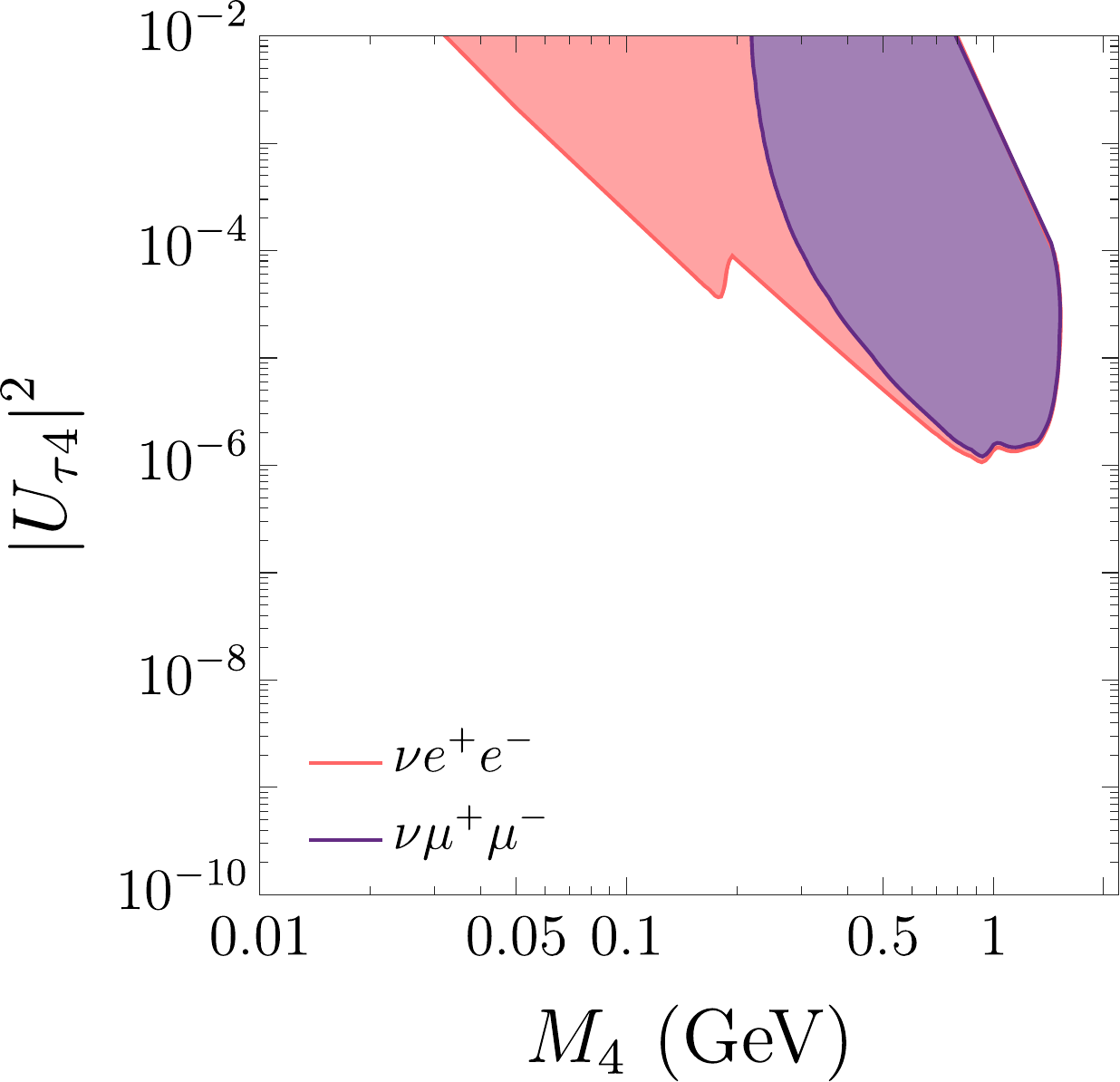} \hspace{-0.52 cm}
\includegraphics[height=5.56cm,keepaspectratio]{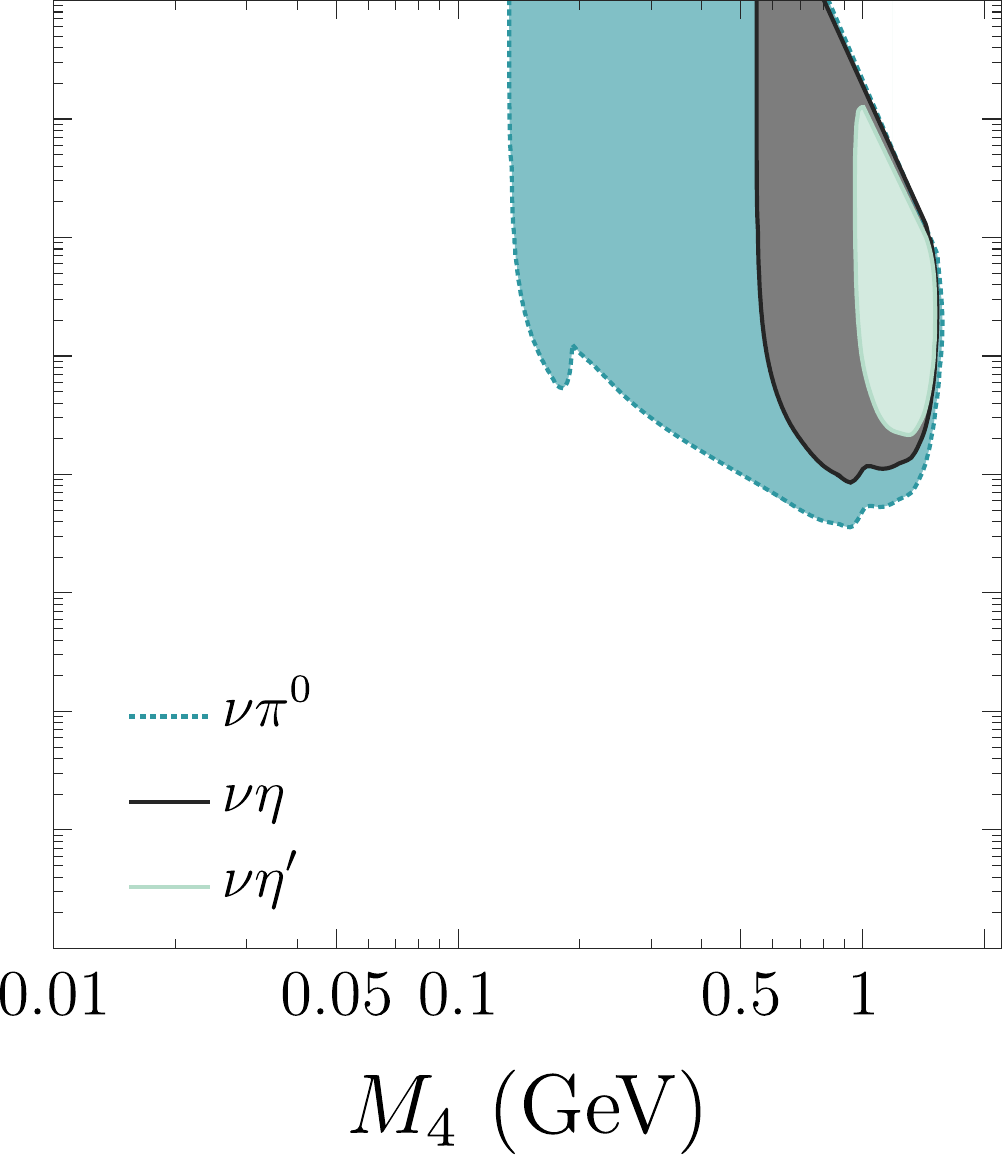} \hspace{-0.52 cm}
\includegraphics[height=5.56cm,keepaspectratio]{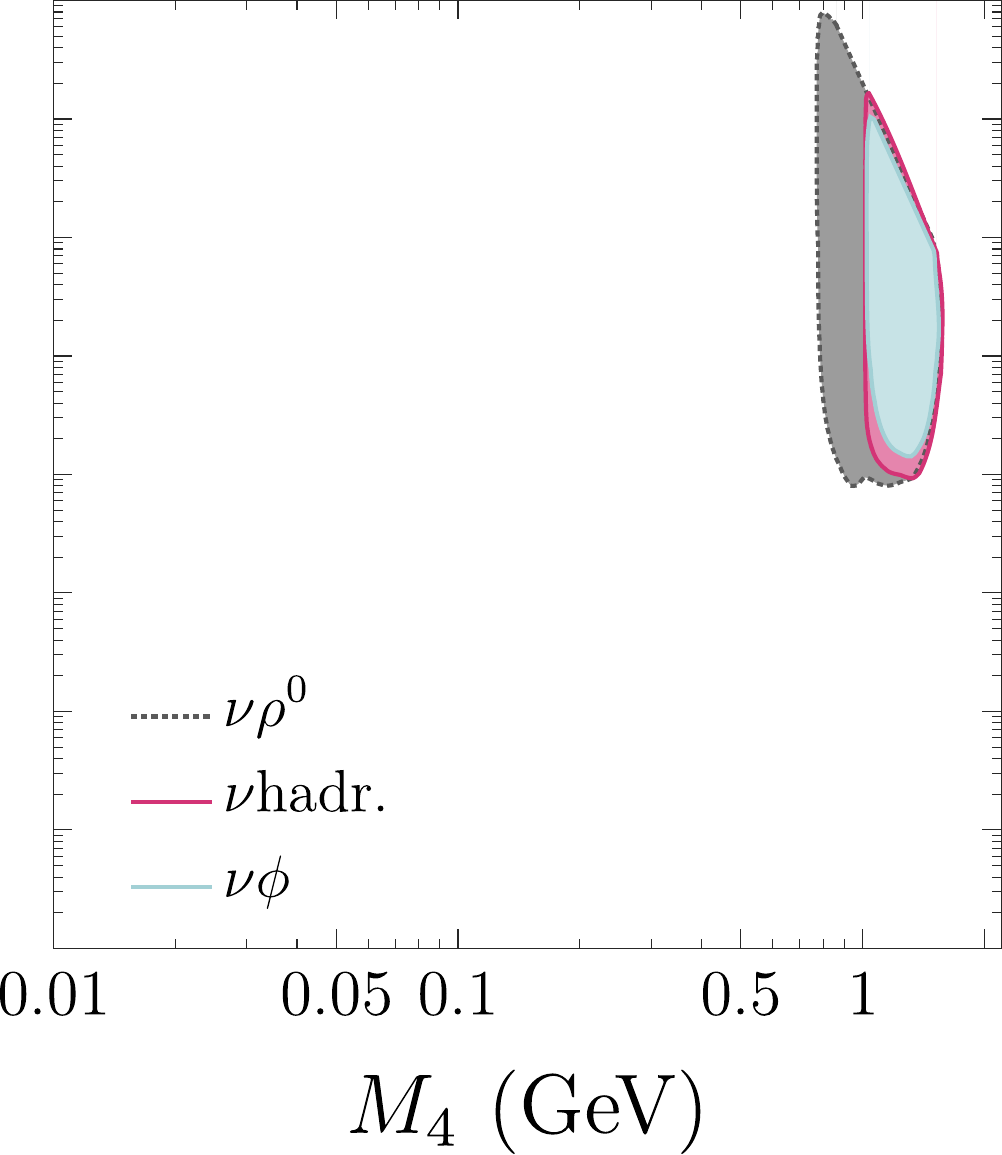}
\caption{Expected DUNE sensitivity (at $90\%$ CL) to the mixing matrix elements $\vert U_{\alpha 4}\vert^2$ as a function of the heavy neutrino mass, for a total of $7.7 \cdot 10^{21}$ PoT collected. In each row, we assume that the HNL only couples to one of the charged leptons as indicated, while the other two mixings are set to zero. The different regions correspond to the results for different final states as indicated by the labels. Left panels correspond to signatures with charged leptons and missing energy, while middle (right) panels correspond to signatures with pseudoscalar (vector) mesons in the final state. In our analysis, we assume a negligible background level after cuts and a signal selection efficiency of 20\%, see text for details. \label{fig:sens}}
\end{figure}

Fig.~\ref{fig:sens} shows the sensitivity contours in the $M_4 - \vert U_{\alpha 4}\vert^2$ plane at 90\% CL, for different decay channels as indicated. The upper, middle and lower panels in the figure show the results assuming that the heavy neutrino mixes predominantly with the $e$, $\mu$ and $\tau$ sectors respectively. 

\begin{figure}[htb!]
\centering
\includegraphics[height=5.742cm,keepaspectratio]{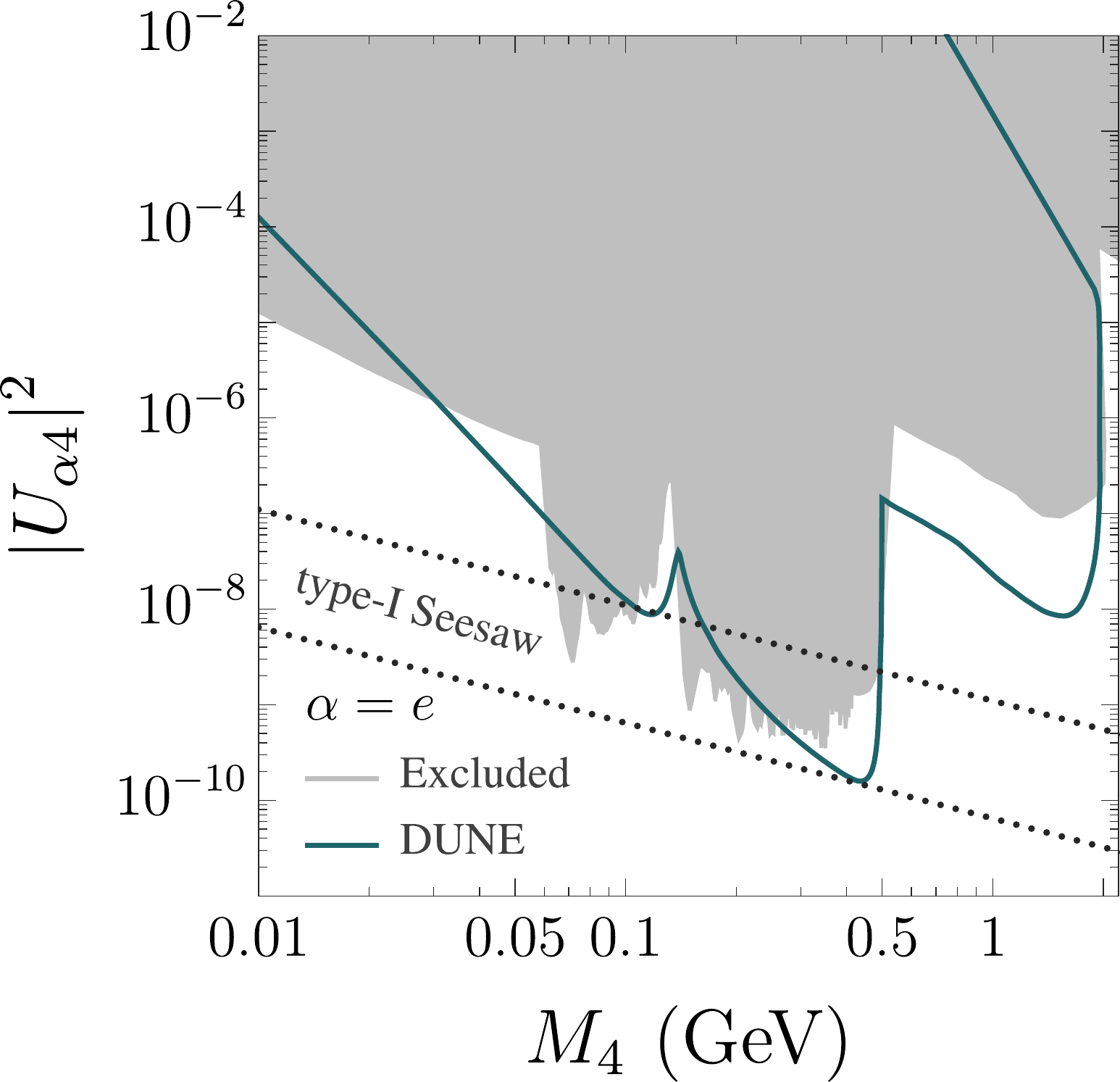} \hspace{-0.52 cm}
\includegraphics[height=5.56cm,keepaspectratio]{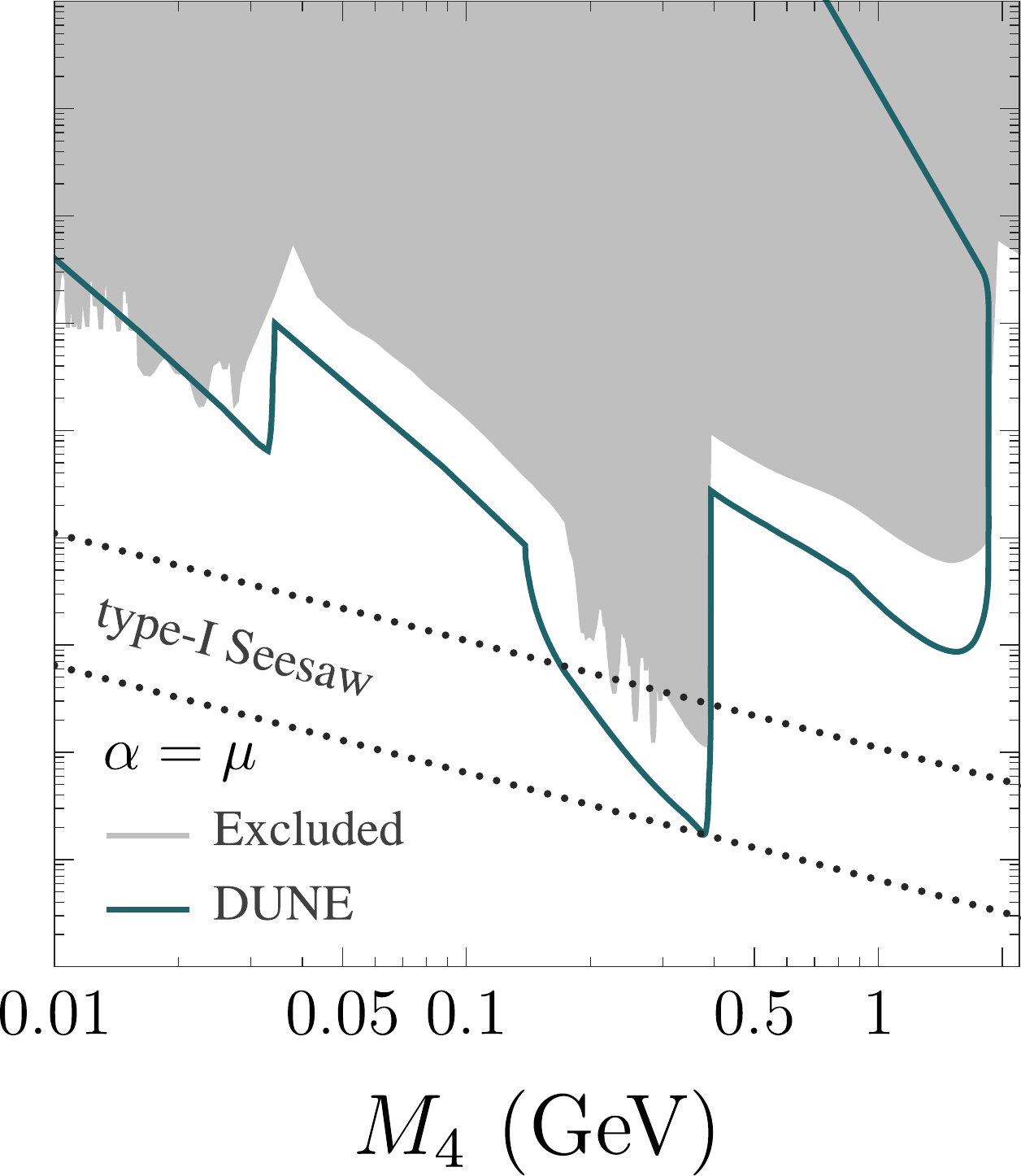} \hspace{-0.52 cm}
\includegraphics[height=5.56cm,keepaspectratio]{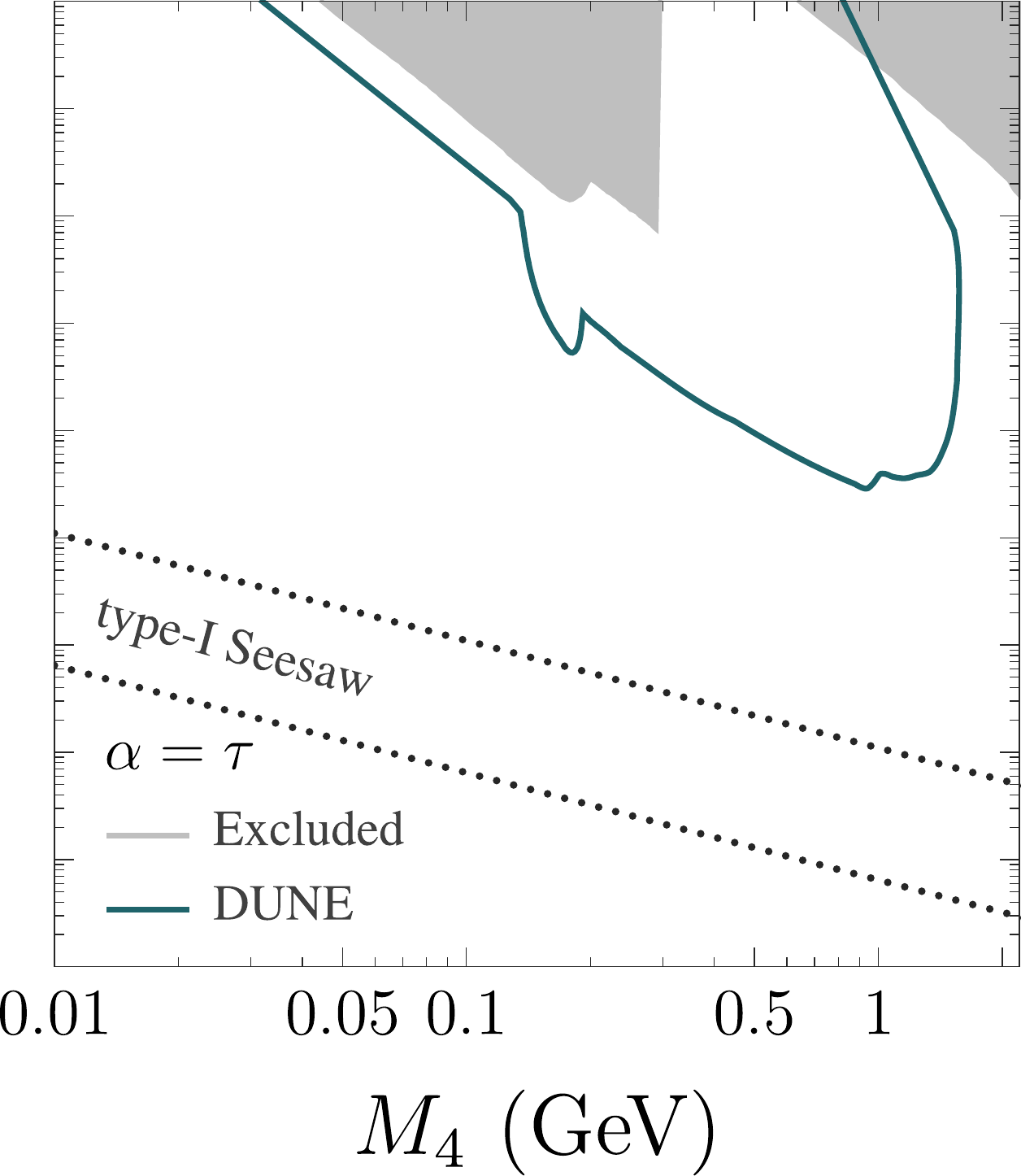}
\caption{Expected DUNE sensitivity (at $90\%$ CL) to the mixing matrix elements $\vert U_{\alpha 4}\vert^2$ as a function of the heavy neutrino mass, for a total of $7.7 \cdot 10^{21}$ PoT collected, combining all possible decay channels for the HNL leading to visible final states in the detector. Results are shown for a HNL coupled to $e$ (left panel), $\mu$ (middle panel), and $\tau$ (right panel). The shaded gray areas are disfavored at 90\% CL by present experiments. The dotted gray lines enclose the region of parameter space where a type-I Seesaw model could generate light neutrino masses in agreement with oscillation experiments and upper bounds coming from $\beta$-decay searches, see text for details. In our analysis, we assume a negligible background level after cuts and a signal selection efficiency of $20\%$. \label{fig:sens_total}}
\end{figure}

Finally, Fig.~\ref{fig:sens_total} summarizes in blue the 90\% CL expected sensitivities at the DUNE near detector to the heavy neutrino mixing $\vert U_{\alpha 4}\vert^2$ as a function of its mass, assuming a Dirac HNL. In the Majorana case, the increase in the number of events would translate into a slightly better sensitivity, although the results would be qualitatively very similar. In this last figure we combine the events from all the channels depicted in Fig.~\ref{fig:sens} under the same assumption of $20\%$ signal efficiency and negligible background, following Ref.~\cite{Abe:2019kgx}. We again estimate the sensitivity following the Feldman and Cousins \cite{Feldman:1997qc} prescription for a Poisson distribution under the hypothesis of no events being observed, which corresponds to the expected total number of signal events combining all channels leading to a visible final state in the detector being smaller than $2.44$. 

For comparison, the shaded gray areas indicate the parameter space disfavored by current experiments (at $90\%$ CL). Relevant bounds on $U_{e4}$ are obtained from results by the TRIUMF \cite{Britton:1992xv}, PIENU \cite{Aguilar-Arevalo:2017vlf}, NA62 \cite{NA62:2020mcv}, T2K \cite{Abe:2019kgx}, CHARM \cite{Bergsma:1985is}, BEBC \cite{CooperSarkar:1985nh} and DELPHI \cite{Abreu:1996pa} collaborations; for $U_{\mu4}$, by PSI \cite{Daum:1987bg}, PIENU \cite{Aguilar-Arevalo:2019owf},  PS191 \cite{Bernardi:1985ny,Bernardi:1987ek},\footnote{The PS191 collaboration did not consider NC-mediated HNL decays; thus,  their bound on $\vert U_{e4}U_{\mu 4}\vert$ can be translated into a bound on $\vert U_{\mu 4}\vert^2$.  This reinterpretation was performed in \cite{Kusenko:2004qc},  which accounted for both NC- and CC-mediated decays of the heavy neutrino.}  E949 \cite{Artamonov:2009sz}, T2K \cite{Abe:2019kgx},  NuTeV \cite{Vaitaitis:1999wq} and DELPHI \cite{Abreu:1996pa}; finally, $U_{\tau 4}$ is much harder to probe experimentally and here the only available constraints come from CHARM \cite{Orloff:2002de} and DELPHI \cite{Abreu:1996pa}\footnote{Note that the bounds from DELPHI are flavor independent.  For low neutrino masses,  the bounds on $U_{\tau 4}$ are looser than those for $U_{e4}$ and $U_{\mu 4}$ due to the kinematic suppression implied by the large mass of the $\tau$. This effect was accounted for in \cite{Atre:2009rg}.}. We find that DUNE is expected to improve over present constraints by several orders of magnitude in a large fraction of the parameter space and, in particular, for HNL masses between the $K$ and $D$ meson thresholds. 

As a target region, we have also indicated in Fig.~\ref{fig:sens_total} the naive expectation for the mixing matrix elements from the Seesaw mechanism: $\vert U_{\alpha 4}\vert^2 \sim m_i/M_4$, where $m_i$ stands for the SM neutrino masses. In particular, we set as the lower end of the band the minimum mass that at least one of the neutrinos must have to correctly reproduce the atmospheric mass splitting as measured in neutrino oscillations $\sqrt{\Delta m^2_\text{atm}}= 0.05$~eV. The upper line has been set using the latest bound of 1.1~eV from the KATRIN experiment~\cite{Aker:2019uuj}. We find that DUNE will be able to start exploring this interesting region, for HNL masses close to the $K$ mass. Notice that this is only a generic expectation from the Seesaw mechanism: individual elements of the mixing matrix could either exceed or fall below these limits.

If sufficiently long lived, HNLs could decay during Big Bang Nucleosynthesis (BBN), altering the prediction for the primordial abundance of light elements. Thus, too small mixings are disfavoured, especially at low masses, as they would imply too long lived HNLs. BBN constraints exclude squared mixings smaller than $\sim10^{-5}$ for a neutrino with a mass of 100 MeV, while the bounds are much looser for larger masses, only disfavouring squared mixings below $10^{-10}$ for an HNL with a mass of 1 GeV \cite{Dolgov:2000pj,Ruchayskiy:2012si,Gelmini:2020ekg,Sabti:2020yrt,Boyarsky:2020dzc}.\footnote{Note that BBN constraints do not significantly depend on the active flavor that dominates the mixing with HNLs.} Nevertheless, since these constraints rely on the cosmological history of the Universe, we choose not to display them together with direct laboratory tests in Fig.~\ref{fig:sens_total}.

The number of HNL events depends on both their production rate and their decay probability inside the detector. At low masses the heavy neutrino production is dominated by pion decay, which is roughly proportional to $\vert U_{\alpha 4}\vert^2M_4^2$ (see Eq.~(\ref{eq:width_P_N_l})). In this region, the most important HNL decay channel is $N_4\to \nu e^+e^-$, which is proportional to $\vert U\vert^2M_4^5$ (see Eq.~(\ref{eq:width_N_la_la})). Thus, according to Eq.~(\ref{eq:prob-approx}), the number of events should scale as $\vert U_{\alpha 4}\vert^4M_4^8$ (an extra $M_4$ power arises due to the $1/\gamma$ factor, proportional to $M_4$). We have indeed verified that the slopes in the low mass regions of Fig.~\ref{fig:sens_total} fit well to $\vert U_{\alpha 4}\vert^2\propto M_4^{-4}$, as expected.

We have also compared our results to similar studies in the literature~\cite{Krasnov:2019kdc,Ballett:2019bgd,Berryman:2019dme}, after the corresponding rescaling of the number of events accounting for the different detector volumes, PoT and efficiencies assumed, we find a rather good agreement between the four estimations of the DUNE sensitivity. We find the best overall agreement with Ref.~\cite{Krasnov:2019kdc}. The main difference is a slightly better sensitivity to the $U_{\tau 4}$ mixing in our results, in the small sensitivity peak we find around $M_4 \sim 1$~GeV, corresponding to the closure of the $\tau \to N_4 \rho^-$ production channel. This peak also seems absent in the other references. Regarding Ref.~\cite{Ballett:2019bgd}, the main differences we find are at the peaks in sensitivity at the kinematic thresholds of the meson masses, where we find better sensitivity. We believe that these differences are due to the effect of the boost factor on the detector acceptance discussed in subsection~\ref{subsec:boost}, which becomes most relevant close to the kinematic thresholds, as shown in Fig.~\ref{fig:events}. We also find that the sensitivity to $U_{\mu 4}$ for values of $M_4$ larger than the Kaon mass, is significantly smaller in Ref.~\cite{Ballett:2019bgd} as compared to the other estimations, which find a similar behavior to that of $U_{e 4}$, as expected from their similar branching ratios. Finally, we also find generally good agreement with Ref.~\cite{Berryman:2019dme}. The main differences are in the areas of parameter space were the HNL decays to $\rho^0$ and especially to $\pi^0$ are most relevant, since these decay modes were not included. The slope of the sensitivity curves is also slightly less steep than the $\vert U_{\alpha 4}\vert^2\propto M_4^{-4}$ found in the other references.

\section{Summary and conclusions}
\label{sec:conclusions}

The addition of at least two nearly-sterile neutrinos (or HNLs) to the SM particle content is the simplest extension of the SM capable of reproducing the observed pattern of neutrino masses and mixing. The Majorana mass scale, unlike the masses of the other elementary particles, is not related to the electroweak scale and is \emph{a priori} a free parameter of the model. The phenomenological consequences due to the existence of such heavy neutrinos would be very diverse depending on its value. In fact, while traditional type-I Seesaw models set their Majorana masses at very high energies (experimentally inaccessible), lower-energy versions (with heavy neutrinos at around the GeV scale) have recently drawn a lot of attention in the community since they are testable, do not worsen the hierarchy problem, and are able to reproduce the observed Baryon Asymmetry of the Universe. In such low-scale Seesaw models, the new singlets may form a pseudo-Dirac pair and lepton number is approximately preserved in the theory.  

The most promising avenues to look for MeV- to GeV-scale neutrinos are peak searches in meson decays, and searches for displaced vertices in fixed target experiments (produced when the neutrino travels a macroscopic distance before decaying back to SM particles). In both cases, an effective theory describing the interactions at low energies between mesons, neutrinos and charged leptons, obtained after the electroweak bosons have been integrated out, is the most suitable description. While most relevant vertices of the effective theory had been partially derived in previous literature, several inconsistencies remained. In this work, we have systematically derived all effective vertices involving mesons with masses of up to 2~GeV with significant branching ratios into HNLs. This allowed us to derive analytic expressions for the decay widths of the heavy neutrino into the different channels, and to clarify the inconsistencies found in previous literature (summarized in Sec.~\ref{subsec:discrepancies}). For convenience, Tab.~\ref{tab:couplings} summarizes the Feynman rules for the effective vertices involving charged leptons, mesons and neutrinos.
\begin{table}
\begin{center}{
\renewcommand{\arraystretch}{1.8}
\begin{tabular}{ | c | cc | cc | }
\hline
 & \multicolumn{2}{c |}{ \textbf{Neutral mesons}   } & \multicolumn{2}{c| }{ \textbf{Charged mesons} }  \\ \hline\hline
\multirow{4}{*}{\begin{sideways}\phantom{$H_H^H$} \textbf{Pseudoscalars} \phantom{$H_H^H$}  \end{sideways}} 
									& $n_i n_j \pi^ 0$ &  $ - C_{ij} G_F  f_\pi \slashed{p} P_L$   
									& $\ell_\alpha n_i \pi^\pm$    &   $ - \sqrt{2} U_{\alpha i}G_F V_{ud}  f_\pi \slashed{p} P_L $  \cr 
                                 & $n_i  n_j \eta$ &    $ - C_{ij} G_F \left[ \frac{\cos\theta_8 f_8}{\sqrt{3}} + \frac{ \sin\theta_0 f_0}{\sqrt{6}} \right] \slashed{p} P_L $    
                                 &  $\ell_\alpha n_i K^\pm$    &   $ - \sqrt{2} U_{\alpha i} G_F V_{us}  f_K \slashed{p} P_L$   \cr
                                 & $n_i n_j \eta^\prime$ &   $ -C_{ij} G_F \left[ \frac{\sin\theta_8 f_8}{\sqrt{3}} - \frac{ \cos\theta_0 f_0}{\sqrt{6}} \right] \slashed{p}  P_L$    
                                 &  $\ell_\alpha n_i D^\pm$    &   $ - \sqrt{2} U_{\alpha i} G_F V_{cd}  f_D \slashed{p}  P_L$    \cr
                                 &  &    
                                 &  $\ell_\alpha n_i D_s^\pm$    &   $ - \sqrt{2} U_{\alpha i} G_F V_{cs}  f_{D_{s}} \slashed{p}  P_L$    \\ \hline
\multirow{3}{*}{\begin{sideways}\phantom{$H_H^H$}  \textbf{Vectors} \phantom{$H_H^H$} \end{sideways}} 
									& $n_i n_j \rho^0_\mu $ &  $-i C_{ij} G_F (1 - 2s_w^2) f_\rho \gamma_\mu  P_L $  
                                 &  $\ell_\alpha n_i \rho_\mu^\pm$    &     $ - i \sqrt{2} U_{\alpha i}G_F V_{ud}  f_\rho \gamma_\mu  P_L $  \cr
                                 & $n_i n_j \omega_\mu$ &   $i C_{ij} G_F \frac{2}{3} s_w^2 f_\omega \gamma_\mu  P_L$  
									&  $\ell_\alpha n_i K_\mu^{*,\pm} $    &  $ - i \sqrt{2} U_{\alpha i} G_F V_{us}  f_{K^*} \gamma_\mu  P_L$   \cr
                                 & $n_i n_j \phi_\mu$ &     $ i C_{ij} G_F \sqrt{2}\left(\frac{1}{2} - \frac{2}{3} s_w^2\right) f_\phi \gamma_\mu  P_L$ 
                                 &   &       \\ \hline                                 
\end{tabular}}
\end{center}
\caption{\label{tab:couplings} List of Feynman rules for the effective vertices involving neutrinos and mesons, where $p$ is the 4-momentum of the corresponding pseudoscalar meson. Here, latin (greek) indices refer to the mass (flavor) basis. The Feynman rule for the vertex involving two pseudoscalar mesons, a neutrino and a charged lepton can be derived from Eq.~(\ref{eq:semilep}).  Numerical values for the meson decay constants (as well as for $\theta_0$ and $\theta_8$) can be found in Tab.~\ref{tab:Fpi}. If all particles are on-shell, further simplifications can be performed to these rules using Dirac's equation, see Sec.~\ref{sec:eft} for details. }
\end{table}

Our results have been made publicly available as FeynRules models~\cite{Alloul:2013bka} so that not only the total widths, but also fully differential event distributions, can be computed using Monte Carlo generators such as MadGraph5~\cite{Alwall:2014hca}. This has been done separately for Dirac and Majorana HNLs. Moreover, note that, while the present work focuses on the low-energy theory, our FeynRules implementation is more general and includes an option to replace all mesons with quarks, so they may also be used to study HNL phenomenology in collider searches at higher energies. 

To illustrate the applicability of the effective theory and its FeynRules implementation, we have performed numerical simulations to obtain the expected heavy neutrino flux that would reach the DUNE near detector (ND), as well as the expected number of HNL decays inside the detector into several decay channels. The very high beam intensity, combined with the availability of a ND located at a distance $L\sim\mathcal{O}(500)$~m, puts the DUNE experiment in an ideal position to search for the decay signals of HNLs produced from meson decays. We have shown how a proper treatment of the boost of the heavy neutrino, accounting for its mass, leads to an increased detector acceptance for the heavy neutrino flux when compared to the light neutrino case, see Figs.~\ref{fig:energy_spectrum} and~\ref{fig:events}.  

Finally, while the computation of the expected sensitivity at DUNE eventually needs a detailed detector simulation to address background rejection, it has been shown that, applying proper kinematic cuts to the particles observed in the final state, it is possible to reduce the background to a negligible level while keeping most of the signal events. Under this assumption, we have estimated in Sec.~\ref{sec:dune} the expected sensitivities to the model as a function of the heavy neutrino mass. We find that DUNE is expected to reach sensitivities comparable to or even better than those of fixed target experiments (see Fig.~\ref{fig:sens_total}). We also find that DUNE will be expected to start exploring the region of parameter space where neutrino masses can be explained using a type-I Seesaw model, for HNL masses around the $K$ mass scale.

\section*{Acknowledgments}

The authors warmly thank Mattias Blennow, Kyrylo Bondarenko, Andrea Caputo, Claudia Garcia-Garcia, Gregorio Herdoiza, Pilar Hernandez, Matheus Hostert and Carlos Pena for very illuminating discussions, and Justo Martin-Albo for collaboration during the early stages of this work. They are also grateful to Olivier Mattelaer for support in the use of MadGraph5, Haifa Rejeb Sfar for her help with the treatment of pion decays, and Kevin Kelly and Albert de Roeck for useful discussions. PC acknowledges support from the Spanish MICINN through the ``Ram\'on y Cajal'' program with grants RYC2018-024240-I. JHG acknowledges support by the grant K125105 of the National Research, Development and Innovation Fund in Hungary. ZP has been supported by Fermi Research Alliance, LLC under Contract No. DE-AC02-07CH11359 with the U.S. Department of Energy, Office of Science, Office of High Energy Physics. The authors acknowledge the support of the Spanish Agencia
Estatal de Investigacion and the EU ``Fondo Europeo de Desarrollo Regional'' (FEDER) through the projects PID2019-108892RB-I00/AEI/10.13039/501100011033 and FPA2016-78645-P as well as the``IFT Centro de Excelencia
Severo Ochoa SEV-2016-0597''. They also acknowledge use of the HPC facilities at the IFT (Hydra cluster). This work was partially supported by grants PROMETEO/2019/083, and the European projects H2020-MSCA-ITN-2015//674896- ELUSIVES and 690575-InvisiblesPlus-H2020-MSCA- RISE-2015.


\newpage
\appendix
\section{Generators of SU(3)}
\label{app:generators}

As outlined in Sec.~\ref{sec:eft}, the normalization for the SU(3) generators has been chosen to satisfy the trace conditions 
\begin{equation}
{\rm Tr}\left\lbrace \lambda_a \lambda_b \right\rbrace = \frac{\delta_{ab}}{2} \, .
\end{equation}
For convenience, we provide explicit expressions for the SU(3) generators below:
\vspace{4mm}
\noindent
\begin{center}
\begin{tabular}{ccc}
$\lambda_1 = \dfrac{1}{2} \left( 
\begin{array}{ccc}
0 & 1 & 0 \\
1 & 0 & 0 \\
0 & 0 & 0
\end{array}\right)$ 
, &
$\lambda_2 = \dfrac{1}{2} \left( 
\begin{array}{ccc}
0 & -i & 0 \\
i & 0 & 0 \\
0 & 0 & 0
\end{array}\right)$ 
, &
$\lambda_3 = \dfrac{1}{2} \left( 
\begin{array}{ccc}
1 & 0 & 0 \\
0 & -1 & 0 \\
0 & 0 & 0
\end{array}\right)$ 
, \\[8mm]
$\lambda_4 = \dfrac{1}{2} \left( 
\begin{array}{ccc}
0 & 0 & 1 \\
0 & 0 & 0 \\
1 & 0 & 0
\end{array}\right)$ 
, &
$\lambda_5 = \dfrac{1}{2} \left( 
\begin{array}{ccc}
0 & 0 & -i \\
0 & 0 & 0 \\
i & 0 & 0
\end{array}\right)$ 
, &
$\lambda_6 = \dfrac{1}{2} \left( 
\begin{array}{ccc}
0 & 0 & 0 \\
0 & 0 & 1 \\
0 & 1 & 0
\end{array}\right)$ 
, \\[8mm]
$\lambda_7 = \dfrac{1}{2} \left( 
\begin{array}{ccc}
0 & 0 & 0 \\
0 & 0 & -i \\
0 & i & 0
\end{array}\right)$ 
, &
$\lambda_8 = \dfrac{1}{2\sqrt{3}} \left( 
\begin{array}{ccc}
1 & 0 & 0 \\
0 & 1 & 0 \\
0 & 0 & -2
\end{array}\right)$ 
, &
$\lambda_0 = \dfrac{1}{\sqrt{6}} \left( 
\begin{array}{ccc}
1 & 0 & 0 \\
0 & 1 & 0 \\
0 & 0 & 1
\end{array}\right)$ 
. 
\end{tabular}
\end{center}

\section{Mixing matrices in a 3 + 1 scenario}
\label{app:matrices}

It can be interesting to consider a case in which only one heavy neutrino is light enough or exhibits a sufficiently large mixing to play a role in the relevant phenomenology\footnote{The masses of the light neutrinos can be neglected for phenomenological purposes here.}. In this case, the model parameters will be four: the three leptonic Yukawa couplings $Y_{\nu,\alpha}$ and the heavy mass $M$, defined in Eqs.~\eqref{eq:type1-lagrangian} or~\eqref{eq:inverse-lagrangian} (for the type-I and inverse Seesaw models, respectively). It is possible to write a $4 \times 4$ mixing matrix $U$ in terms of these parameters, which rotates from the flavor basis to the mass one. The shape of such matrix will depend on whether neutrinos are either Majorana or Dirac fermions.

For the Dirac case, the mixing matrix $U$ relates the 4 left-handed neutrinos $(\nu_{L,e}$, $\nu_{L,\mu}, \nu_{L,\tau}, N_L)$ to the 4 mass eigenstates $(n_1, n_2, n_3, N_4)$.  In terms of the parameters mentioned above, the mixing matrix reads:
\begin{equation}
U = \left( \begin{array}{cccc}
1-\frac{\left(r-1\right)|\theta_e|^2}{r \theta^2} & -\frac{\left(r-1\right)\theta_e \theta_\mu^*}{r \theta^2} & -\frac{\left(r-1\right)\theta_e \theta_\tau^*}{r \theta^2} & \frac{\theta_e}{r} \\[2mm]
-\frac{\left(r-1\right)\theta_\mu \theta_e^*}{r \theta^2} & 1-\frac{\left(r-1\right)|\theta_\mu|^2}{r \theta^2} & -\frac{\left(r-1\right)\theta_\mu \theta_\tau^*}{r \theta^2} & \frac{\theta_\mu}{r}\\[2mm]
-\frac{\left(r-1\right)\theta_\tau \theta_e^*}{r \theta^2} & -\frac{\left(r-1\right)\theta_\tau \theta_\mu^*}{r \theta^2} & 1-\frac{\left(r-1\right)|\theta_\tau|^2}{r \theta^2} & \frac{\theta_\tau}{r}\\[2mm]
-\frac{\theta_e^*}{r} & -\frac{\theta_\mu^*}{r} & -\frac{\theta_\tau^*}{r} & \frac{1}{r} 
\end{array} \right) \, ,
\end{equation}
where $\theta_{\alpha} \equiv Y_{\nu,\alpha} v/\sqrt{2}M$, $\theta^2 \equiv |\theta_e|^2+|\theta_\mu|^2+|\theta_\tau|^2$ and $r \equiv \sqrt{1+\theta^2}$. In this case, only $N_4$ is massive, with a Dirac mass $M_4 = r M$. In the limit in which all the mixing parameters $\theta_\alpha$ are small, $r\sim 1$ and the mass of the heavy neutrino will be approximately $M$.

On the other hand, in the Majorana case the flavor eigenstates are $(\nu_{L,e}, \nu_{L,\mu}, \nu_{L,\tau}, N^c_R)$. The mixing matrix now takes the form:
\begin{equation}
U = \left( \begin{array}{cccc}
\frac{\theta_\tau-\theta_\mu}{\sqrt{3 \theta^2 - |\theta_s|^2}} & \frac{\theta_s \theta_e^* -\theta^2}{\theta \sqrt{3 \theta^2 - |\theta_s|^2}} & -i\frac{\sqrt{1+\rho}\theta_e^*}{\sqrt{2}\theta} & \frac{\sqrt{2}\rho\theta_e^*}{\sqrt{1+\rho}} \\[2mm]
\frac{\theta_e-\theta_\tau}{\sqrt{3 \theta^2 - |\theta_s|^2}} & \frac{\theta_s \theta_\mu^* -\theta^2}{\theta \sqrt{3 \theta^2 - |\theta_s|^2}} & -i\frac{\sqrt{1+\rho}\theta_\mu^*}{\sqrt{2}\theta} & \frac{\sqrt{2}\rho\theta_\mu^*}{\sqrt{1+\rho}} \\[2mm]
\frac{\theta_\mu-\theta_e}{\sqrt{3 \theta^2 - |\theta_s|^2}} & \frac{\theta_s \theta_\tau^* -\theta^2}{\theta \sqrt{3 \theta^2 - |\theta_s|^2}} & -i\frac{\sqrt{1+\rho}\theta_\tau^*}{\sqrt{2}\theta} & \frac{\sqrt{2}\rho\theta_\tau^*}{\sqrt{1+\rho}} \\[2mm]
0 & 0 & i\sqrt{\frac{1-\rho}{2}} & \sqrt{\frac{1+\rho}{2}} 
\end{array} \right) \, ,
\end{equation}
with $\theta_s \equiv \theta_e+\theta_\mu+\theta_\tau$ and $\rho\equiv 1/\sqrt{1+4\theta^2}$. The $i$ factors are chosen to obtain positive masses when diagonalizing the mass matrix. Now only two mass eigenstates, $n_1$ and $n_2$, are massless, while $n_3$ and $N_4$ have Majorana masses of $\frac{M}{2} \vert 1\mp \rho^{-1}\vert$ respectively. If all the mixing parameters $\theta_\alpha$ are small, then $\rho \sim 1$ , so the mass of $n_3$ is negligible and that of $N_4$ is approximately equal to $M$.

\section{Determination of the vector meson decay constants}
\label{app:decayconstants}

Unlike pseudoscalar mesons, vector meson resonances are wide and unstable under QCD. Thus, the determination of their decay constants is generally challenging, with more variability among different estimations in the literature. In order to bypass this issue, a possibility is to compute the width for a decay channel mediated by the electroweak interaction that has been precisely measured, comparing the result to the experimental values from Ref.~\cite{Tanabashi:2018oca}. This way the corresponding value of the decay constant can be directly extracted for each of the resonances under consideration, ensuring that the notation and normalization conventions used are consistent. 

\subsection{Neutral vector mesons}

In this case, a good choice is the decay channel $V \to e^+ e^-$, which has been precisely measured and is dominated by photon exchange. Thus, we decompose the electromagnetic (EM) current 
\[
j_{\mathrm{EM}, \mu}^{V} = i \sum_q e Q^q \bar{q} \gamma_\mu q \, 
\]
as a linear combination of the meson currents, as we did for the $Z$ current in Sec.~\ref{sec:eft}: 
\begin{equation}
\label{eq:EM-current}
j_{\mathrm{EM}, \mu}^{V} = i e\left[ j_{\rho, \mu}^{V}  + \frac{1}{3} j_{\omega, \mu}^{V} -\frac{\sqrt{2}}{3}  j_{\phi, \mu}^{V}  \right] \, .
\end{equation}
This allows to compute the width for the vector meson decays into $e^- e^+$ pairs mediated by a photon, as
\begin{eqnarray}
\label{eq:rho-ee}
\Gamma(\rho \to e^+ e^-) & = & \frac{2\pi}{3}\frac{\alpha^2 f_\rho^2}{m_\rho^3} \, , \\
\label{eq:omega-ee}
\Gamma(\omega \to e^+ e^-) & = & \frac{2\pi}{27}\frac{\alpha^2 f_\omega^2}{m_\omega^3} \, , \\
\label{eq:phi-ee}
\Gamma(\phi \to e^+ e^-) & = & \frac{4\pi}{27}\frac{\alpha^2 f_\phi^2}{m_\phi^3} \, .
\end{eqnarray}
Comparing these results to the corresponding measurements \cite{Tanabashi:2018oca, webmesons}, we find the values for the decay constants $f_V$ listed in Tab.~\ref{tab:Fpi}.

\subsection{Charged vector mesons}

For the $\rho^\pm$ mesons, we will use the $\rho^0$ constant, already determined, since the isospin breaking corrections should be negligible.

However, for the $K^{*,\pm}$ meson we must compute the decay width for an electroweak process, extracting the decay constant from there as we did for the neutral vector mesons.
In this case, a good choice is the process $\tau^- \to K^{*,-} \nu_\tau$. The authors of Ref.~\cite{Maris:1999nt} perform such calculation and report the value of the ratio between the $\rho$ and $K^*$ decay constants:
\begin{equation}
\frac{f_{K^*}}{f_{\rho}} = 1.042 \, .
\end{equation}
Therefore, using $f_\rho = 0.171~\mathrm{GeV}^2$, we obtain $f_{K^*} = 0.178~\mathrm{GeV}^2$ as listed in Tab.~\ref{tab:Fpi}.

\section{Implementation of semileptonic form factors into FeynRules}
\label{app:feynrules} 

The form factors involved in semileptonic meson decays include a dependence on the squared momentum transfer between the involved mesons, $q^2$, which is not trivial to implement in a UFO model. For this reason, we have included two different implementation choices into our FeynRules models: a simpler option, which neglects the $q^2$ dependence and has been tuned to approximately reproduce the correct branching ratios for semileptonic decay channels; and a more sophisticated one, which includes the correct $q^2$ dependence as described in Sec.~\ref{subsec:formfactors}. 

As a first option, our FeynRules model files include by default constant form factors, evaluated at an average value of the (squared) momentum transfer, $\langle q^2 \rangle$. This average value is determined by imposing that the correct total decay width is obtained, and depends mildly on the heavy neutrino mass $M_4$ and the charged lepton mass.  We perform a cubic fit for the function $\langle q^2\rangle (M_4)$ in the kinematically allowed range for the heavy neutrino mass. Furthermore, the coefficients of such a fit depend on the flavor of the involved charged lepton, in order to account for the correct kinematics. The accuracy of this approximation relies on the fact that the dependence of the form factors on the momentum transfer is very mild in the allowed kinematic range. 

As a second option, together with the FeynRules model files we provide a Python script that, upon running, modifies the relevant files in the output UFO. This way, the correct energy dependence of the vertices, according to the linear and pole parametrizations of the form factors described in Sec.~\ref{subsec:formfactors}, can be implemented, allowing for precise event generation in MadGraph5. 

In short, if the provided Python script is run after generating the UFO file with FeynRules, the correct energy dependence of the form factors can be fully incorporated into MadGraph5. Otherwise, the constant form factors evaluated at $\langle q^2 \rangle$ in the default FeynRules model allow for a good approximation also for the other formats into which the model may be exported to.


\newpage
\bibliographystyle{utphys_Josu}

\begin{thebibliography}{100}

\bibitem{Minkowski:1977sc}
P.~Minkowski, \textit{$\mu \to e\gamma$ at a Rate of One Out of $10^{9}$ Muon
  Decays?},
\href{http://dx.doi.org/10.1016/0370-2693(77)90435-X}{{Phys. Lett.} {\bfseries
  67B} (1977) 421--428}.

\bibitem{Mohapatra:1979ia}
R.~N. Mohapatra and G.~Senjanovic, \textit{Neutrino Mass and Spontaneous Parity
  Nonconservation},
  \href{http://dx.doi.org/10.1103/PhysRevLett.44.912}{{Phys. Rev. Lett.}
  {\bfseries 44} (1980) 912}.

\bibitem{Yanagida:1979as}
T.~Yanagida, \textit{Horizontal gauge symmetry and masses of neutrinos},
{Conf. Proc.} {\bfseries C7902131} (1979) 95--99.

\bibitem{GellMann:1980vs}
M.~Gell-Mann, P.~Ramond, and R.~Slansky, \textit{Complex Spinors and Unified
  Theories}, {Conf. Proc.} {\bfseries C790927} (1979) 315--321,
\href{http://arxiv.org/abs/1306.4669}{{\ttfamily arXiv:1306.4669 [hep-th]}}.

\bibitem{Fukugita:1986hr}
M.~Fukugita and T.~Yanagida, \textit{Baryogenesis Without Grand Unification},
  \href{http://dx.doi.org/10.1016/0370-2693(86)91126-3}{{Phys. Lett. B}
  {\bfseries 174} (1986) 45--47}.

\bibitem{Vissani:1997ys}
F.~Vissani, \textit{Do experiments suggest a hierarchy problem?},
  \href{http://dx.doi.org/10.1103/PhysRevD.57.7027}{{Phys. Rev. D} {\bfseries
  57} (1998) 7027--7030}, \href{http://arxiv.org/abs/hep-ph/9709409}{{\ttfamily
  arXiv:hep-ph/9709409}}.

\bibitem{Casas:2004gh}
J.~Casas, J.~Espinosa, and I.~Hidalgo, \textit{Implications for new physics from
  fine-tuning arguments. 1. Application to SUSY and seesaw cases},
  \href{http://dx.doi.org/10.1088/1126-6708/2004/11/057}{{JHEP} {\bfseries 11}
  (2004) 057}, \href{http://arxiv.org/abs/hep-ph/0410298}{{\ttfamily
  arXiv:hep-ph/0410298}}.

\bibitem{Mohapatra:1986aw}
R.~N. Mohapatra, \textit{Mechanism for Understanding Small Neutrino Mass in
  Superstring Theories},
\href{http://dx.doi.org/10.1103/PhysRevLett.56.561}{{Phys. Rev. Lett.}
  {\bfseries 56} (1986) 561--563}.

\bibitem{Mohapatra:1986bd}
R.~N. Mohapatra and J.~W.~F. Valle, \textit{Neutrino Mass and Baryon Number
  Nonconservation in Superstring Models},
  \href{http://dx.doi.org/10.1103/PhysRevD.34.1642}{{Phys. Rev.} {\bfseries
  D34} (1986) 1642}.

\bibitem{Bernabeu:1987gr}
J.~Bernabeu, A.~Santamaria, J.~Vidal, A.~Mendez, and J.~W.~F. Valle, \textit{Lepton
  Flavor Nonconservation at High-Energies in a Superstring Inspired Standard
  Model},
\href{http://dx.doi.org/10.1016/0370-2693(87)91100-2}{{Phys. Lett.} {\bfseries
  B187} (1987) 303--308}.

\bibitem{Malinsky:2005bi}
M.~Malinsky, J.~C. Romao, and J.~W.~F. Valle, \textit{Novel supersymmetric SO(10)
  seesaw mechanism},
  \href{http://dx.doi.org/10.1103/PhysRevLett.95.161801}{{Phys. Rev. Lett.}
  {\bfseries 95} (2005) 161801},
\href{http://arxiv.org/abs/hep-ph/0506296}{{\ttfamily arXiv:hep-ph/0506296 [hep-ph]}}.

\bibitem{Branco:1988ex}
G.~C. Branco, W.~Grimus, and L.~Lavoura, \textit{The Seesaw Mechanism in the Presence of a Conserved Lepton Number},
\href{http://dx.doi.org/10.1016/0550-3213(89)90304-0}{{Nucl. Phys.} {\bfseries
  B312} (1989) 492}.

\bibitem{Kersten:2007vk}
J.~Kersten and A.~{\relax Yu}. Smirnov, \textit{Right-Handed Neutrinos at CERN LHC
  and the Mechanism of Neutrino Mass Generation},
  \href{http://dx.doi.org/10.1103/PhysRevD.76.073005}{{Phys. Rev.} {\bfseries
  D76} (2007) 073005},
\href{http://arxiv.org/abs/0705.3221}{{\ttfamily arXiv:0705.3221 [hep-ph]}}.

\bibitem{Abada:2007ux}
A.~Abada, C.~Biggio, F.~Bonnet, M.~B. Gavela, and T.~Hambye, \textit{Low energy
  effects of neutrino masses},
  \href{http://dx.doi.org/10.1088/1126-6708/2007/12/061}{{JHEP} {\bfseries 12}
  (2007) 061},
\href{http://arxiv.org/abs/0707.4058}{{\ttfamily arXiv:0707.4058}}.

\bibitem{Akhmedov:1998qx}
E.~K. Akhmedov, V.~Rubakov, and A.~Smirnov, \textit{Baryogenesis via neutrino
  oscillations}, \href{http://dx.doi.org/10.1103/PhysRevLett.81.1359}{{Phys.
  Rev. Lett.} {\bfseries 81} (1998) 1359--1362},
  \href{http://arxiv.org/abs/hep-ph/9803255}{{\ttfamily arXiv:hep-ph/9803255}}.

\bibitem{Asaka:2005pn}
T.~Asaka and M.~Shaposhnikov, \textit{The $\nu$MSM, dark matter and baryon asymmetry
  of the universe},
  \href{http://dx.doi.org/10.1016/j.physletb.2005.06.020}{{Phys. Lett. B}
  {\bfseries 620} (2005) 17--26},
  \href{http://arxiv.org/abs/hep-ph/0505013}{{\ttfamily arXiv:hep-ph/0505013}}.

\bibitem{Shaposhnikov:2008pf}
M.~Shaposhnikov, \textit{The nuMSM, leptonic asymmetries, and properties of singlet
  fermions}, \href{http://dx.doi.org/10.1088/1126-6708/2008/08/008}{{JHEP}
  {\bfseries 08} (2008) 008}, \href{http://arxiv.org/abs/0804.4542}{{\ttfamily
  arXiv:0804.4542 [hep-ph]}}.

\bibitem{FernandezMartinez:2007ms}
E.~Fernandez-Martinez, M.~Gavela, J.~Lopez-Pavon, and O.~Yasuda,
  \textit{CP-violation from non-unitary leptonic mixing},
  \href{http://dx.doi.org/10.1016/j.physletb.2007.03.069}{{Phys. Lett. B}
  {\bfseries 649} (2007) 427--435},
  \href{http://arxiv.org/abs/hep-ph/0703098}{{\ttfamily arXiv:hep-ph/0703098}}.

\bibitem{Antusch:2009pm}
S.~Antusch, M.~Blennow, E.~Fernandez-Martinez, and J.~Lopez-Pavon, \textit{Probing
  non-unitary mixing and CP-violation at a Neutrino Factory},
  \href{http://dx.doi.org/10.1103/PhysRevD.80.033002}{{Phys. Rev. D} {\bfseries
  80} (2009) 033002}, \href{http://arxiv.org/abs/0903.3986}{{\ttfamily
  arXiv:0903.3986 [hep-ph]}}.

\bibitem{Parke:2015goa}
S.~Parke and M.~Ross-Lonergan, \textit{Unitarity and the three flavor neutrino
  mixing matrix}, \href{http://dx.doi.org/10.1103/PhysRevD.93.113009}{{Phys.
  Rev. D} {\bfseries 93}  (2016) 113009},
  \href{http://arxiv.org/abs/1508.05095}{{\ttfamily arXiv:1508.05095 [hep-ph]}}.

\bibitem{Miranda:2016wdr}
O.~Miranda, M.~Tortola, and J.~Valle, \textit{New ambiguity in probing CP violation
  in neutrino oscillations},
  \href{http://dx.doi.org/10.1103/PhysRevLett.117.061804}{{Phys. Rev. Lett.}
  {\bfseries 117}  (2016) 061804},
  \href{http://arxiv.org/abs/1604.05690}{{\ttfamily arXiv:1604.05690
  [hep-ph]}}.

\bibitem{Ge:2016xya}
S.-F. Ge, P.~Pasquini, M.~Tortola, and J.~Valle, \textit{Measuring the leptonic CP
  phase in neutrino oscillations with nonunitary mixing},
  \href{http://dx.doi.org/10.1103/PhysRevD.95.033005}{{Phys. Rev. D} {\bfseries
  95}  (2017) 033005}, \href{http://arxiv.org/abs/1605.01670}{{\ttfamily
  arXiv:1605.01670 [hep-ph]}}.

\bibitem{Blennow:2016jkn}
M.~Blennow, P.~Coloma, E.~Fernandez-Martinez, J.~Hernandez-Garcia, and
  J.~Lopez-Pavon, \textit{Non-Unitarity, sterile neutrinos, and Non-Standard
  neutrino Interactions},
  \href{http://dx.doi.org/10.1007/JHEP04(2017)153}{{JHEP} {\bfseries 04} (2017)
  153}, \href{http://arxiv.org/abs/1609.08637}{{\ttfamily arXiv:1609.08637
  [hep-ph]}}.

\bibitem{Escrihuela:2016ube}
F.~Escrihuela, D.~Forero, O.~Miranda, M.~T\'ortola, and J.~Valle, \textit{Probing CP
  violation with non-unitary mixing in long-baseline neutrino oscillation
  experiments: DUNE as a case study},
  \href{http://dx.doi.org/10.1088/1367-2630/aa79ec}{{New J. Phys.} {\bfseries
  19} (2017) 093005}, \href{http://arxiv.org/abs/1612.07377}{{\ttfamily
  arXiv:1612.07377 [hep-ph]}}.

\bibitem{Kosmas:2017zbh}
T.~Kosmas, D.~Papoulias, M.~Tortola, and J.~Valle, \textit{Probing light sterile
  neutrino signatures at reactor and Spallation Neutron Source neutrino
  experiments}, \href{http://dx.doi.org/10.1103/PhysRevD.96.063013}{{Phys.
  Rev. D} {\bfseries 96}  (2017) 063013},
  \href{http://arxiv.org/abs/1703.00054}{{\ttfamily arXiv:1703.00054
  [hep-ph]}}.

\bibitem{Miranda:2020syh}
O.~Miranda, D.~Papoulias, O.~Sanders, M.~T\'ortola, and J.~Valle, \textit{Future
  CEvNS experiments as probes of lepton unitarity and light-sterile
  neutrinos}, \href{http://arxiv.org/abs/2008.02759}{{\ttfamily
  arXiv:2008.02759 [hep-ph]}}.

\bibitem{delAguila:2008cj}
F.~del Aguila and J.~Aguilar-Saavedra, \textit{Distinguishing seesaw models at LHC
  with multi-lepton signals},
  \href{http://dx.doi.org/10.1016/j.nuclphysb.2008.12.029}{{Nucl. Phys. B}
  {\bfseries 813} (2009) 22--90},
  \href{http://arxiv.org/abs/0808.2468}{{\ttfamily arXiv:0808.2468 [hep-ph]}}.

\bibitem{Atre:2009rg}
A.~Atre, T.~Han, S.~Pascoli, and B.~Zhang, \textit{The Search for Heavy Majorana
  Neutrinos}, \href{http://dx.doi.org/10.1088/1126-6708/2009/05/030}{{JHEP}
  {\bfseries 05} (2009) 030}, \href{http://arxiv.org/abs/0901.3589}{{\ttfamily
  arXiv:0901.3589 [hep-ph]}}.

\bibitem{Antusch:2015mia}
S.~Antusch and O.~Fischer, \textit{Testing sterile neutrino extensions of the
  Standard Model at future lepton colliders},
  \href{http://dx.doi.org/10.1007/JHEP05(2015)053}{{JHEP} {\bfseries 05} (2015)
  053}, \href{http://arxiv.org/abs/1502.05915}{{\ttfamily arXiv:1502.05915
  [hep-ph]}}.

\bibitem{Deppisch:2015qwa}
F.~F. Deppisch, P.~Bhupal~Dev, and A.~Pilaftsis, \textit{Neutrinos and Collider
  Physics}, \href{http://dx.doi.org/10.1088/1367-2630/17/7/075019}{{New J.
  Phys.} {\bfseries 17}  (2015) 075019},
  \href{http://arxiv.org/abs/1502.06541}{{\ttfamily arXiv:1502.06541
  [hep-ph]}}.

\bibitem{Antusch:2016ejd}
S.~Antusch, E.~Cazzato, and O.~Fischer, \textit{Sterile neutrino searches at future
  $e^-e^+$, $pp$, and $e^-p$ colliders},
  \href{http://dx.doi.org/10.1142/S0217751X17500786}{{Int. J. Mod. Phys. A}
  {\bfseries 32}  (2017) 1750078},
  \href{http://arxiv.org/abs/1612.02728}{{\ttfamily arXiv:1612.02728
  [hep-ph]}}.

\bibitem{Das:2017nvm}
A.~Das and N.~Okada, \textit{Bounds on heavy Majorana neutrinos in type-I seesaw and
  implications for collider searches},
  \href{http://dx.doi.org/10.1016/j.physletb.2017.09.042}{{Phys. Lett. B}
  {\bfseries 774} (2017) 32--40},
  \href{http://arxiv.org/abs/1702.04668}{{\ttfamily arXiv:1702.04668
  [hep-ph]}}.

\bibitem{Das:2017zjc}
A.~Das, P.~S.~B. Dev, and C.~Kim, \textit{Constraining Sterile Neutrinos from
  Precision Higgs Data},
  \href{http://dx.doi.org/10.1103/PhysRevD.95.115013}{{Phys. Rev. D} {\bfseries
  95}  (2017) 115013}, \href{http://arxiv.org/abs/1704.00880}{{\ttfamily
  arXiv:1704.00880 [hep-ph]}}.

\bibitem{Cai:2017mow}
Y.~Cai, T.~Han, T.~Li, and R.~Ruiz, \textit{Lepton Number Violation: Seesaw Models
  and Their Collider Tests},
  \href{http://dx.doi.org/10.3389/fphy.2018.00040}{{Front. in Phys.} {\bfseries
  6} (2018) 40}, \href{http://arxiv.org/abs/1711.02180}{{\ttfamily
  arXiv:1711.02180 [hep-ph]}}.

\bibitem{Das:2018hph}
A.~Das, \textit{Searching for the minimal Seesaw models at the LHC and beyond},
  \href{http://dx.doi.org/10.1155/2018/9785318}{{Adv. High Energy Phys.}
  {\bfseries 2018} (2018) 9785318},
  \href{http://arxiv.org/abs/1803.10940}{{\ttfamily arXiv:1803.10940
  [hep-ph]}}.

\bibitem{Dev:2018kpa}
P.~Bhupal~Dev and Y.~Zhang, \textit{Displaced vertex signatures of doubly charged
  scalars in the type-II seesaw and its left-right extensions},
  \href{http://dx.doi.org/10.1007/JHEP10(2018)199}{{JHEP} {\bfseries 10} (2018)
  199}, \href{http://arxiv.org/abs/1808.00943}{{\ttfamily arXiv:1808.00943
  [hep-ph]}}.

\bibitem{Pascoli:2018heg}
S.~Pascoli, R.~Ruiz, and C.~Weiland, \textit{Heavy neutrinos with dynamic jet
  vetoes: multilepton searches at $ \sqrt{s}=14 $ , 27, and 100 TeV},
  \href{http://dx.doi.org/10.1007/JHEP06(2019)049}{{JHEP} {\bfseries 06} (2019)
  049}, \href{http://arxiv.org/abs/1812.08750}{{\ttfamily arXiv:1812.08750
  [hep-ph]}}.

\bibitem{Liu:2019qfa}
N.~Liu, Z.-G. Si, L.~Wu, H.~Zhou, and B.~Zhu, \textit{Top quark as a probe of heavy
  Majorana neutrino at the LHC and future colliders},
  \href{http://dx.doi.org/10.1103/PhysRevD.101.071701}{{Phys. Rev. D}
  {\bfseries 101}  (2020) 071701},
  \href{http://arxiv.org/abs/1910.00749}{{\ttfamily arXiv:1910.00749
  [hep-ph]}}.

\bibitem{Shrock:1980ct}
R.~E. Shrock, \textit{General Theory of Weak Leptonic and Semileptonic Decays. 1.
  Leptonic Pseudoscalar Meson Decays, with Associated Tests For, and Bounds on,
  Neutrino Masses and Lepton Mixing},
\href{http://dx.doi.org/10.1103/PhysRevD.24.1232}{{Phys. Rev.} {\bfseries D24}
  (1981) 1232}.

\bibitem{Shrock:1981wq}
R.~E. Shrock, \textit{General Theory of Weak Processes Involving Neutrinos. 2. Pure
  Leptonic Decays},
\href{http://dx.doi.org/10.1103/PhysRevD.24.1275}{{Phys. Rev.} {\bfseries D24}
  (1981) 1275}.

\bibitem{Langacker:1988ur}
P.~Langacker and D.~London, \textit{Mixing Between Ordinary and Exotic Fermions},
\href{http://dx.doi.org/10.1103/PhysRevD.38.886}{{Phys.Rev.} {\bfseries D38}
  (1988) 886}.

\bibitem{Tommasini:1995ii}
D.~Tommasini, G.~Barenboim, J.~Bernabeu, and C.~Jarlskog, \textit{Nondecoupling of
  heavy neutrinos and lepton flavor violation},
  \href{http://dx.doi.org/10.1016/0550-3213(95)00201-3}{{Nucl. Phys. B}
  {\bfseries 444} (1995) 451--467},
  \href{http://arxiv.org/abs/hep-ph/9503228}{{\ttfamily arXiv:hep-ph/9503228}}.

\bibitem{Antusch:2006vwa}
S.~Antusch, C.~Biggio, E.~Fernandez-Martinez, M.~Gavela, and J.~Lopez-Pavon,
  \textit{Unitarity of the Leptonic Mixing Matrix},
  \href{http://dx.doi.org/10.1088/1126-6708/2006/10/084}{{JHEP} {\bfseries 10}
  (2006) 084}, \href{http://arxiv.org/abs/hep-ph/0607020}{{\ttfamily
  arXiv:hep-ph/0607020}}.

\bibitem{Antusch:2008tz}
S.~Antusch, J.~P. Baumann, and E.~Fernandez-Martinez, \textit{Non-Standard Neutrino
  Interactions with Matter from Physics Beyond the Standard Model},
  \href{http://dx.doi.org/10.1016/j.nuclphysb.2008.11.018}{{Nucl. Phys. B}
  {\bfseries 810} (2009) 369--388},
  \href{http://arxiv.org/abs/0807.1003}{{\ttfamily arXiv:0807.1003 [hep-ph]}}.

\bibitem{Forero:2011pc}
D.~Forero, S.~Morisi, M.~Tortola, and J.~Valle, \textit{Lepton flavor violation and
  non-unitary lepton mixing in low-scale type-I seesaw},
  \href{http://dx.doi.org/10.1007/JHEP09(2011)142}{{JHEP} {\bfseries 09} (2011)
  142}, \href{http://arxiv.org/abs/1107.6009}{{\ttfamily arXiv:1107.6009
  [hep-ph]}}.

\bibitem{Antusch:2014woa}
S.~Antusch and O.~Fischer, \textit{Non-unitarity of the leptonic mixing matrix:
  Present bounds and future sensitivities},
  \href{http://dx.doi.org/10.1007/JHEP10(2014)094}{{JHEP} {\bfseries 10} (2014)
  094}, \href{http://arxiv.org/abs/1407.6607}{{\ttfamily arXiv:1407.6607
  [hep-ph]}}.

\bibitem{Fernandez-Martinez:2016lgt}
E.~Fernandez-Martinez, J.~Hernandez-Garcia, and J.~Lopez-Pavon, \textit{Global
  constraints on heavy neutrino mixing},
  \href{http://dx.doi.org/10.1007/JHEP08(2016)033}{{JHEP} {\bfseries 08} (2016)
  033}, \href{http://arxiv.org/abs/1605.08774}{{\ttfamily arXiv:1605.08774
  [hep-ph]}}.

\bibitem{Coutinho:2019aiy}
A.~M. Coutinho, A.~Crivellin, and C.~A. Manzari, \textit{Global Fit to Modified
  Neutrino Couplings and the Cabibbo-Angle Anomaly},
  \href{http://arxiv.org/abs/1912.08823}{{\ttfamily arXiv:1912.08823
  [hep-ph]}}.

\bibitem{Gorbunov:2007ak}
D.~Gorbunov and M.~Shaposhnikov, \textit{How to find neutral leptons of the
  $\nu$MSM?}, \href{http://dx.doi.org/10.1088/1126-6708/2007/10/015}{{JHEP}
  {\bfseries 10} (2007) 015}, \href{http://arxiv.org/abs/0705.1729}{{\ttfamily
  arXiv:0705.1729 [hep-ph]}}. [Erratum: JHEP 11, 101 (2013)].

\bibitem{Abada:2016plb}
A.~Abada, D.~Be\v{c}irevi\'c, O.~Sumensari, C.~Weiland, and
  R.~Zukanovich~Funchal, \textit{Sterile neutrinos facing kaon physics
  experiments}, \href{http://dx.doi.org/10.1103/PhysRevD.95.075023}{{Phys.
  Rev. D} {\bfseries 95}  (2017) 075023},
  \href{http://arxiv.org/abs/1612.04737}{{\ttfamily arXiv:1612.04737
  [hep-ph]}}.

\bibitem{Abada:2018nio}
A.~Abada and A.~M. Teixeira, \textit{Heavy neutral leptons and high-intensity
  observables}, \href{http://dx.doi.org/10.3389/fphy.2018.00142}{{Front. in
  Phys.} {\bfseries 6} (2018) 142},
  \href{http://arxiv.org/abs/1812.08062}{{\ttfamily arXiv:1812.08062
  [hep-ph]}}.

\bibitem{Abada:2018sfh}
A.~Abada, N.~Bernal, M.~Losada, and X.~Marcano, \textit{Inclusive Displaced Vertex
  Searches for Heavy Neutral Leptons at the LHC},
  \href{http://dx.doi.org/10.1007/JHEP01(2019)093}{{JHEP} {\bfseries 01} (2019)
  093}, \href{http://arxiv.org/abs/1807.10024}{{\ttfamily arXiv:1807.10024
  [hep-ph]}}.

\bibitem{Ballett:2019bgd}
P.~Ballett, T.~Boschi, and S.~Pascoli, \textit{Heavy Neutral Leptons from low-scale
  seesaws at the DUNE Near Detector},
  \href{http://dx.doi.org/10.1007/JHEP03(2020)111}{{JHEP} {\bfseries 20} (2020)
  111}, \href{http://arxiv.org/abs/1905.00284}{{\ttfamily arXiv:1905.00284
  [hep-ph]}}.

\bibitem{Berryman:2019dme}
J.~M. Berryman, A.~de~Gouvea, P.~J. Fox, B.~J. Kayser, K.~J. Kelly, and J.~L.
  Raaf, \textit{Searches for Decays of New Particles in the DUNE Multi-Purpose Near
  Detector}, \href{http://dx.doi.org/10.1007/JHEP02(2020)174}{{JHEP}
  {\bfseries 02} (2020) 174}, \href{http://arxiv.org/abs/1912.07622}{{\ttfamily
  arXiv:1912.07622 [hep-ph]}}.

\bibitem{Abada:2019bac}
A.~Abada, C.~Hati, X.~Marcano, and A.~Teixeira, \textit{Interference effects in LNV
  and LFV semileptonic decays: the Majorana hypothesis},
  \href{http://dx.doi.org/10.1007/JHEP09(2019)017}{{JHEP} {\bfseries 09} (2019)
  017}, \href{http://arxiv.org/abs/1904.05367}{{\ttfamily arXiv:1904.05367
  [hep-ph]}}.

\bibitem{Krasnov:2019kdc}
I.~Krasnov, \textit{DUNE prospects in the search for sterile neutrinos},
  \href{http://dx.doi.org/10.1103/PhysRevD.100.075023}{{Phys. Rev. D}
  {\bfseries 100}  (2019) 075023},
  \href{http://arxiv.org/abs/1902.06099}{{\ttfamily arXiv:1902.06099
  [hep-ph]}}.

\bibitem{Bryman:2019ssi}
D.~Bryman and R.~Shrock, \textit{Improved Constraints on Sterile Neutrinos in the
  MeV to GeV Mass Range},
  \href{http://dx.doi.org/10.1103/PhysRevD.100.053006}{{Phys. Rev. D}
  {\bfseries 100} (2019) 053006},
  \href{http://arxiv.org/abs/1904.06787}{{\ttfamily arXiv:1904.06787
  [hep-ph]}}.

\bibitem{Bryman:2019bjg}
D.~Bryman and R.~Shrock, \textit{Constraints on Sterile Neutrinos in the MeV to GeV
  Mass Range}, \href{http://dx.doi.org/10.1103/PhysRevD.100.073011}{{Phys.
  Rev. D} {\bfseries 100} (2019) 073011},
  \href{http://arxiv.org/abs/1909.11198}{{\ttfamily arXiv:1909.11198
  [hep-ph]}}.

\bibitem{Bondarenko:2019yob}
K.~Bondarenko, A.~Boyarsky, M.~Ovchynnikov, and O.~Ruchayskiy, \textit{Sensitivity
  of the intensity frontier experiments for neutrino and scalar portals:
  analytic estimates},
  \href{http://dx.doi.org/10.1007/JHEP08(2019)061}{{JHEP} {\bfseries 08} (2019)
  061}, \href{http://arxiv.org/abs/1902.06240}{{\ttfamily arXiv:1902.06240
  [hep-ph]}}.

\bibitem{Drewes:2015iva}
M.~Drewes and B.~Garbrecht, \textit{Combining experimental and cosmological
  constraints on heavy neutrinos},
  \href{http://dx.doi.org/10.1016/j.nuclphysb.2017.05.001}{{Nucl. Phys. B}
  {\bfseries 921} (2017) 250--315},
  \href{http://arxiv.org/abs/1502.00477}{{\ttfamily arXiv:1502.00477
  [hep-ph]}}.

\bibitem{Chrzaszcz:2019inj}
M.~Chrzaszcz, M.~Drewes, T.~E. Gonzalo, J.~Harz, S.~Krishnamurthy, and
  C.~Weniger, \textit{A frequentist analysis of three right-handed neutrinos with
  GAMBIT}, \href{http://dx.doi.org/10.1140/epjc/s10052-020-8073-9}{{Eur.
  Phys. J. C} {\bfseries 80} (2020) 569},
  \href{http://arxiv.org/abs/1908.02302}{{\ttfamily arXiv:1908.02302
  [hep-ph]}}.

\bibitem{Gorbunov:2020rjx}
D.~Gorbunov, I.~Krasnov, Y.~Kudenko, and S.~Suvorov, \textit{Heavy Neutral Leptons
  from kaon decays in the SHiP experiment},
  \href{http://dx.doi.org/10.1016/j.physletb.2020.135817}{{Phys. Lett. B}
  {\bfseries 810} (2020) 135817},
  \href{http://arxiv.org/abs/2004.07974}{{\ttfamily arXiv:2004.07974
  [hep-ph]}}.

\bibitem{Hernandez:2016kel}
P.~Hernandez, M.~Kekic, J.~Lopez-Pavon, J.~Racker, and J.~Salvado, \textit{Testable
  Baryogenesis in Seesaw Models},
  \href{http://dx.doi.org/10.1007/JHEP08(2016)157}{{JHEP} {\bfseries 08} (2016)
  157}, \href{http://arxiv.org/abs/1606.06719}{{\ttfamily arXiv:1606.06719
  [hep-ph]}}.

\bibitem{Abada:2018oly}
A.~Abada, G.~Arcadi, V.~Domcke, M.~Drewes, J.~Klaric, and M.~Lucente,
  \textit{Low-scale leptogenesis with three heavy neutrinos},
  \href{http://dx.doi.org/10.1007/JHEP01(2019)164}{{JHEP} {\bfseries 01} (2019)
  164}, \href{http://arxiv.org/abs/1810.12463}{{\ttfamily arXiv:1810.12463
  [hep-ph]}}.

\bibitem{Ghiglieri:2019kbw}
J.~Ghiglieri and M.~Laine, \textit{Sterile neutrino dark matter via GeV-scale
  leptogenesis?}, \href{http://dx.doi.org/10.1007/JHEP07(2019)078}{{JHEP}
  {\bfseries 07} (2019) 078}, \href{http://arxiv.org/abs/1905.08814}{{\ttfamily
  arXiv:1905.08814 [hep-ph]}}.

\bibitem{Bondarenko:2018ptm}
K.~Bondarenko, A.~Boyarsky, D.~Gorbunov, and O.~Ruchayskiy, \textit{Phenomenology of
  GeV-scale Heavy Neutral Leptons},
  \href{http://dx.doi.org/10.1007/JHEP11(2018)032}{{JHEP} {\bfseries 11} (2018)
  032}, \href{http://arxiv.org/abs/1805.08567}{{\ttfamily arXiv:1805.08567
  [hep-ph]}}.

\bibitem{Alloul:2013bka}
A.~Alloul, N.~D. Christensen, C.~Degrande, C.~Duhr, and B.~Fuks, \textit{FeynRules
  2.0 - A complete toolbox for tree-level phenomenology},
  \href{http://dx.doi.org/10.1016/j.cpc.2014.04.012}{{Comput. Phys. Commun.}
  {\bfseries 185} (2014) 2250--2300},
  \href{http://arxiv.org/abs/1310.1921}{{\ttfamily arXiv:1310.1921 [hep-ph]}}.

\bibitem{Alwall:2014hca}
J.~Alwall, R.~Frederix, S.~Frixione, V.~Hirschi, F.~Maltoni, O.~Mattelaer,
  H.~S. Shao, T.~Stelzer, P.~Torrielli, and M.~Zaro, \textit{The automated
  computation of tree-level and next-to-leading order differential cross
  sections, and their matching to parton shower simulations},
  \href{http://dx.doi.org/10.1007/JHEP07(2014)079}{{JHEP} {\bfseries 07} (2014)
  079}, \href{http://arxiv.org/abs/1405.0301}{{\ttfamily arXiv:1405.0301
  [hep-ph]}}.

\bibitem{Tanabashi:2018oca}
{\bfseries Particle Data Group} Collaboration, M.~Tanabashi {et~al.}, \textit{Review
  of Particle Physics (2019 update)},
  \href{http://dx.doi.org/10.1103/PhysRevD.98.030001}{{Phys. Rev.} {\bfseries
  D98}  (2018) 030001}.

\bibitem{Escribano:2015yup}
R.~Escribano, S.~Gonzàlez-Solís, P.~Masjuan, and P.~Sanchez-Puertas,
  \textit{$\eta^\prime$ transition form factor from space- and timelike experimental
  data}, \href{http://dx.doi.org/10.1103/PhysRevD.94.054033}{{Phys. Rev.}
  {\bfseries D94}  (2016) 054033},
\href{http://arxiv.org/abs/1512.07520}{{\ttfamily arXiv:1512.07520 [hep-ph]}}.

\bibitem{Lubicz:2017syv}
{\bfseries ETM} Collaboration, V.~Lubicz, L.~Riggio, G.~Salerno, S.~Simula, and
  C.~Tarantino, \textit{Scalar and vector form factors of $D \to \pi(K) \ell \nu$
  decays with $N_f=2+1+1$ twisted fermions},
  \href{http://dx.doi.org/10.1103/PhysRevD.96.054514,
  10.1103/PhysRevD.99.099902}{{Phys. Rev.} {\bfseries D96} (2017)
  054514}, \href{http://arxiv.org/abs/1706.03017}{{\ttfamily arXiv:1706.03017
  [hep-lat]}}.
[erratum: Phys. Rev.D99, 099902(2019)].

\bibitem{Bijnens:1994me}
J.~Bijnens, G.~Colangelo, G.~Ecker, and J.~Gasser, \textit{Semileptonic kaon
  decays},  {{2nd DAPHNE Physics Handbook: 315-389 (1994)}}
\href{http://arxiv.org/abs/hep-ph/9411311}{{\ttfamily arXiv:hep-ph/9411311
  [hep-ph]}}.
\newblock

\bibitem{Helo:2010cw}
J.~C. Helo, S.~Kovalenko, and I.~Schmidt, \textit{Sterile neutrinos in lepton number
  and lepton flavor violating decays},
  \href{http://dx.doi.org/10.1016/j.nuclphysb.2011.07.020}{{Nucl. Phys. B}
  {\bfseries 853} (2011) 80--104},
  \href{http://arxiv.org/abs/1005.1607}{{\ttfamily arXiv:1005.1607 [hep-ph]}}.

\bibitem{Abada:2013aba}
A.~Abada, A.~Teixeira, A.~Vicente, and C.~Weiland, \textit{Sterile neutrinos in
  leptonic and semileptonic decays},
  \href{http://dx.doi.org/10.1007/JHEP02(2014)091}{{JHEP} {\bfseries 02} (2014)
  091}, \href{http://arxiv.org/abs/1311.2830}{{\ttfamily arXiv:1311.2830
  [hep-ph]}}.

\bibitem{Gorishnii:1990vf}
S.~Gorishnii, A.~Kataev, and S.~Larin, \textit{The $O(\alpha^{3}_{s})$-corrections
  to $\sigma_{tot}(e^{+}e^{-}\rightarrow hadrons)$ and $\Gamma(\tau^{-}
  \rightarrow \nu_{\tau} + hadrons)$ in QCD},
  \href{http://dx.doi.org/10.1016/0370-2693(91)90149-K}{{Phys. Lett. B}
  {\bfseries 259} (1991) 144--150}.

\bibitem{SHiP:2018xqw}
{\bfseries SHiP} Collaboration, C.~Ahdida {et~al.}, \textit{Sensitivity of the SHiP
  experiment to Heavy Neutral Leptons},
  \href{http://dx.doi.org/10.1007/JHEP04(2019)077}{{JHEP} {\bfseries 04} (2019)
  077}, \href{http://arxiv.org/abs/1811.00930}{{\ttfamily arXiv:1811.00930
  [hep-ph]}}.

\bibitem{Abi:2020wmh}
{\bfseries DUNE} Collaboration, B.~Abi {et~al.}, \textit{Deep Underground Neutrino
  Experiment (DUNE), Far Detector Technical Design Report, Volume I
  Introduction to DUNE},
\href{http://arxiv.org/abs/2002.02967}{{\ttfamily arXiv:2002.02967
  [physics.ins-det]}}.

\bibitem{neardet}
T.~A. Mohayai, \textit{High-Pressure Gas TPC for DUNE near detector} (2018).  Available online, \href{https://zenodo.org/record/2642360#.YHWPUy0lP0q}{https://zenodo.org/record/2642360.YHWPUy0lP0q}

\bibitem{tilt}
J.~Martin-Albo.
\newblock Private communication.

\bibitem{Agostinelli:2002hh}
{\bfseries GEANT4} Collaboration, S.~Agostinelli {et~al.}, \textit{GEANT4--a
  simulation toolkit},
  \href{http://dx.doi.org/10.1016/S0168-9002(03)01368-8}{{Nucl. Instrum. Meth.
  A} {\bfseries 506} (2003) 250--303}.

\bibitem{Allison:2006ve}
J.~Allison {et~al.}, \textit{Geant4 developments and applications},
  \href{http://dx.doi.org/10.1109/TNS.2006.869826}{{IEEE Trans. Nucl. Sci.}
  {\bfseries 53} (2006) 270}.

\bibitem{Allison:2016lfl}
J.~Allison {et~al.}, \textit{Recent developments in Geant4},
  \href{http://dx.doi.org/10.1016/j.nima.2016.06.125}{{Nucl. Instrum. Meth. A}
  {\bfseries 835} (2016) 186--225}.

\bibitem{Sjostrand:2014zea}
T.~Sjöstrand, S.~Ask, J.~R. Christiansen, R.~Corke, N.~Desai, P.~Ilten,
  S.~Mrenna, S.~Prestel, C.~O. Rasmussen, and P.~Z. Skands, \textit{An introduction
  to PYTHIA 8.2}, \href{http://dx.doi.org/10.1016/j.cpc.2015.01.024}{{Comput.
  Phys. Commun.} {\bfseries 191} (2015) 159--177},
  \href{http://arxiv.org/abs/1410.3012}{{\ttfamily arXiv:1410.3012 [hep-ph]}}.

\bibitem{Abe:2019kgx}
{\bfseries T2K} Collaboration, K.~Abe {et~al.}, \textit{Search for heavy neutrinos
  with the T2K near detector ND280},
  \href{http://dx.doi.org/10.1103/PhysRevD.100.052006}{{Phys. Rev.} {\bfseries
  D100}  (2019) 052006},
\href{http://arxiv.org/abs/1902.07598}{{\ttfamily arXiv:1902.07598 [hep-ex]}}.

\bibitem{Feldman:1997qc}
G.~J. Feldman and R.~D. Cousins, \textit{A Unified approach to the classical
  statistical analysis of small signals},
  \href{http://dx.doi.org/10.1103/PhysRevD.57.3873}{{Phys. Rev. D} {\bfseries
  57} (1998) 3873--3889},
  \href{http://arxiv.org/abs/physics/9711021}{{\ttfamily
  arXiv:physics/9711021}}.

\bibitem{Britton:1992xv}
D.~Britton {et~al.}, \textit{Improved search for massive neutrinos in $\pi^+\to e^+$ neutrino decay}, \href{http://dx.doi.org/10.1103/PhysRevD.46.R885}{{Phys.
  Rev. D} {\bfseries 46} (1992) 885--887}.

\bibitem{Aguilar-Arevalo:2017vlf}
{\bfseries PIENU} Collaboration, A.~Aguilar-Arevalo {et~al.}, \textit{Improved
  search for heavy neutrinos in the decay $\pi\rightarrow e\nu$},
  \href{http://dx.doi.org/10.1103/PhysRevD.97.072012}{{Phys. Rev. D} {\bfseries
  97}  (2018) 072012}, \href{http://arxiv.org/abs/1712.03275}{{\ttfamily
  arXiv:1712.03275 [hep-ex]}}.

\bibitem{NA62:2020mcv}
{\bfseries NA62} Collaboration, E.~Cortina~Gil {et~al.}, \textit{Search for heavy
  neutral lepton production in $K^+$ decays to positrons},
  \href{http://arxiv.org/abs/2005.09575}{{\ttfamily arXiv:2005.09575
  [hep-ex]}}.

\bibitem{Bergsma:1985is}
{\bfseries CHARM} Collaboration, F.~Bergsma {et~al.}, \textit{A Search for Decays of
  Heavy Neutrinos in the Mass Range 0.5-{GeV} to 2.8-{GeV}},
  \href{http://dx.doi.org/10.1016/0370-2693(86)91601-1}{{Phys. Lett. B}
  {\bfseries 166} (1986) 473--478}.

\bibitem{CooperSarkar:1985nh}
{\bfseries WA66} Collaboration, A.~M. Cooper-Sarkar {et~al.}, \textit{Search for
  Heavy Neutrino Decays in the {BEBC} Beam Dump Experiment},
  \href{http://dx.doi.org/10.1016/0370-2693(85)91493-5}{{Phys. Lett. B}
  {\bfseries 160} (1985) 207--211}.

\bibitem{Abreu:1996pa}
{\bfseries DELPHI} Collaboration, P.~Abreu {et~al.}, \textit{Search for neutral
  heavy leptons produced in Z decays},
  \href{http://dx.doi.org/10.1007/s002880050370}{{Z. Phys. C} {\bfseries 74}
  (1997) 57--71}. [Erratum: Z.Phys.C 75, 580 (1997)].

\bibitem{Daum:1987bg}
M.~Daum, B.~Jost, R.~Marshall, R.~Minehart, W.~Stephens, and K.~Ziock,
  \textit{Search for Admixtures of Massive Neutrinos in the Decay $\pi^+ \to \mu^+$
  Neutrino}, \href{http://dx.doi.org/10.1103/PhysRevD.36.2624}{{Phys. Rev. D}
  {\bfseries 36} (1987) 2624}.

\bibitem{Aguilar-Arevalo:2019owf}
{\bfseries PIENU} Collaboration, A.~Aguilar-Arevalo {et~al.}, \textit{Search for
  heavy neutrinos in $\pi \to \mu\nu$ decay},
  \href{http://dx.doi.org/10.1016/j.physletb.2019.134980}{{Phys. Lett. B}
  {\bfseries 798} (2019) 134980},
  \href{http://arxiv.org/abs/1904.03269}{{\ttfamily arXiv:1904.03269
  [hep-ex]}}.

\bibitem{Bernardi:1985ny}
G.~Bernardi {et~al.}, \textit{Search for Neutrino Decay},
  \href{http://dx.doi.org/10.1016/0370-2693(86)91602-3}{{Phys. Lett. B}
  {\bfseries 166} (1986) 479--483}.

\bibitem{Bernardi:1987ek}
G.~Bernardi {et~al.}, \textit{Further Limits on Heavy Neutrino Couplings},
  \href{http://dx.doi.org/10.1016/0370-2693(88)90563-1}{{Phys. Lett. B}
  {\bfseries 203} (1988) 332--334}.

\bibitem{Kusenko:2004qc}
A.~Kusenko, S.~Pascoli, and D.~Semikoz, \textit{New bounds on MeV sterile neutrinos
  based on the accelerator and Super-Kamiokande results},
  \href{http://dx.doi.org/10.1088/1126-6708/2005/11/028}{{JHEP} {\bfseries 11}
  (2005) 028}, \href{http://arxiv.org/abs/hep-ph/0405198}{{\ttfamily
  arXiv:hep-ph/0405198}}.

\bibitem{Artamonov:2009sz}
{\bfseries BNL-E949} Collaboration, A.~Artamonov {et~al.}, \textit{Study of the
  decay $K^+\to\pi^+\nu \bar\nu$ in the momentum region $140 < P_\pi < 199$
  MeV/c}, \href{http://dx.doi.org/10.1103/PhysRevD.79.092004}{{Phys. Rev. D}
  {\bfseries 79} (2009) 092004},
  \href{http://arxiv.org/abs/0903.0030}{{\ttfamily arXiv:0903.0030 [hep-ex]}}.

\bibitem{Vaitaitis:1999wq}
{\bfseries NuTeV, E815} Collaboration, A.~Vaitaitis {et~al.}, \textit{Search for
  neutral heavy leptons in a high-energy neutrino beam},
  \href{http://dx.doi.org/10.1103/PhysRevLett.83.4943}{{Phys. Rev. Lett.}
  {\bfseries 83} (1999) 4943--4946},
  \href{http://arxiv.org/abs/hep-ex/9908011}{{\ttfamily arXiv:hep-ex/9908011}}.

\bibitem{Orloff:2002de}
J.~Orloff, A.~N. Rozanov, and C.~Santoni, \textit{Limits on the mixing of tau
  neutrino to heavy neutrinos},
  \href{http://dx.doi.org/10.1016/S0370-2693(02)02769-7}{{Phys. Lett. B}
  {\bfseries 550} (2002) 8--15},
  \href{http://arxiv.org/abs/hep-ph/0208075}{{\ttfamily arXiv:hep-ph/0208075}}.

\bibitem{Aker:2019uuj}
{\bfseries KATRIN} Collaboration, M.~Aker {et~al.}, \textit{Improved Upper Limit on
  the Neutrino Mass from a Direct Kinematic Method by KATRIN},
  \href{http://dx.doi.org/10.1103/PhysRevLett.123.221802}{{Phys. Rev. Lett.}
  {\bfseries 123}  (2019) 221802},
  \href{http://arxiv.org/abs/1909.06048}{{\ttfamily arXiv:1909.06048
  [hep-ex]}}.

\bibitem{Dolgov:2000pj}
A.~Dolgov, S.~Hansen, G.~Raffelt, and D.~Semikoz, \textit{Cosmological and
  astrophysical bounds on a heavy sterile neutrino and the KARMEN anomaly},
  \href{http://dx.doi.org/10.1016/S0550-3213(00)00203-0}{{Nucl. Phys. B}
  {\bfseries 580} (2000) 331--351},
  \href{http://arxiv.org/abs/hep-ph/0002223}{{\ttfamily arXiv:hep-ph/0002223}}.

\bibitem{Ruchayskiy:2012si}
O.~Ruchayskiy and A.~Ivashko, \textit{Restrictions on the lifetime of sterile
  neutrinos from primordial nucleosynthesis},
  \href{http://dx.doi.org/10.1088/1475-7516/2012/10/014}{{JCAP} {\bfseries 10}
  (2012) 014}, \href{http://arxiv.org/abs/1202.2841}{{\ttfamily arXiv:1202.2841
  [hep-ph]}}.

\bibitem{Gelmini:2020ekg}
G.~B. Gelmini, M.~Kawasaki, A.~Kusenko, K.~Murai, and V.~Takhistov, \textit{Big Bang
  Nucleosynthesis constraints on sterile neutrino and lepton asymmetry of the
  Universe}, \href{http://arxiv.org/abs/2005.06721}{{\ttfamily
  arXiv:2005.06721 [hep-ph]}}.

\bibitem{Sabti:2020yrt}
N.~Sabti, A.~Magalich, and A.~Filimonova, \textit{An Extended Analysis of Heavy
  Neutral Leptons during Big Bang Nucleosynthesis},
  \href{http://dx.doi.org/10.1088/1475-7516/2020/11/056}{{JCAP} {\bfseries 11}
  (2020) 056}, \href{http://arxiv.org/abs/2006.07387}{{\ttfamily
  arXiv:2006.07387 [hep-ph]}}.

\bibitem{Boyarsky:2020dzc}
A.~Boyarsky, M.~Ovchynnikov, O.~Ruchayskiy, and V.~Syvolap, \textit{Improved BBN
  Constraints on Heavy Neutral Leptons},
  \href{http://arxiv.org/abs/2008.00749}{{\ttfamily arXiv:2008.00749
  [hep-ph]}}.

\bibitem{webmesons}
\url{http://pdg.lbl.gov/2019/tables/rpp2019-sum-mesons.pdf}.

\bibitem{Maris:1999nt}
P.~Maris and P.~C. Tandy, \textit{Bethe-Salpeter study of vector meson masses and
  decay constants},
  \href{http://dx.doi.org/10.1103/PhysRevC.60.055214}{{Phys. Rev.} {\bfseries
  C60} (1999) 055214},
\href{http://arxiv.org/abs/nucl-th/9905056}{{\ttfamily arXiv:nucl-th/9905056
  [nucl-th]}}.

\bibitem{Aoki:2019cca}
{\bfseries Flavour Lattice Averaging Group} Collaboration, S.~Aoki {et~al.},
  \textit{FLAG Review 2019},
\href{http://arxiv.org/abs/1902.08191}{{\ttfamily arXiv:1902.08191 [hep-lat]}}.

\end{thebibliography}

\providecommand{\href}[2]{#2}\begingroup\raggedright\endgroup

\end{document}